\documentclass[12pt,preprint]{aastex61}
\usepackage{graphicx}
\usepackage{natbib}

\newcommand{\twCO}{$^{12}$CO}  \newcommand{\thCO}{$^{13}$CO}

\newcommand{\co}{$^{12}$CO}                             
\newcommand{\xco}{$^{13}$CO}                            
\newcommand{\xxco}{C$^{18}$O}                           
\newcommand{\kms}{km\,s$^{\rm -1}$}
\newcommand{\kkms}{K\,km\,s$^{\rm -1}$}

\begin{document}

\title{Molecular clouds in the extreme outer Galaxy between $l=34.75^{\circ}$ to $45.25^{\circ}$}

\correspondingauthor{Yan Sun}
\email{yansun@pmo.ac.cn}
\author{Yan Sun}
\affil{Purple Mountain Observatory and Key Laboratory of Radio Astronomy,
Chinese Academy of Sciences, Nanjing 210008, China}

\author[0000-0002-0197-470X]{Yang Su}
\affil{Purple Mountain Observatory and Key Laboratory of Radio Astronomy,
Chinese Academy of Sciences, Nanjing 210008, China}

\author{Shao-Bo Zhang}
\affil{Purple Mountain Observatory and Key Laboratory of Radio Astronomy,
Chinese Academy of Sciences, Nanjing 210008, China}

\author{Ye Xu}
\affil{Purple Mountain Observatory and Key Laboratory of Radio Astronomy,
Chinese Academy of Sciences, Nanjing 210008, China}

\author{Xue-Peng Chen}
\affil{Purple Mountain Observatory and Key Laboratory of Radio Astronomy,
Chinese Academy of Sciences, Nanjing 210008, China}

\author{Ji Yang}
\affil{Purple Mountain Observatory and Key Laboratory of Radio Astronomy,
Chinese Academy of Sciences, Nanjing 210008, China}

\author{Zhi-Bo Jiang}
\affil{Purple Mountain Observatory and Key Laboratory of Radio Astronomy,
Chinese Academy of Sciences, Nanjing 210008, China}

\author{Min Fang}
\affil{Purple Mountain Observatory and Key Laboratory of Radio Astronomy,
Chinese Academy of Sciences, Nanjing 210008, China}



\begin{abstract}
We present the result of an unbiased CO survey in Galactic range 
of 34.75$^{\circ}\leq$$l$$\leq$ 45.25$^{\circ}$ and
-5.25$^{\circ}\leq$$b$$\leq$ 5.25$^{\circ}$, and velocity range beyond the Outer arm. 
A total of 168 molecular clouds (MCs) are identified within the Extreme Outer Galaxy~(EOG) 
region, and 31 of these MCs are associated with \xco~ emission. However, none of them 
show significant \xxco~ emission under current detection limit. The typical size and mass of these MCs 
are 5~pc and 3$\times$10$^{\rm 3}M_{\sun}$, implying the lack of large and massive MCs in the EOG region.
Similar to MCs in the outer Galaxy, the velocity dispersions of EOG clouds are also correlated with 
their sizes, however, are well displaced below the scaling relationship defined by the inner 
Galaxy MCs. These MCs with a median Galactocentric 
radius of 12.6 kpc, show very different distributions from those of the MCs in the Outer arm published in
our previous paper, while roughly follow the Outer Scutum-Centaurus arm defined by \citet{2011ApJ...734L..24D}. 
This result may provide a robust evidence for the existence of the Outer Scutum-Centaurus arm. 
The lower limit of the total mass of this segment is about 2.7$\times$10$^5$ $M_{\sun}$, which is about
one magnitude lower than that of the Outer arm. 
The mean thickness of gaseous disk is about 1.45$^{\circ}$ or 450~pc, and the scale height is about 
1.27$^{\circ}$ or 400~pc above the $b=0^{\circ}$ plane. 
The warp traced by CO emission is very obvious in the EOG region and its amplitude is consistent with 
the predictions by other warp models using different tracers, such as dust, HI and stellar components of 
our Galaxy.
\end{abstract}


\section{INTRODUCTION~\label{sec:intro}}
The Extreme Outer Galaxy (EOG), is defined as the region of the Milky Way at Galactic radii ($R$) 
greater than 2$R_{0}$~($R_{0}$ is the distance from the Sun to the Galactic center), 
or simply as the region outside the Outer arm~\citep{1994ApJ...422...92D}.
Molecular clouds~(MCs) in the EOG regions not only delineate the spiral structure 
and warping of our Galaxy, but they also serve as an excellent 
laboratory for studying the star-formation process in a physical environment that 
is very different from that of the solar neighborhood~\citep[e.g.,][]{1996ApJ...458..653R,2000ApJ...532..423K,
2008ApJ...683..178K,2006ApJ...649..753Y,2008ApJ...675..443Y,2014ApJ...795...66I}. 
The first few detections of EOG clouds were the results of a targeted CO survey
towards the HI emission peaks in the EOG region in the second Galactic quadrant~\citep{1994ApJ...422...92D}.
A few more EOG clouds were revealed by the subsequent unbiased CO survey of the 
Outer Galaxy using Five College Radio Astronomy Observatory 14 meter 
telescope~\citep{1998ApJS..115..241H, 2003ApJS..144...47B}.

Unlike the second quadrant with few unbiased survey therein, many surveys were carried out 
in the first quadrant of the Galaxy~\citep[e.g.,][]{1985ApJ...297..751D,1986ApJ...305..892D,1986ApJS...60..297C,
1988ApJ...327..139C,1987ApJS...63..821S,1988A&A...195...93J,1989ApJ...339..919S, 2006ApJS..163..145J,
2010ApJ...723..492R}. However, limited by 
the sensitivity, the latitude coverage, and the velocity coverage of previous
CO surveys, all those studies were merely concentrated on the inner Galaxy. 
Therefore, the CO emission lying beyond the solar circle in the first quadrant 
of the Galaxy is less studied, let alone beyond the Outer arm~\citep{2001ApJ...547..792D,
2015ARA&A..53..583H}. 
It was not until 2011, however, that the Outer Scutum-Centaurus~(hereafter Outer Scu-Cen) arm 
was found by \cite{2011ApJ...734L..24D}
lying beyond the Outer arm in the first Galactic quadrant. Ten EOG clouds~(with brightness 
temperature ranging from 0.11 to 0.47~K) were detected in the 
interval of 13$^{\circ}<$~$l$~$<$ 55$^{\circ}$ in the new CO observations with the CfA 1.2 m telescope towards poorly 
resolved clumps in the HI~\citep{2011ApJ...734L..24D}. 

Applying different identification and analysis methods to the same data sets leads to 
different MCs catalogs. Using a Dendrogram technique and 
a hierarchical cluster identification method applied to the \citet{2001ApJ...547..792D} CfA 1.2 m data, 
\citet{2016ApJ...822...52R} and \citet{2017ApJ...834...57M} recently obtained a catalog of 1064 massive MCs and 
a catalog of 8107 MCs, respectively. These catalogs provide the most complete distribution of Milky Way MCs. 
Based on the l-v distribution of these MCs, none of the 1064 massive MCs lies beyond the Outer arm,  
and about ten out of 8107 MCs probably lie beyond the Outer arm in the first quadrant.    

Because of the far distances involved and the warp of the Galactic plane
from the solar radius outwards~\citep[e.g.,][]{1957AJ.....62...90B,1958MNRAS.118..379O,
1982ApJ...263..116H}, an unbiased Galactic plane CO survey with the combination of high-sensitivity, 
wide latitude coverage, and wide velocity coverage is essential for the discovery of more EOG clouds.
The Milky Way Imaging Scroll Painting (MWISP) project is a high resolution ($50''$) $J$ = 1-0
\co, \xco, and \xxco~ survey of the northern Galactic Plane, performed with the Purple
Mountain Observatory Delingha 13.7 m telescope. The survey started in 2011, and will
cover Galactic longitudes from -10.25$^{\circ}$ to 250.25$^{\circ}$ and latitudes from
-5.25$^{\circ}$ to 5.25$^{\circ}$ over a period of $\sim$10 years. The high-sensitivity data
from MWISP survey will provide us a unique opportunity to systematically study the spiral 
structures and star-formation activities at the edge of the Milky Way. Using our new CO data of 
the MWISP, we have found dozens of new EOG clouds, which delineate a new segment of a spiral 
arm between Galactocentric radii of 15 and 19 kpc in the second Galactic quadrant~\citep{2015ApJ...798L..27S}.

In the first Galactic quadrant, the MWISP survey so far has completely covered the regions 
between 34.75$^{\circ}\leq~l~\leq$ 45.25$^{\circ}$ and -5.25$^{\circ}\leq~b~\leq$ 5.25$^{\circ}$. 
The mapping at least covered emission from four arms in this direction, including the  
Scu-Cen tangent, Sagittarius, Perseus, Outer, and Outer Scu-Cen arms that mentioned above. 
The results of the Outer arm within the 
velocity interval of -1.6 $l$ + 13.2 $\le$ $V_{\rm LSR}$ $\le$ -10~\kms~ in this Galactic  
range have been presented by \citet{2016ApJ...828...59S}. 
The Outer Scu-Cen arm lying beyond the Outer arm
is the outermost spiral arm, in other words, represents the edge
of the Galactic plane in this direction.
Here, we report on the results of emission from this arm 
according to the distribution of the CO gas between $l = 34.75^{\circ}$ to $45.25^{\circ}$, 
$b = -5.25^{\circ}$ to $5.25^{\circ}$.
First, the observations and data reduction are described in Sect. 2, and in Sect. 3.1 and 3.2
the identification, and parameters of the EOG clouds are presented.  
Next, in Sect. 3.3 the revised scaling relations of Larson are examined in 
the EOG clouds. In Sect. 3.4, the distributions of the EOG clouds are explored. In Sect. 3.4, 
the spiral Arm  and warp of the Galactic plane traced by CO emission
are presented. In Sect. 4 a summary of the results is given. 

\section{CO OBSERVATIONS AND DATA REDUCTION}
The observations were conducted during November 2011 to
March 2015 using the 13.7~m millimeter-wavelength telescope of the Purple Mountain 
Observatory~(PMO) in Delingha, China.
The nine-beam Superconducting Spectroscopic Array Receiver (SSAR) 
system was used as front end, and each Fast Fourier transform 
spectrometer (FFTS) with a bandwidth of 1 GHz provides 16384 channels 
and a spectral resolution of 61 kHz \citep[see the details in][]{Shan}.
The molecular lines of \twCO\ ($J$=1--0) in the upper sideband, and \thCO\ ($J$=1--0) 
and C$^{18}$O ($J$=1--0) in the lower sideband were observed simultaneously. 
Typical system temperature is
150-200 K in the lower sideband and 250-300 K in the upper sideband.

Regions within the Galactic range of 34.75$^{\circ}\leq$ $l$ $\leq$ 45.25$^{\circ}$, 
and -5.25$^{\circ}\leq b \leq$ 5.25$^{\circ}$ are divided into 439 patches in ($l$, $b$) 
grid for mapping, each with a size of 30$\arcmin$$\times$30$\arcmin$. The on-the-fly~(OTF) mode was applied
with typical sample steps of 10$''$-15$''$. The OTF raw data are re-gridded to a FITS cube with pixel 
size of 30$''$ using the GILDAS software package\footnote{http://www.iram.fr/IRAMFR/GILDAS}. 
The nominal sensitivity in the survey is set for 0.3 K in \xco~ and \xxco~ at the resolution of 0.17~\kms, and 0.5 K in 
\co~ at the resolution of 0.16~\kms. Each patch with a dimension 30$'\times30'$
was scanned at least in two orthogonal directions, along the Galactic 
longitude and the Galactic latitude, in order to reduce the fluctuation 
of noise. The half-power beam width (HPBW) of the telescope
was about $50''$ at 115 GHz and the pointing accuracy was greater
than $5''$ in the observing epoch.
It should be noted that any results presented in the figures and tables are on the 
brightness temperature scale (T$_{\rm R}^∗$), corrected for beam efficiencies, 
using $T_{\rm mb}$ = $T_{\rm A}^{\rm *}$ /$\eta_{\rm mb}$, where $\eta_{\rm mb}$ is typically 
0.51 for \xco~ and \xxco~, and 0.46 for \co~(see the status report of the 
telescope\footnote{http://www.radioast.nsdc.cn/mwisp.php}).

\section{RESULTS AND DISCUSSION}
\subsection{Molecular Clouds Identification}

\begin{figure}
\includegraphics[angle=0,scale=0.8]{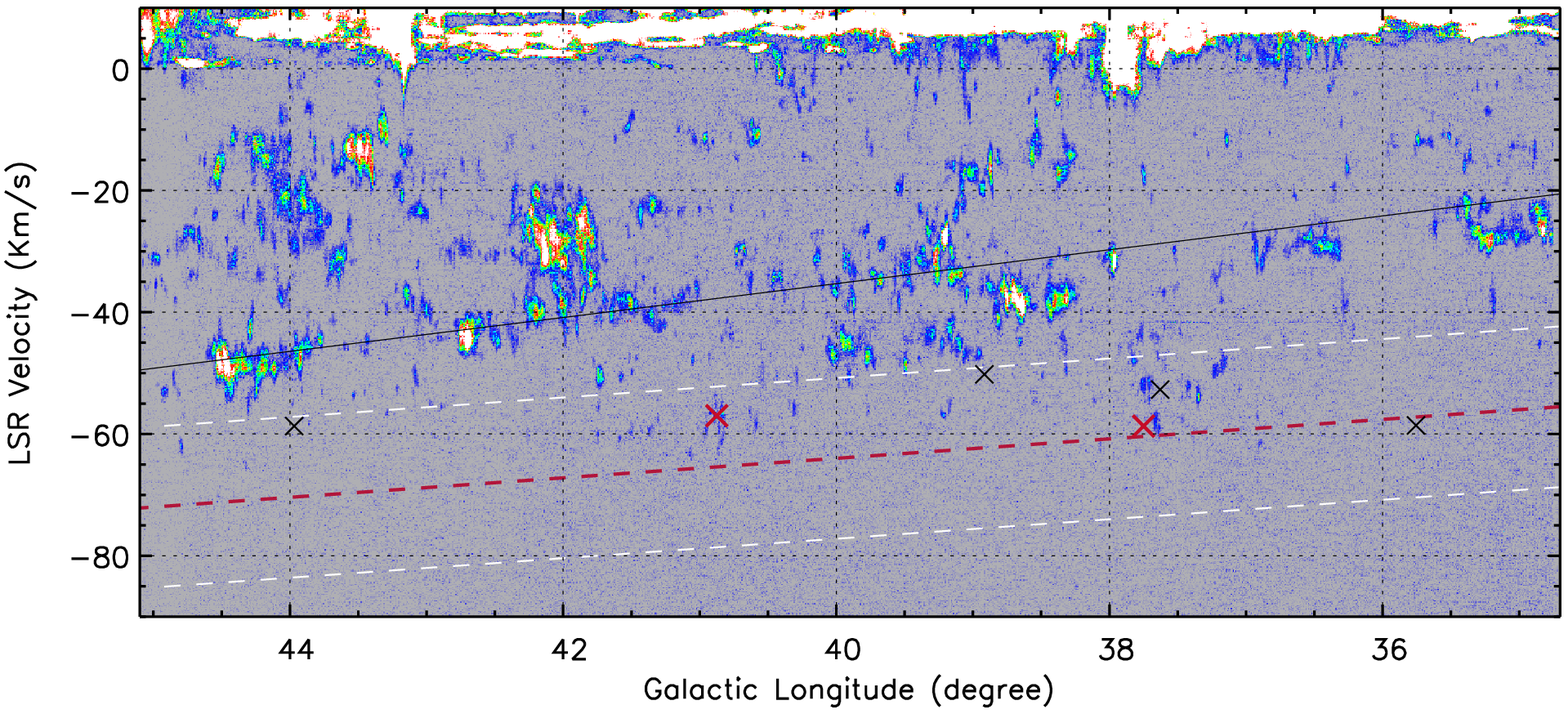}
\caption{Longitude-velocity diagram of CO, integrated over latitudes from
-1.5$^{\circ}$ to 3.2$^{\circ}$. The black and red lines indicate the longitude-velocity
relations of the Outer arm~\citep{2016ApJ...828...59S} and the Outer Scu-Cen 
arm~\citep{2011ApJ...734L..24D}, respectively. The white dashed lines indicate
the velocity range of the Outer Scu-Cen arm adopted by us. The crosses~
\citep[red and black, ][]{2011ApJ...734L..24D, 2017ApJ...834...57M} indicate the previously 
known EOG clouds in this velocity range.\label{fig:lv1}} 
\end{figure}

Longitude-velocity diagram of the CO emission beyond the Solar circle is shown 
in Figure~\ref{fig:lv1}. Note that the integrated latitude range is from -1.5$^{\circ}$ to 3.2$^{\circ}$, 
since the CO emission from the Outer Galaxy is concentrated in
this latitude range. In addition, only those pixels with at least 3 continuous 
channels above 3$\sigma$ are averaged. 
The black-solid line~\citep[$V_{\rm LSR}$ = -2.78~$l$ + 75.9~\kms;][]{2016ApJ...828...59S} 
and red-dashed line~\citep[$V_{\rm LSR}$ = -1.6~$l$~\kms;][]{2011ApJ...734L..24D} indicate the longitude-velocity
relations of the Outer arm and the Outer Scu-Cen arm, respectively. The white-dashed lines~($V_{\rm LSR}$ = 
$-1.6~l \pm 13.2$~\kms) indicate the assumed velocity range of the Outer Scu-Cen arm.
Only six EOG clouds indicated by red crosses~\citep{2011ApJ...734L..24D} 
and black crosses~\citep{2017ApJ...834...57M} in Figure~\ref{fig:lv1} are known MCs in this velocity range. 
However, our high-sensitivity CO data reveal much more 
features distributed beyond the Outer arm at a maximum negative velocity. 

Firstly, we compiled an automatic procedure to list all the positions with emission greater than 
0.92 K($\sim$2$\sigma$) and minimum number of adjacent pixels of 12 in the ($l$, $b$, $V$) three-dimensional~(3D) 
\co~ J = 1-0 data cube within velocity range of $-1.6~l \pm 13.2$~\kms. Then, we checked both 
the list and data cube and identified the clouds by eye. Here we should point out that we did 
not separate the isolated MCs into small pieces of molecular clumps. In other words, some of the 
MCs have more than one significant emission peak, only the strongest emission peak was 
chosen for the position of the cloud, which was the case for less than 20 per cen of the MCs.  

\begin{figure}
\includegraphics[angle=0,scale=0.8]{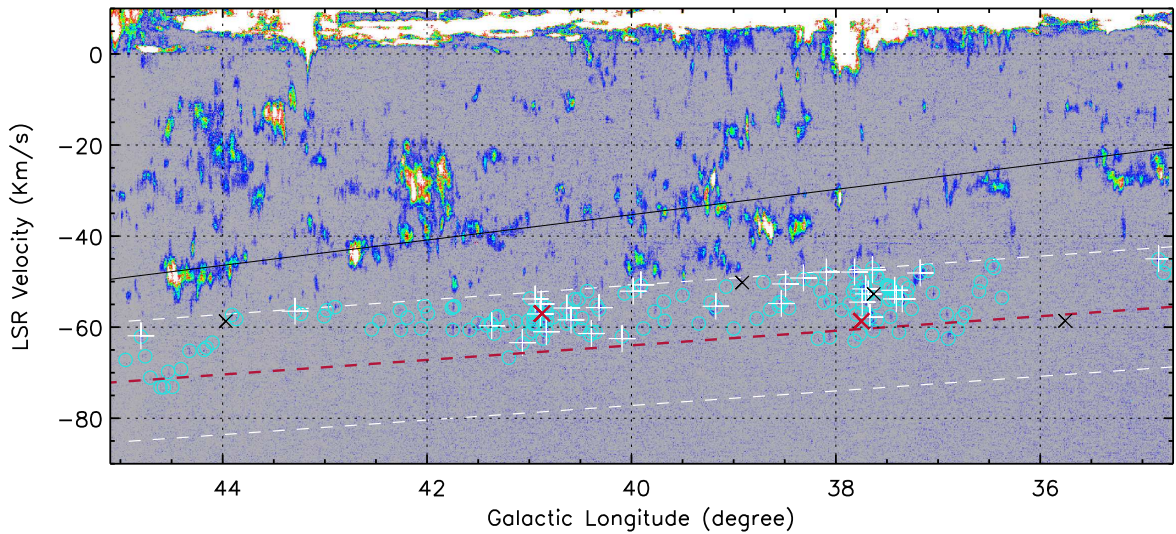}
\caption{Same as Fig.~\ref{fig:lv1}. The circles and pluses mark the MCs 
identified by MWISP survey in \co~ and \xco, respectively.\label{fig:lv2}} 
\end{figure}

Finally, we detected 174 MCs, and the peak emissions of them are marked with circles 
in the longitude-velocity diagram~(Figure~\ref{fig:lv2}). Besides, 23 other MCs are 
marginally detected.  
We note that 168 out of the 174 MCs 
with the $V_{\rm LSR} <$ -1.6 $l$ + 13.2~\kms~ might be located beyong the Outer arm, 
the remaining six clouds with $V_{\rm LSR} >$ -1.6 $l$ + 13.2~\kms~ 
might be located in the Outer arm or in the inter-arm. Seven of the 174 clouds were 
also cataloged by \citet{2016ApJ...828...59S}, but were identified by CLUMPFIND algorithm. 
Since the algorithm is different from this study, the seven sources are also
cataloged. 
It is worthwhile to note that the two known EOG clouds identified 
by \citet{2011ApJ...734L..24D} are also detected by us. 
Condering the beam size of the CfA 1.2 m telescope and the size of the MCs, three out 
of the four EOG clouds identified by \citet{2017ApJ...834...57M}
are also detected by our new observations, the other one is marginally detected
by us.

We also checked \xco~ and \xxco~ emissions in these MCs. 
Apparent \xco~ emissions were detected in 25 of these clouds. Once we average the images over an 
$\pm$30'' spatial extent of the emission peak to improve sensitivity, 35 dense clumps of 
34 MCs showed \xco~ emission. 31 of them with $V_{\rm LSR} <$ -1.6 $l$ + 13.2~\kms~ might 
be located beyond the Outer arm. None of them show significant \xxco~ emission under current 
detection limit. 

\begin{figure}
\includegraphics[angle=0,scale=0.49]{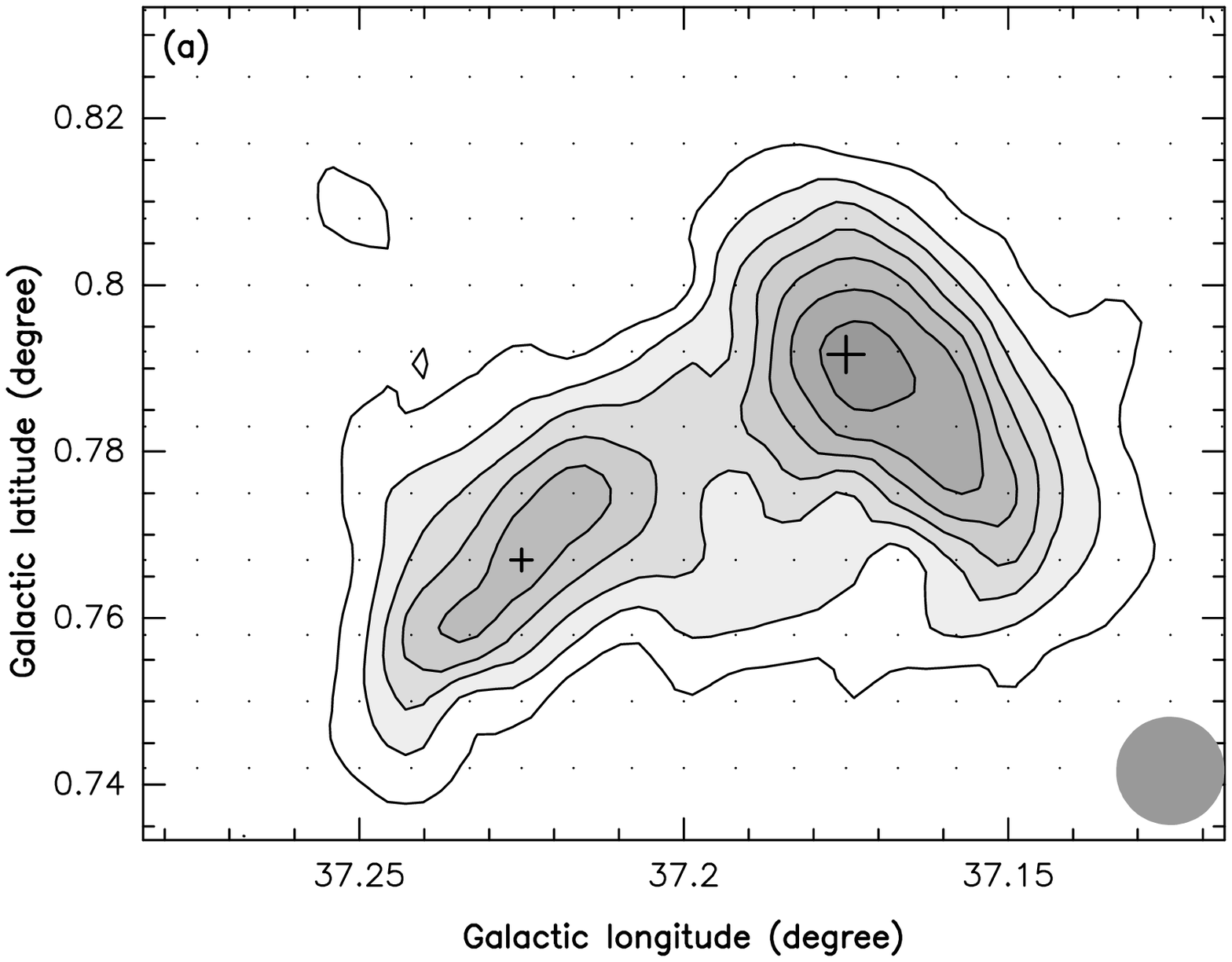}
\includegraphics[angle=0,scale=0.42]{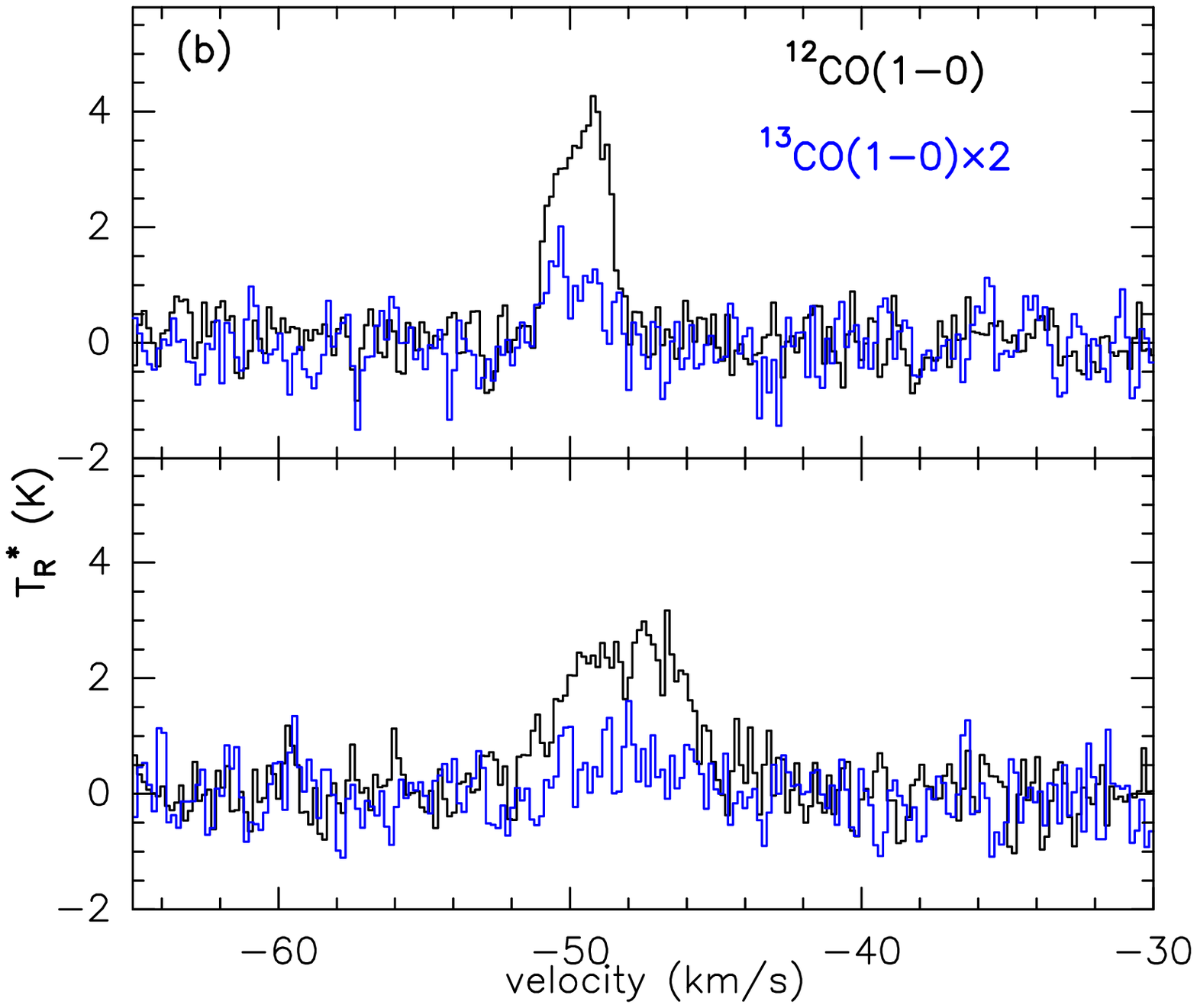}
\caption{(a) Velocity-integrated intensity of MC MWISP~37.175$+$0.792,  
obtained by integrating CO emission from -52.5 to -44.6 \kms.
Contour levels start at 3$\sigma$ and increase by 3$\sigma$ at each interval.
The beam-size is indicated on the right hand side. 
(b) Spectra of the two emission peaks that marked with pluses in panel (a). 
\label{fig:cloud}}
\end{figure}

As an example, Figure~\ref{fig:cloud} presents the integrated intensity map of one MC, MWISP~G37.175$+$0.792, 
which shows two emission peaks. The lowest contour level is 3$\sigma$ and increases in step of 3$\sigma$.
\co~ and \xco~ spectra of the two emission peaks that marked with pluses in Figure~\ref{fig:cloud}a, 
are plotted as black and blue histograms, respectively. The intensity of \xco~ was 
multiplied by a factor of two for a better comparison. We find that the structures 
of the MC can be partially resolved by our current observation with spatial resolution 
of about 1$\arcmin$.

\subsection{Molecular Clouds Parameters}
Table~\ref{tab:12co} displays a summary of the 174 MCs that identified in this study.
The first column of the table contains the name of each MC, designated as MWISP 
(for CO survey project name) followed by the Galactic longitude and latitude ($l$ and $b$) 
coordinates of the peak of the emission in degrees, e.g., MWISP~G37.175$+$0.792.
Cols. 2-5 provide the single Gaussian fitting results of the brightest CO spectrum of the MCs, 
including LSR velocity ($V_{\rm LSR}$), line width FWHM ($\Delta$$V$), integrated intensity ($I_{\rm ^{12}CO}$), 
and peak intensity ($T_{\rm peak}$). The angular area of the MCs defined by the 3$\sigma$ limits is given 
in Col. 6. 

We also estimated the physical parameters of these clouds. Distance is one of the most 
important parameters for determining the properties of the MCs, the spiral structure and warping 
of the Galactic plane. 
Since the extreme outer region of the Galactic plane is rarely 
studied in particularly in the first Galactic quadrant, none of the MCs have parallax or 
photometry distances, we then adopt kinematic distance in this study.  

Fortunately, the Outer Scu-Cen arm lies beyond the solar circle,                          
thus the molecular gas traced by CO emission does not suffer the 
kinematic distance ambiguity encountered in the inner Galaxy.
The derived kinematic distance depends upon the choice of rotation curve and 
Solar motion parameters. We compared models of \citet{2014ApJ...783..130R}~(hereafter 
Reid model) and \citet{1993A&A...275...67B}, which represent two often used models for kinematic distance.
Generally, the derived kinematic distances of these MCs are insensitive to the exact 
choice of rotation curve, and are about 1~kpc~(or 5 per cen) smaller on the assumption of Reid model. 
In this study, the rotation curve and Solar motion parameters of Model A5 of \citet{2014ApJ...783..130R} 
($\Theta_0$ = 240~\kms, $R_0$ = 8.34~kpc, $ \frac{d\Theta}{dR}=-0.2$ km $\rm s^{-1}$ $\rm kpc^{-1}$,
$ U_{\odot} = 10.7$ km $\rm s^{-1}$, $ V_{\odot} = 15.6$ km $\rm s^{-1}$, $ W_{\odot} = 8.9$ km $\rm s^{-1}$,
$ \overline{U}_{s} = 2.9$ km $\rm s^{-1}$, and $ \overline{V}_{s} = -1.6$ km $\rm s^{-1}$),
are used to assign heliocentric and Galactocentric distances to each MC~(Cols. 7 and 8).
The vertical scale height $Z$ is defined by $Z = d~sin~(b)$~(Col. 9).  
The cloud radius given in Col. 10 is obtained after beam deconvolution, 
$r = d~\sqrt{\frac{A}{\pi} - \frac{\theta^{2}_{\mathrm{MB}}}{4}}$, 
where $A$ is the angular size of a cloud, and $\theta_{\mathrm{MB}}$ is the main beam width. 
The mass of a cloud obtained by adopting a constant CO-to-H$_{\rm 2}$ X conversion factor of 2.0$\times$10$^{\rm 20}$ 
cm$^{\rm -2}{\rm (K\,km\,s^{-1})^{-1}}$~\citep{2013ARAA..51..207B}, is listed in Col. 11. To evaluate the role of 
self-gravity, we also calculated the virial mass~(Col. 12), $M_{\rm vir} = 210~R~\Delta$$V^2$~\citep{1992ApJ...395..140B}. 
 The molecular gas surface density~(Col. 13), $\Sigma$, of a cloud is simply the $H_{\rm 2}$ mass 
divided by the projected area, $\Sigma = M/(\pi$$r^{\rm 2})$.
Note that the beam-filling factor is assumed to be 1.

Table~\ref{tab:margin} displays a summary of the 23 MCs that marginally detected 
in this study. A deeper observation is needed to confirm and/or to better constrain their properties.
Table~\ref{tab:13co} summarizes the measured parameters of \xco~ emissions, including 
the Gaussian fitting results of \xco~($V_{\rm LSR}$, $\Delta$$V$, and $I_{\rm ^{13}CO}$), 
the integrated intensity ratio between \co~ and \xco~($I_{\rm ^{12}CO}/I_{\rm ^{13}CO}$), 
and the rms level of each detection with a velocity resolution of 0.166 \kms. Note 
that those with rms values larger than 0.2 K are the rms value of a single pixel, 
otherwise the rms was averaged over nine pixels.

\subsection{Properties of the Molecular Clouds in the Extreme Outer Galaxy}

\begin{figure}
\includegraphics[angle=0,scale=0.37,bb=30 0 460 360]{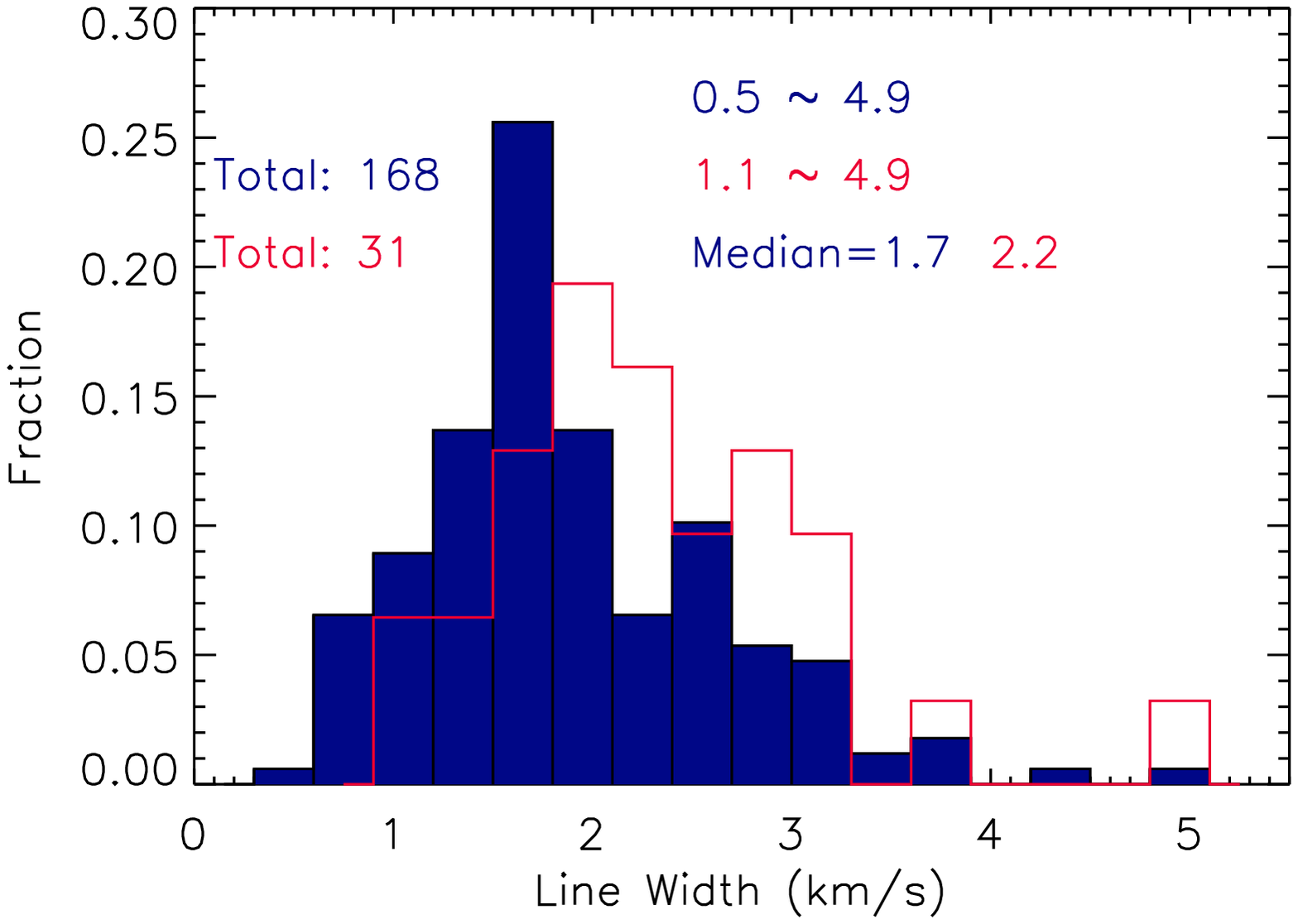}%
\includegraphics[angle=0,scale=0.37,bb=30 0 460 360]{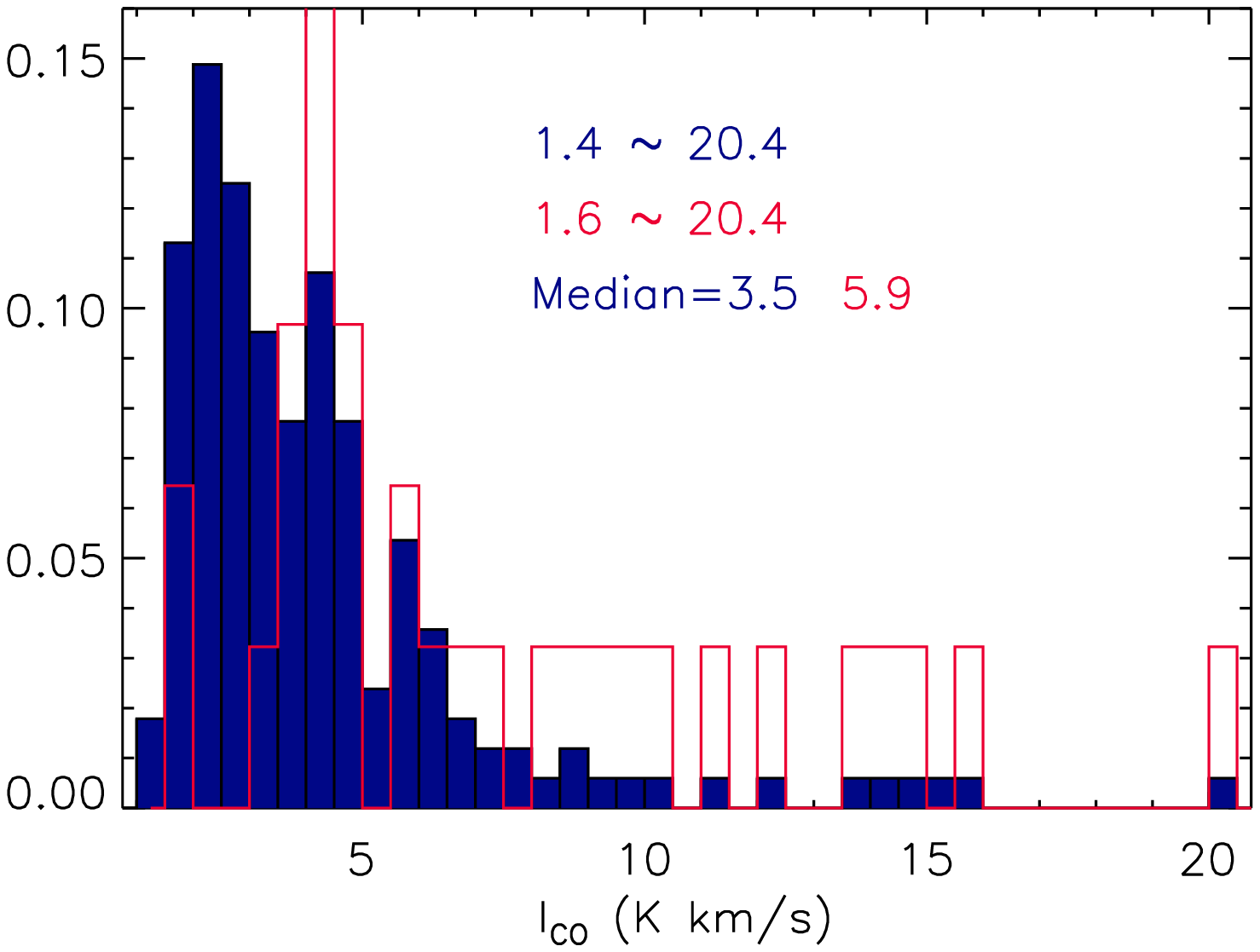}%
\includegraphics[angle=0,scale=0.37,bb=30 0 460 360]{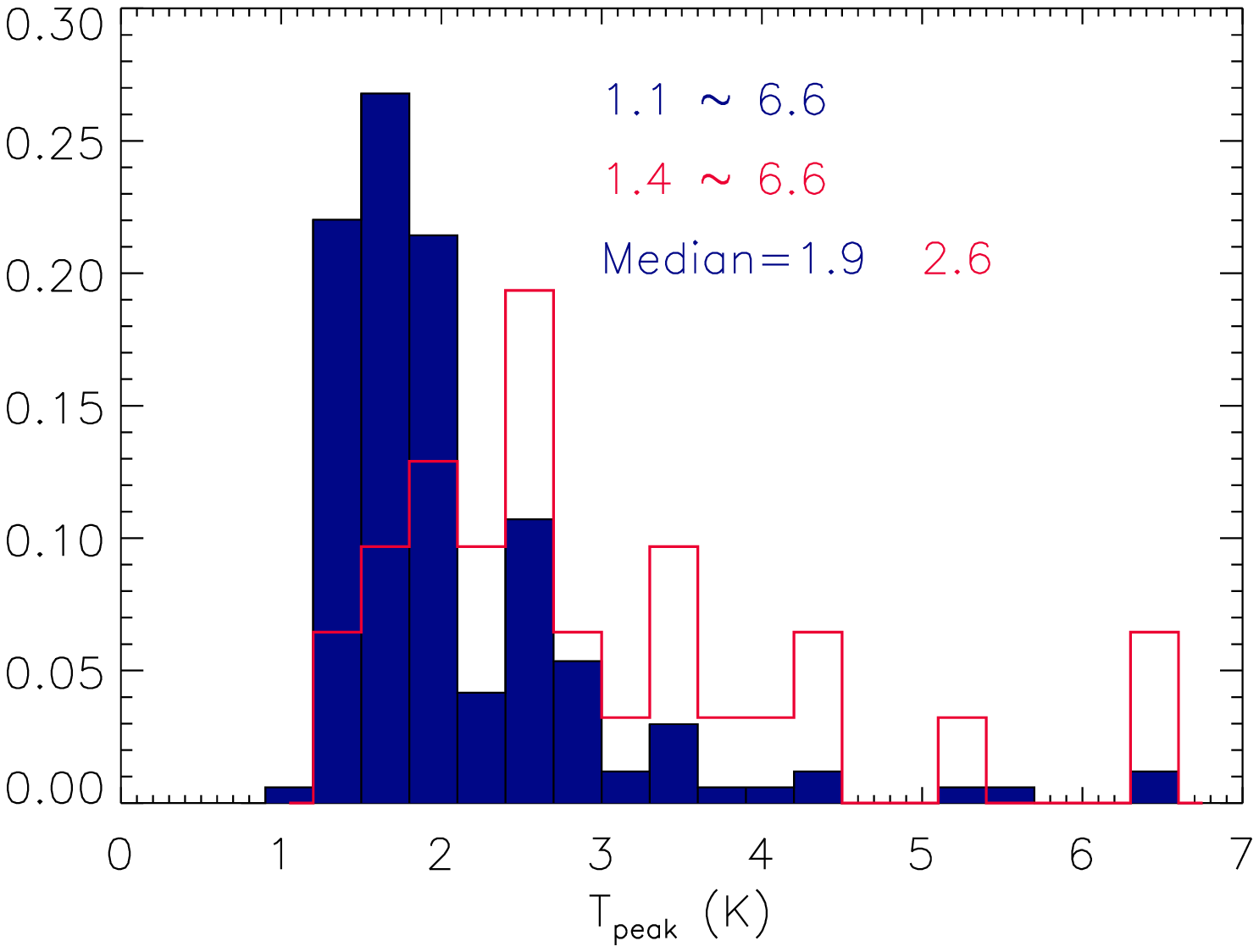}\\
\includegraphics[angle=0,scale=0.37,bb=30 0 460 360]{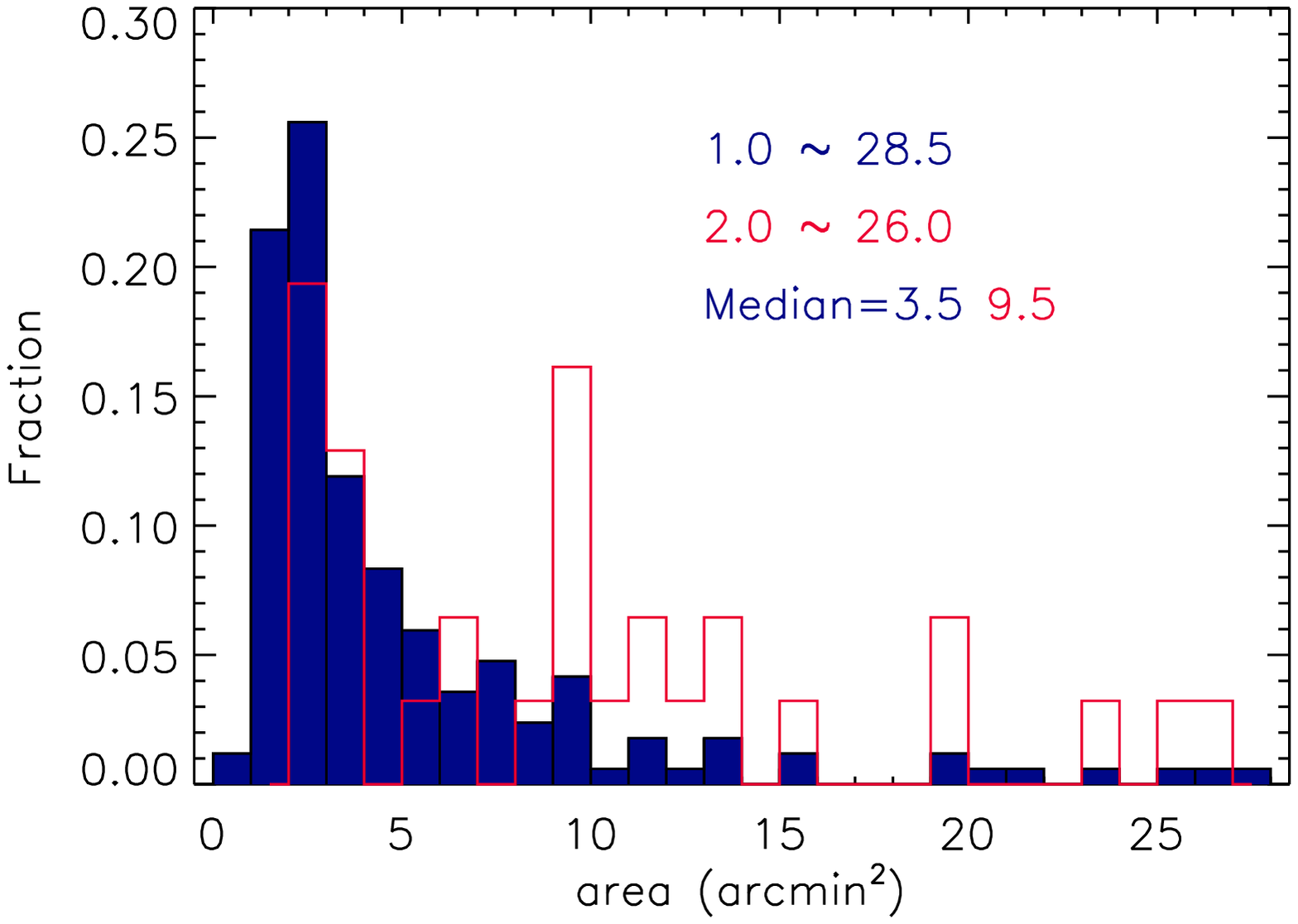}%
\includegraphics[angle=0,scale=0.37,bb=30 0 460 360]{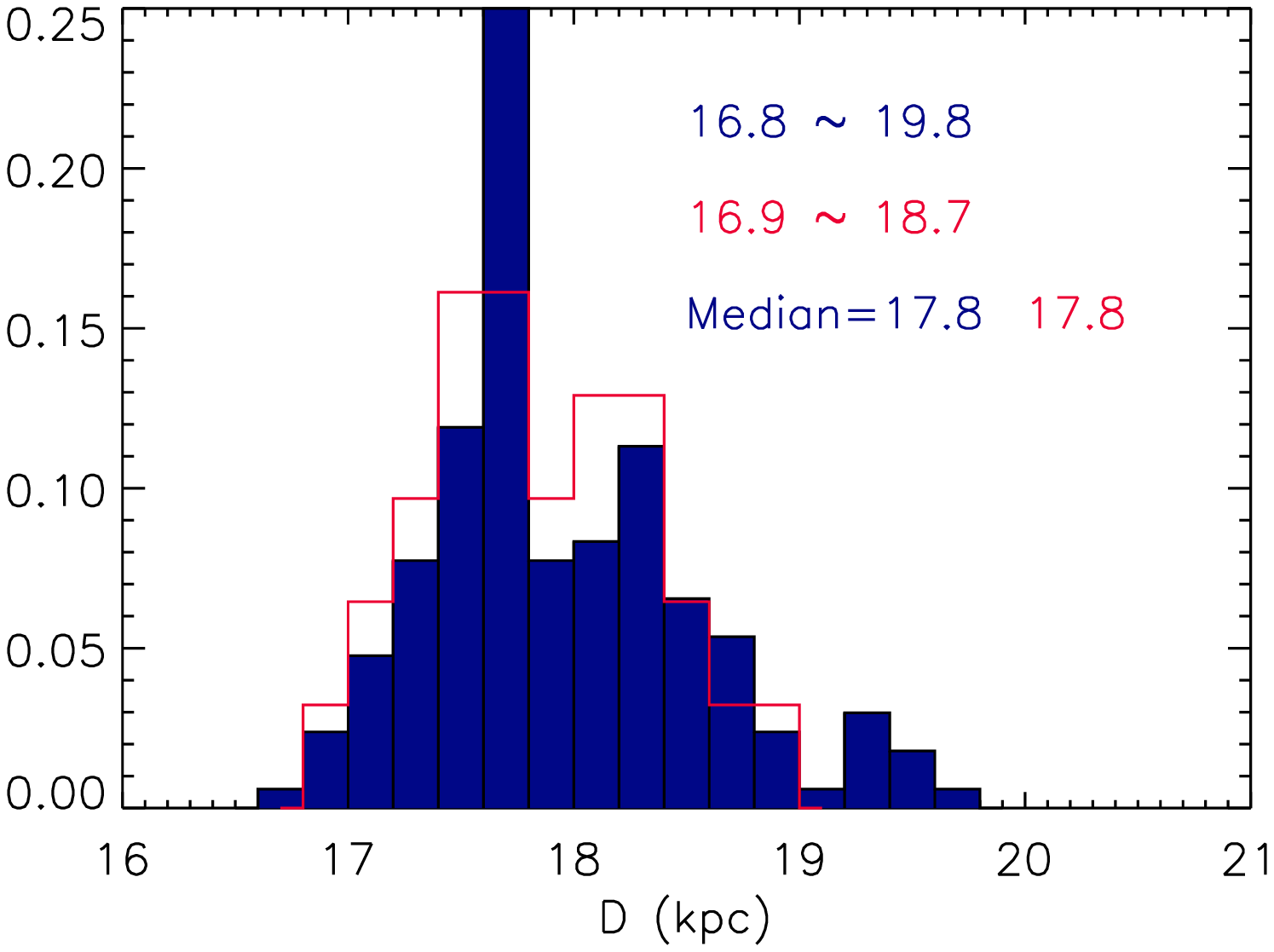}%
\includegraphics[angle=0,scale=0.37,bb=30 0 460 360]{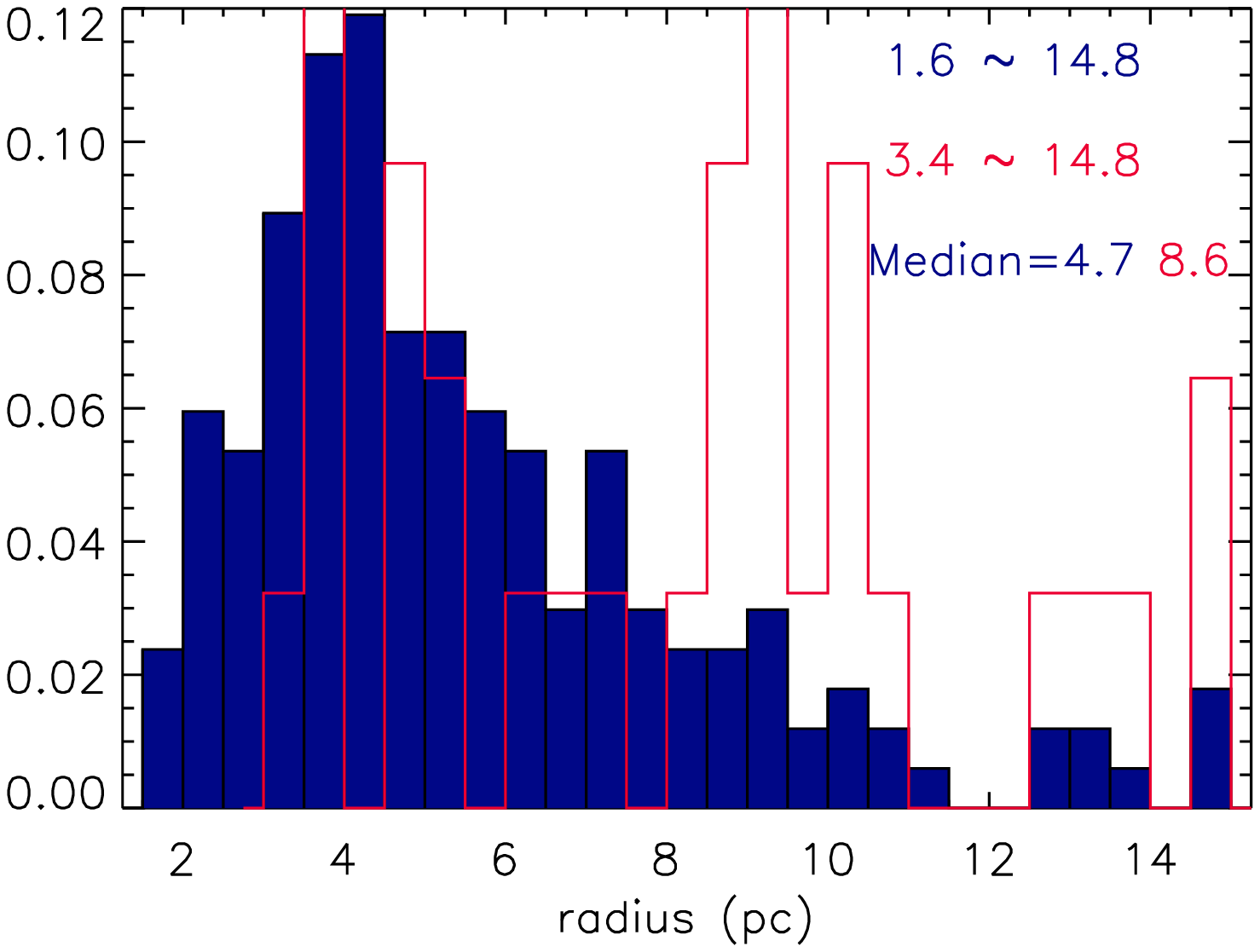}\\
\includegraphics[angle=0,scale=0.37,bb=30 0 460 360]{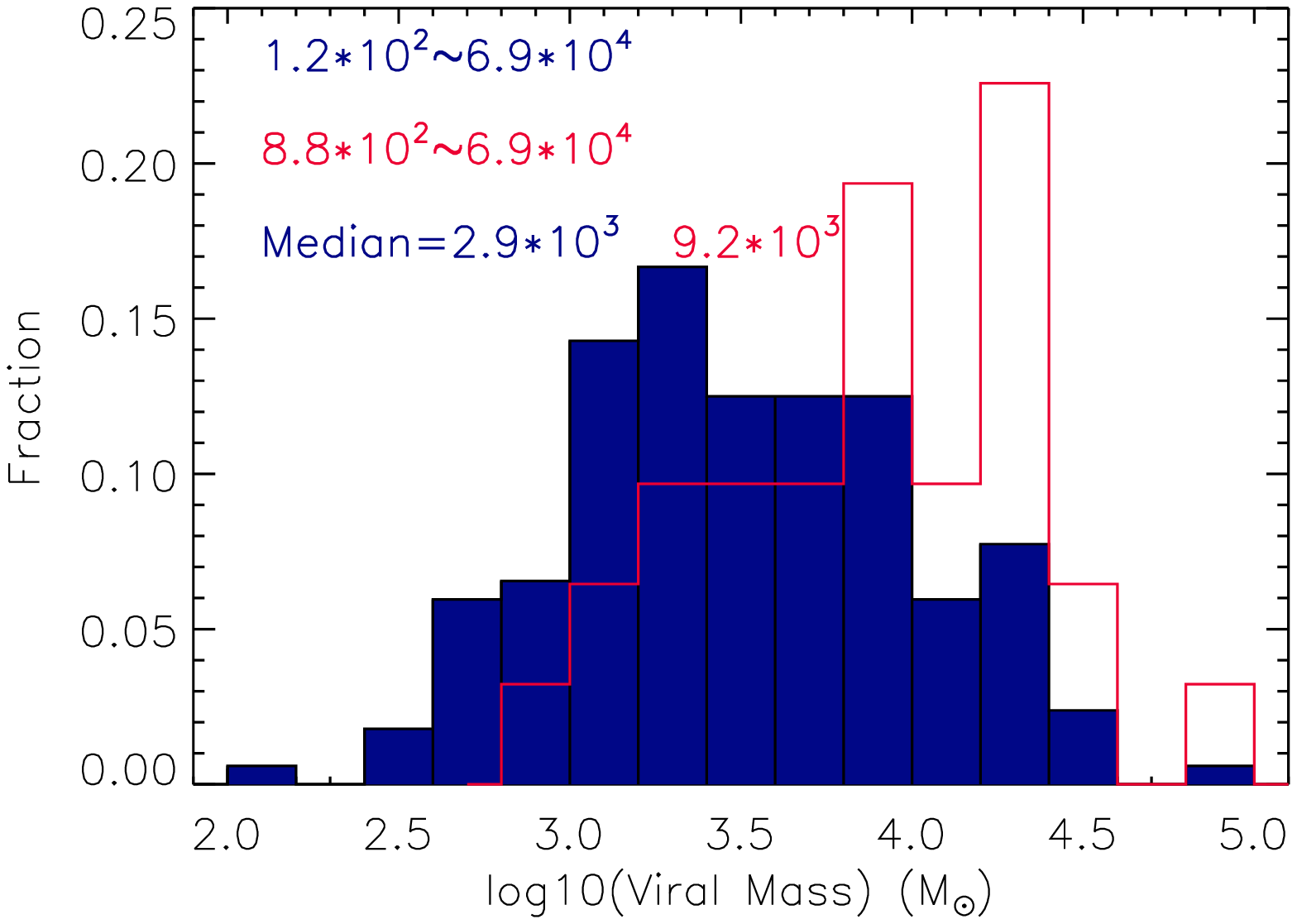}%
\includegraphics[angle=0,scale=0.37,bb=30 0 460 360]{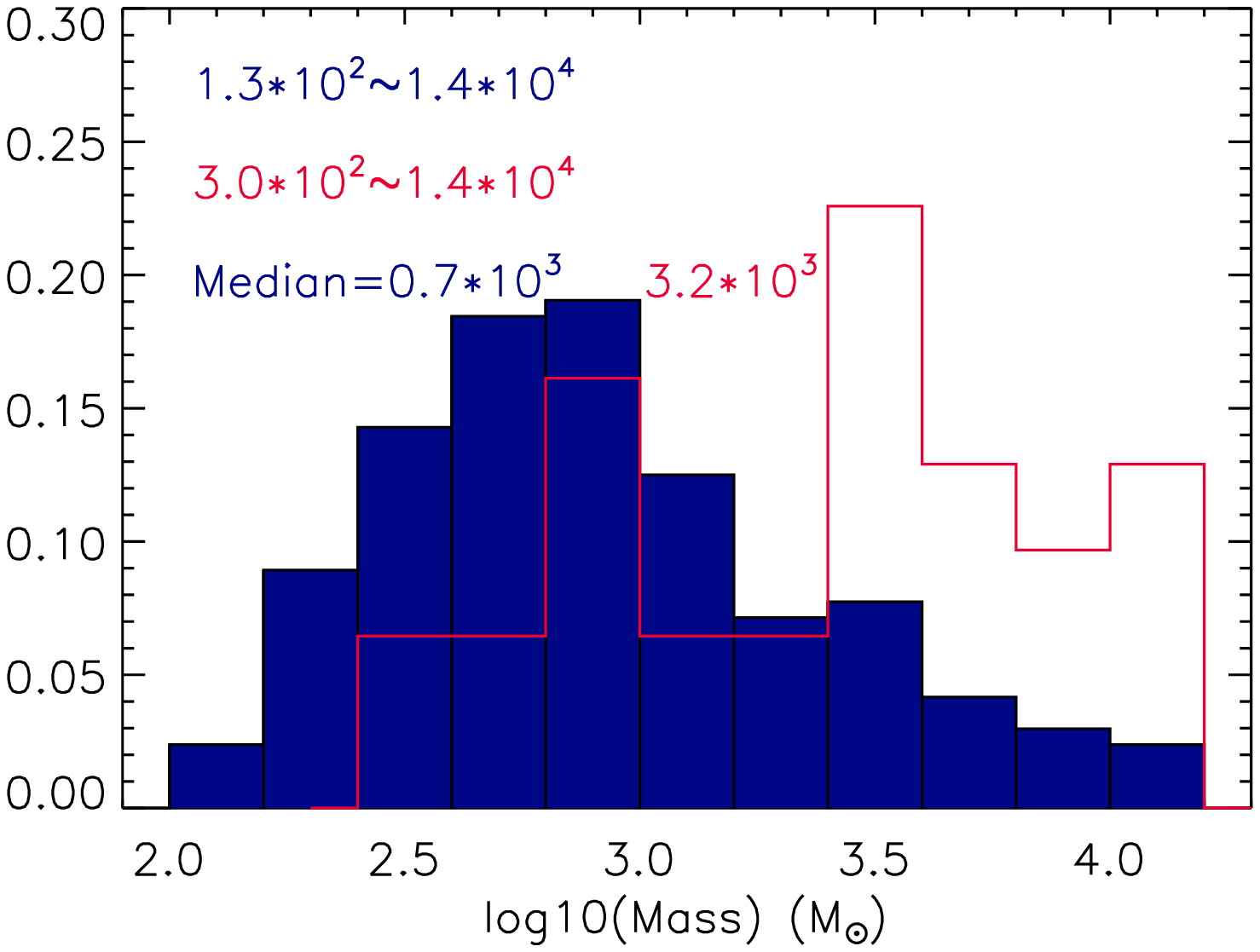}%
\includegraphics[angle=0,scale=0.37,bb=30 0 460 360]{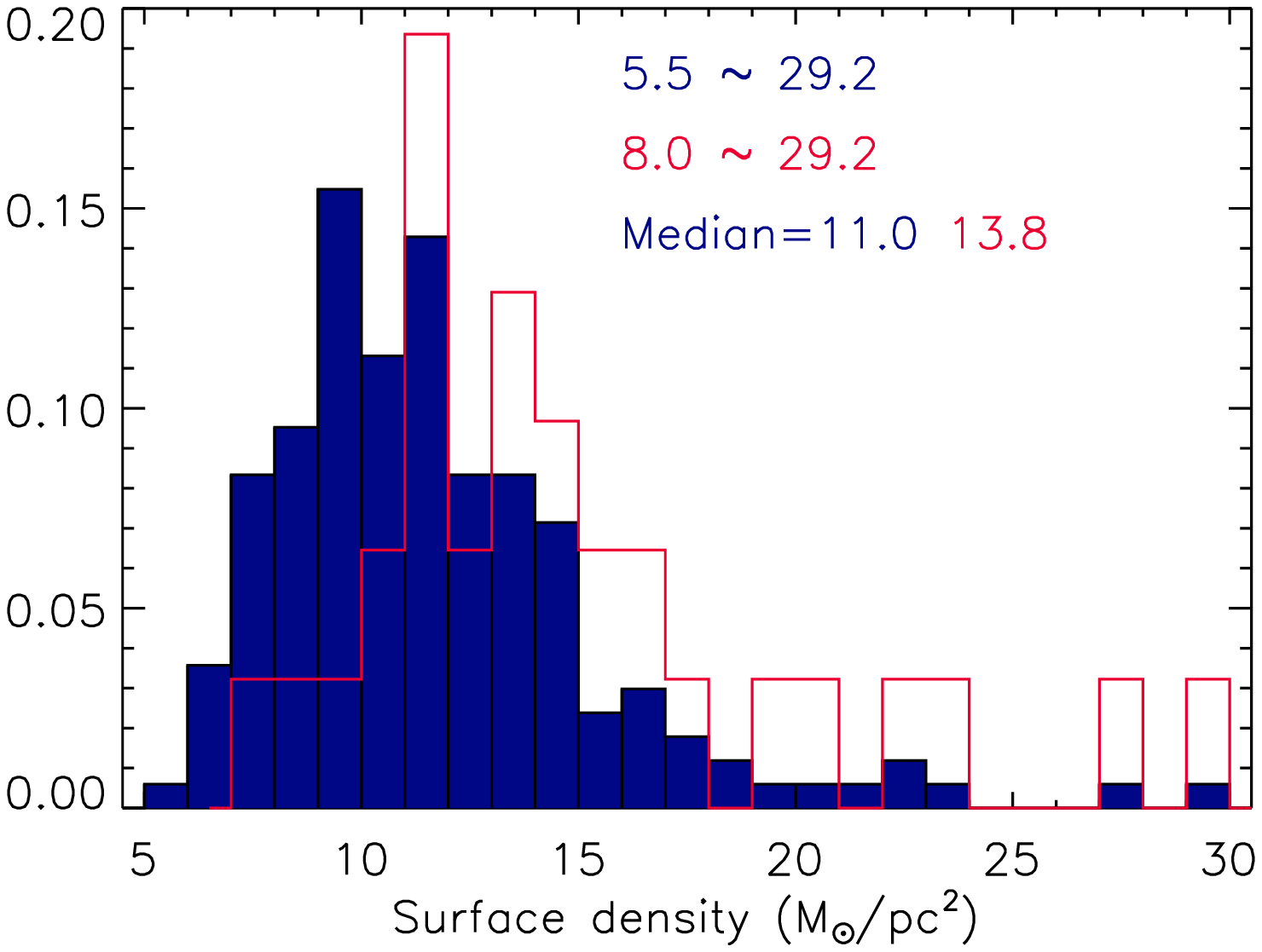}
\caption{Histograms in blue indicate the normalized number distributions of the EOG clouds, including Gaussian 
fitting results towards the emission peak of each cloud (line width, integrated intensity, and $T_{\rm peak}$), 
area of the clouds, heliocentric distance, equivalent radius, viral mass, mass derived from $X=2.0\times10^{20}$, 
and surface gas density. Histograms in red indicate the distributions of the EOG  
clouds with the presence of \xco~ emissions. The range and median values of each 
property are marked on each plot. \label{fig:property}}
\end{figure}

Only these clouds that might lie beyond the Outer arm are discussed in the below sections.
The blue histograms in Figure~\ref{fig:property} show the normalized number distributions for 
the measured \co~ parameters, including the Gaussian fitting results of the emission peak (line width, 
integrated intensity $I_{\rm ^{12}CO}$ and $T_{\rm peak}$), and 
angular size of the EOG clouds, which yield typical parameters of 1.7~\kms, 3.5~\kkms, 1.9~K, 
and 3.5~arcmin$^{\rm 2}$, respectively. The derived properties of these clouds are also plotted in Figure~\ref{fig:property},
including the heliocentric distance, the equivalent radius of the EOG clouds, virial mass, mass derived from
$X$-factor, and surface gas density, which yield typical properties of 17.8~kpc, 
4.7~pc, 2.9$\times$10$^{\rm 3}$ $M_{\sun}$, 0.7$\times$10$^{\rm 3}$ $M_{\sun}$, 
and 11.0~$M_{\sun}{\rm pc}^{\rm -2}$, respectively. At a median distance of 17.8~kpc, these are the most
distant detections of MCs in this direction between $l=34.75^{\circ}$ to $45.25^{\circ}$.
The fact that the sizes and masses of these clouds range from
1.6~pc to 14.8~pc, and 1.3$\times$10$^2$ $M_{\sun}$ to 1.4$\times$10$^4$ $M_{\sun}$ implies the
lack of large and massive clouds in the EOG regions.
The median gas surface density of these clouds obtained here is similar to that of MCs at similar
Galactocentric radius \citep{2017ApJ...834...57M}, however, is about half
value of the Outer arm \citep{2016ApJ...828...59S}.
In addition, it appears that the viral mass is much larger than the mass obtained by $X$ 
factor in the EOG region.
Observations of \citet{1991ApJ...366...95S} in the outer Galaxy also show similar trends.
The subsequent studies by \citet{1992A&A...257..715F} and \citet{2001ApJ...551..852H}
also found virial masses are much larger than CO-derived masses for the clouds with mass less 
than $\sim$1,000 $M_{\sun}$ in the outer Galaxy.

 The distributions for those clouds with \xco~ detections are also overlaid with 
red histograms in each panel in Fig.~\ref{fig:property} for comparisons. The ranges 
and median values of each property are also marked. 
We find that those clouds with the presence of \xco~ emissions are generally
brighter and more massive. The ratios between \co~ and \xco~ are not discussed
in depth, since there are considerable uncertainties in the beam-filling factors for the
two molecular lines. The ratios with median and mean values of 6.1 and 6.5 represent
the upper limits for EOG clouds. Future observations with higher resolution are essential for further study.

The three empirical scaling relations of \citet{1981MNRAS.194..809L} can guide our understanding of the 
internal structure and dynamics of molecular clouds as well as the fundamental nature of the ISM 
environment. These relations were revised by \citet{2009ApJ...699.1092H}. 
They found that equating the virial mass to the mass of the cloud implies that 
$\sigma_v$$ = (\pi$$G/5)^{\rm 1/2}R_{\rm MC}^{\rm 1/2}\Sigma_{\rm MC}^{\rm 1/2}$, where $G$ is the gravitational 
constant, $R_{\rm MC}$ is the cloud radius, and $\Sigma_{\rm MC}$ is the cloud surface density.
This expression implies a scaling exponent of 1/2 for the line width-size 
relationship for constant mass surface density~\citep{2009ApJ...699.1092H}.  

\begin{figure}
\includegraphics[angle=0,scale=0.5,bb=10 0 490 360]{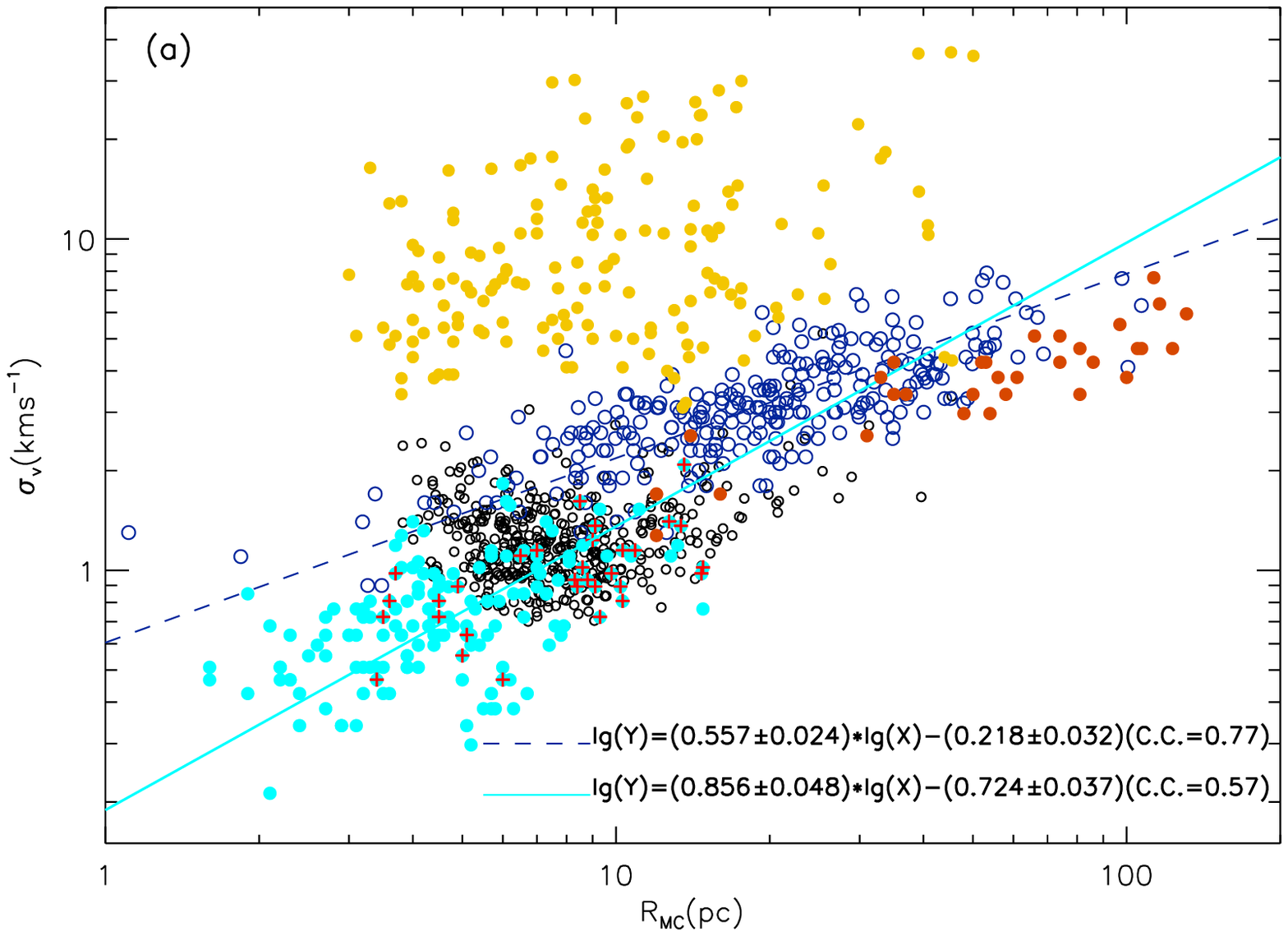}%
\includegraphics[angle=0,scale=0.5,bb=10 0 490 360]{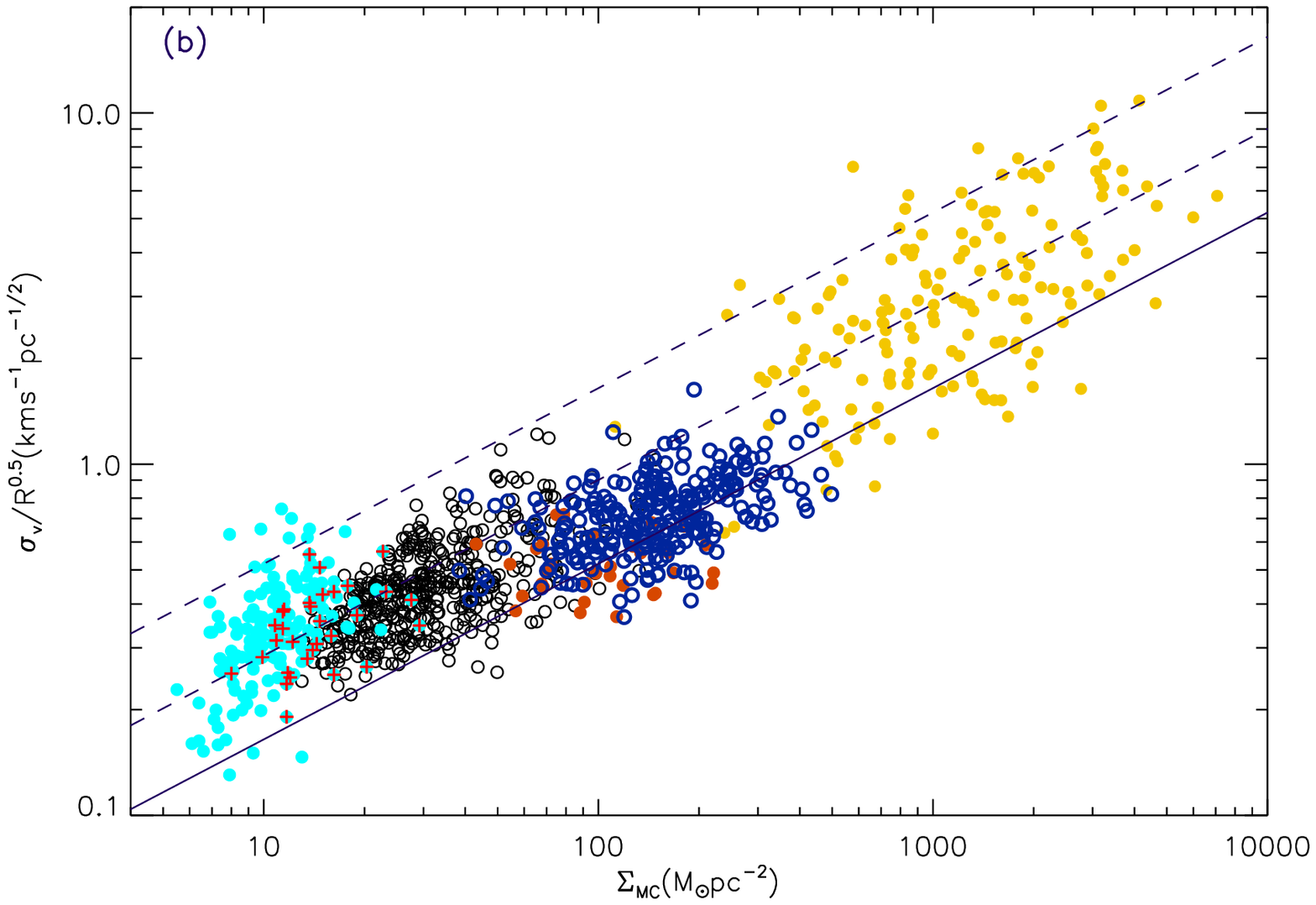}
\caption{(a) Velocity dispersion $\sigma_v$ as a function of size for 
clouds located in EOG region~(cyan, this work), Galactic Center \citep[gold, ][]{2001ApJ...562..348O},
Galactic ring \citep[red and blue, ][]{1987ApJ...319..730S,1986ApJ...305..892D}, and outer 
Galaxy \citep[black, ][]{2001ApJ...551..852H}. The dashed and solid lines indicate the linear fitting results 
to the \citet{1987ApJ...319..730S} points and the EOG clouds, respectively. 
(b) Scaling coefficient, $\sigma_v$/$R^{\rm 1/2}$, as a function of the mass surface density.
 The solid line and two dashed lines show the loci for 
$\alpha_{vir}$=1~(lower), 3~(middle) and 10~(upper), respectively. In both panel $(a)$ and $(b)$, the pluses 
mark the EOG clouds with \xco~ detections.\label{fig:heyer}}
\end{figure}

Similar to \citet{2015ARA&A..53..583H}, the revised scaling Larson relations are examined and 
compared across a wide range of Galactic environments in Figure~\ref{fig:heyer}. 
The molecular cloud velocity dispersion $\sigma_v$ as a function of size for
clouds located in EOG region~(cyan, this work), Galactic Center \citep[gold, ][]{2001ApJ...562..348O},
Galactic ring \citep[red and blue, ][]{1987ApJ...319..730S,1986ApJ...305..892D}, and outer
Galaxy \citep[black, ][]{2001ApJ...551..852H} is shown in Figure~\ref{fig:heyer}a. 
The dashed line indicates a linear fitting to the inner Galaxy clouds from \citet{1987ApJ...319..730S}, 
computed using the least squares bisector. The power index is 0.557$\pm$0.024, and the 
Pearson correlation coefficient of relation is 0.77. 
We find that the EOG clouds are well displaced below the scaling relationship 
defined by the inner Galaxy molecular clouds. The results
is similar to the finding in the outer Galaxy clouds~\citep{2015ARA&A..53..583H}.
The best fitting result of $R_{\rm MC}$-$\sigma_v$ relation for the EOG clouds with a power index 
of 0.856$\pm$0.048 is indicated by solid line in Fig.~\ref{fig:heyer}a.
The Pearson correlation coefficient of this relation is 0.57. 
Generally, the clouds in the outer and extreme outer parts of our Galaxy have a shallower 
$R_{\rm MC}$-$\sigma_v$ relation, which is reflected in the larger uncertainties on the fitting 
results and the relatively lower Pearson correlation coefficient than clouds in the inner Galactic plane. 
In addition, the power index is much steeper for clouds in outer and extreme outer parts of our Galaxy.
This is compatible with the analysis of \citet{2017ApJ...834...57M}, who found that the
scaling coefficient of the $R_{\rm MC}$-$\sigma_v$ relation decreases towards the outer Galaxy (middle panel of their Fig. 13).
The observed variation of the power index of different samples of MCs simply reflects
plausible differences of the amplitude of the turbulent motions between clouds owing to 
Galactic environments~\citep{2017ApJ...834...57M}.

It is important to note that althought these EOG clouds in this study are well detected, but their
properties (e.g., cloud size) are not well determined due to the relatively poor spatial resolution
and sensitivity of our current observations.
The radii might be overestimated for the small EOG clouds, at the lower value side in Fig.~\ref{fig:heyer}a, 
where the radii are approaching our spatial resolution limit~($\sim$2~pc, at a median distance of 17.8~kpc).  
On the other hand, limited by our current detection limit of $\sim$10$^{\rm 2}$ $M_{\sun}$ in 
the EOG regions, the radii might be underestimated for the 
large EOG clouds~(with angular size much larger than our beam size), at the higher value 
side in Fig~\ref{fig:heyer}a. 
Therefore, the power index only represents an upper limit for the EOG clouds.
Future observations with improved resolution and sensitivity are essential for more accurate
$R_{\rm MC}$-$\sigma_v$ relation for MCs in the EOG regions. 

The $R_{\rm MC}$-$\sigma_v$ relation shows significant dispersion when clouds from different 
environments are compiled together, as revealed in Figure \ref{fig:heyer}a in this study and Figure 9 of 
\citet{2015ARA&A..53..583H}. \citet{2009ApJ...699.1092H} firstly noticed that the dispersion is 
significantly reduced when the surface density is added. Figure~\ref{fig:heyer}b shows the 
variation of the scaling coefficient, $\sigma_v$/$R^{\rm 1/2}$, with the mass surface density. 
The dependence of $\sigma_v$/$R^{\rm 1/2}$ on surface density for clouds within both the 
Galactic disk or Galactic center is evident in Fig.~\ref{fig:heyer}b, as implied by the 
above expression of size, line width, and surface density, that under the assumption of 
virial equilibrium. A similar correlation between $\sigma_v$/$R^{\rm 1/2}$ and $\Sigma$
is also present in the properties of Galactic MCs derived from \xco~\citep{2009ApJ...699.1092H} 
and extragalactic GMCs~\citep{2008ApJ...686..948B}. The results of these different analyses for 
large numbers of targets traced by different molecular lines, certify a necessary modification
to Larson's scaling relationships, as suggested by \citet{2009ApJ...699.1092H} that $\sigma_v$ 
depends on both $R_{\rm MC}$ and $\Sigma$.
Note, however, the scatter is larger for clouds within these extreme locations of the 
Galaxy, both the center of the Galaxy and the edge of the Galactic disk~(EOG region).
The solid line and two dashed lines in Fig.~\ref{fig:heyer}b show the loci for
$\alpha_{vir}$=1~(lower), 3~(middle) and 10~(upper), respectively. 
Interestingly, almost no clouds within the inner Galactic disk have virial ratio $\alpha_{\rm vir}>3$,
small portion of clouds in the outer Galaxy have $\alpha_{\rm vir}>3$, while 
most of the EOG clouds have $\alpha_{\rm vir}>3$. This is consistent with the finding of \citet{2017ApJ...834...57M}  
that MCs in the outer Galaxy exhibit higher value of $\alpha_{\rm vir}$ than MCs in the inner Galaxy 
with 3~$<$~$R_{\rm GC}$~$<$~7~kpc.

Here, we should keep in mind that the masses of all MCs
are derived by assuming a constant X conversion factor that measured in the Solar neighbourhood.
If we believe that MCs are gravitationally bound objects, then 
the distributions of $\alpha_{\rm vir}$ may suggest   
 that $X$ factor in the EOG region should have a 
somewhat higher value than in the inner Galaxy. In fact, 
various theoretical and observational investigations have indicated high values of $X$ factor in the outer Galaxy~
\citep[e.g.][]{1995A&A...303..851B,2001ApJ...551..852H,2010ApJ...710..133A,2011MNRAS.412.1686S}.
These are not difficult to understand. Since the increasing dominance of the CO-faint
molecular gas associated probably with decreasing metallicities in the outer and/or
extreme outer disk~\citep{2010ApJ...716.1191W,2013A&A...554A.103P,2014A&A...561A.122L}, will naturally result in a higher value of $X$ 
factor~\citep{2013ARAA..51..207B}.
An alternative explanation could be that the
EOG clouds are predominantly gravitationally unbound.  
In recent research, accumulated evidence seems to indicate that 
molecular clumps within MCs~(the site for star formation to take place) should be 
gravitationally bound, however, it is not neccessary that the MCs as a whole 
to be bound~\citep{2005MNRAS.359..809C,2008MNRAS.386....3C,2011MNRAS.411...65B}. 
\citet{2011MNRAS.413.2935D} had addressed the question of how MCs can remain unbound.
They found that it is a natural consequence of a scenario of cloud-cloud collisions and 
stellar feedback. And in this scenario, MCs are short-lived, typically a few Myr~\citep{2011MNRAS.413.2935D}.   

In both panel (a) and (b) of Fig.~\ref{fig:heyer}, the pluses mark the EOG clouds with \xco~ 
detections. It appears that there is no significant difference in the $R_{\rm MC}-\sigma_{\rm v}$ 
and $\Sigma_{\rm MC}-\sigma_v$/R$^{1/2}$ 
relations between the two groups with or without the presence of \xco~ emission.

\subsection{Spatial Distributions of the Molecular Clouds in the Extreme Outer Galaxy}

\begin{figure}
\includegraphics[angle=0,scale=0.37,bb=30 0 460 360]{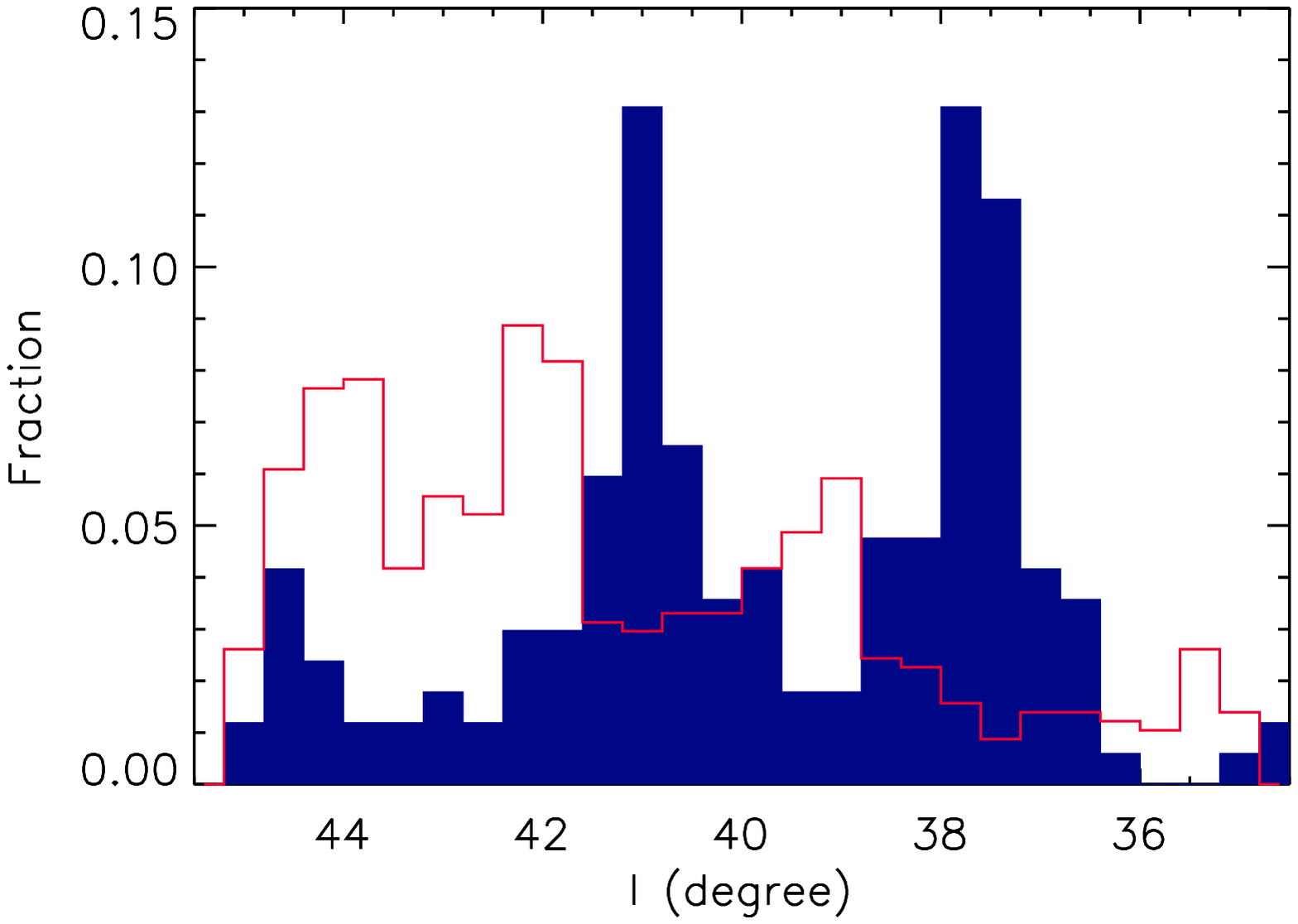}%
\includegraphics[angle=0,scale=0.37,bb=30 0 460 360]{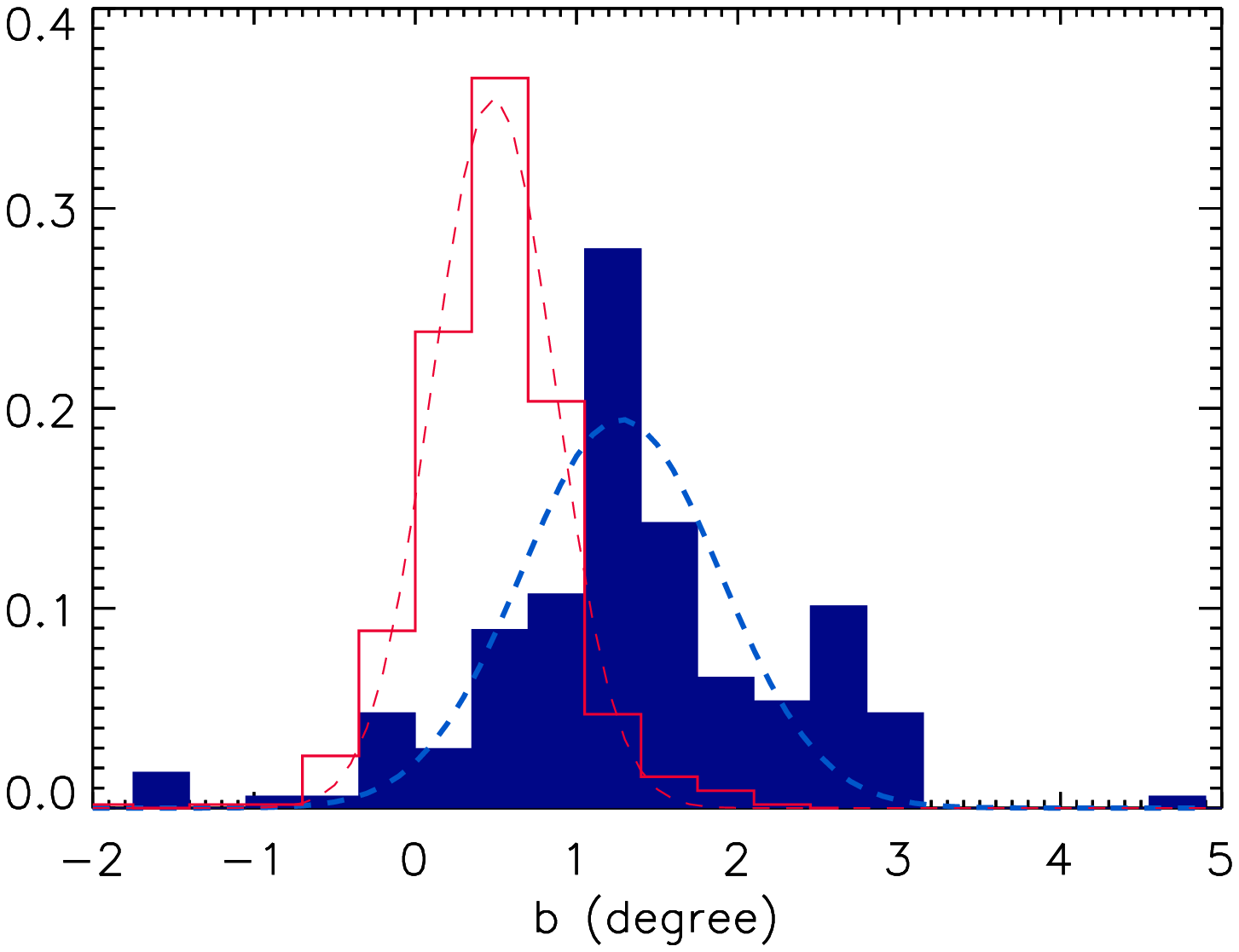}%
\includegraphics[angle=0,scale=0.37,bb=30 0 460 360]{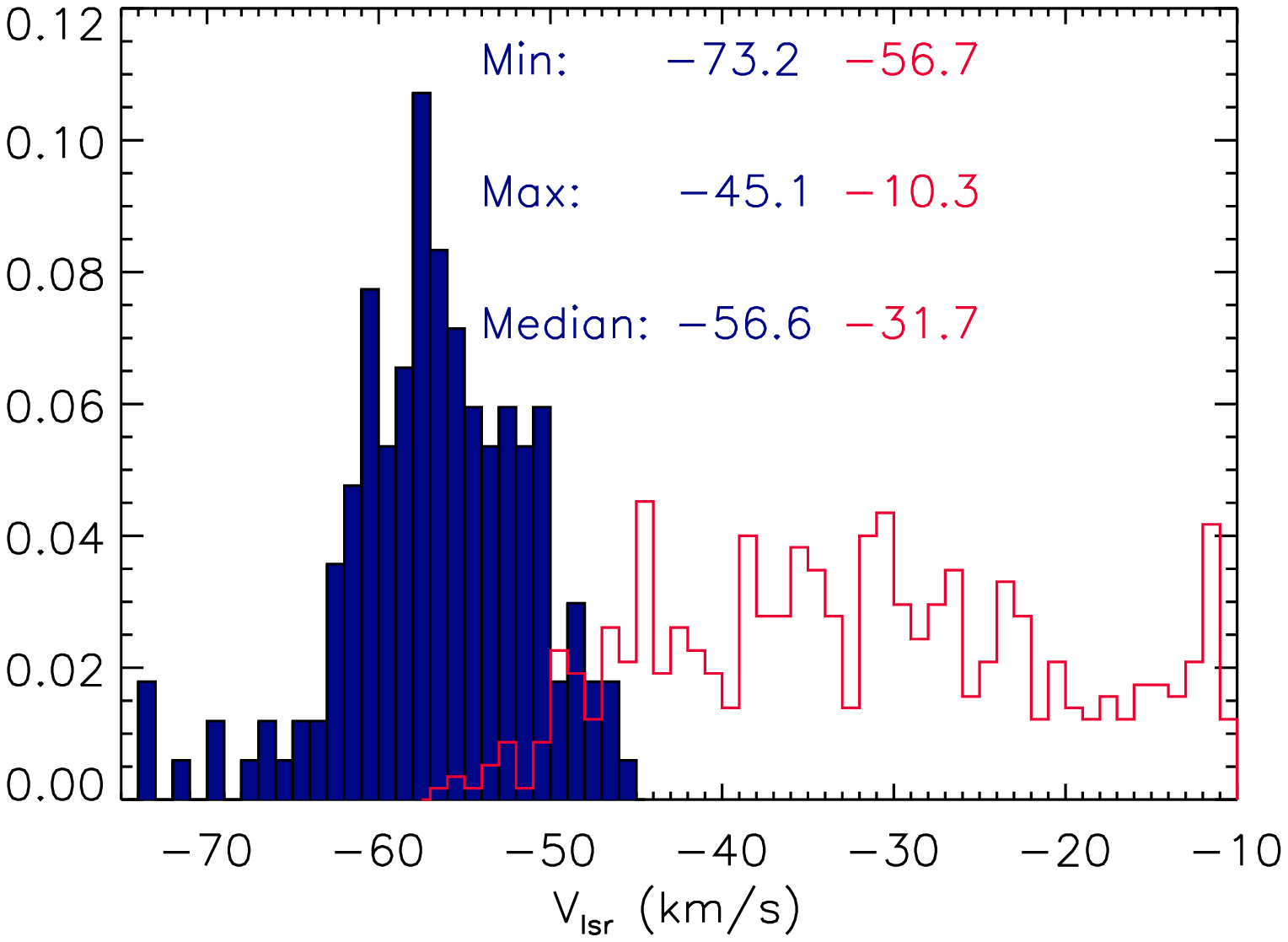}\\
\includegraphics[angle=0,scale=0.37,bb=30 0 460 360]{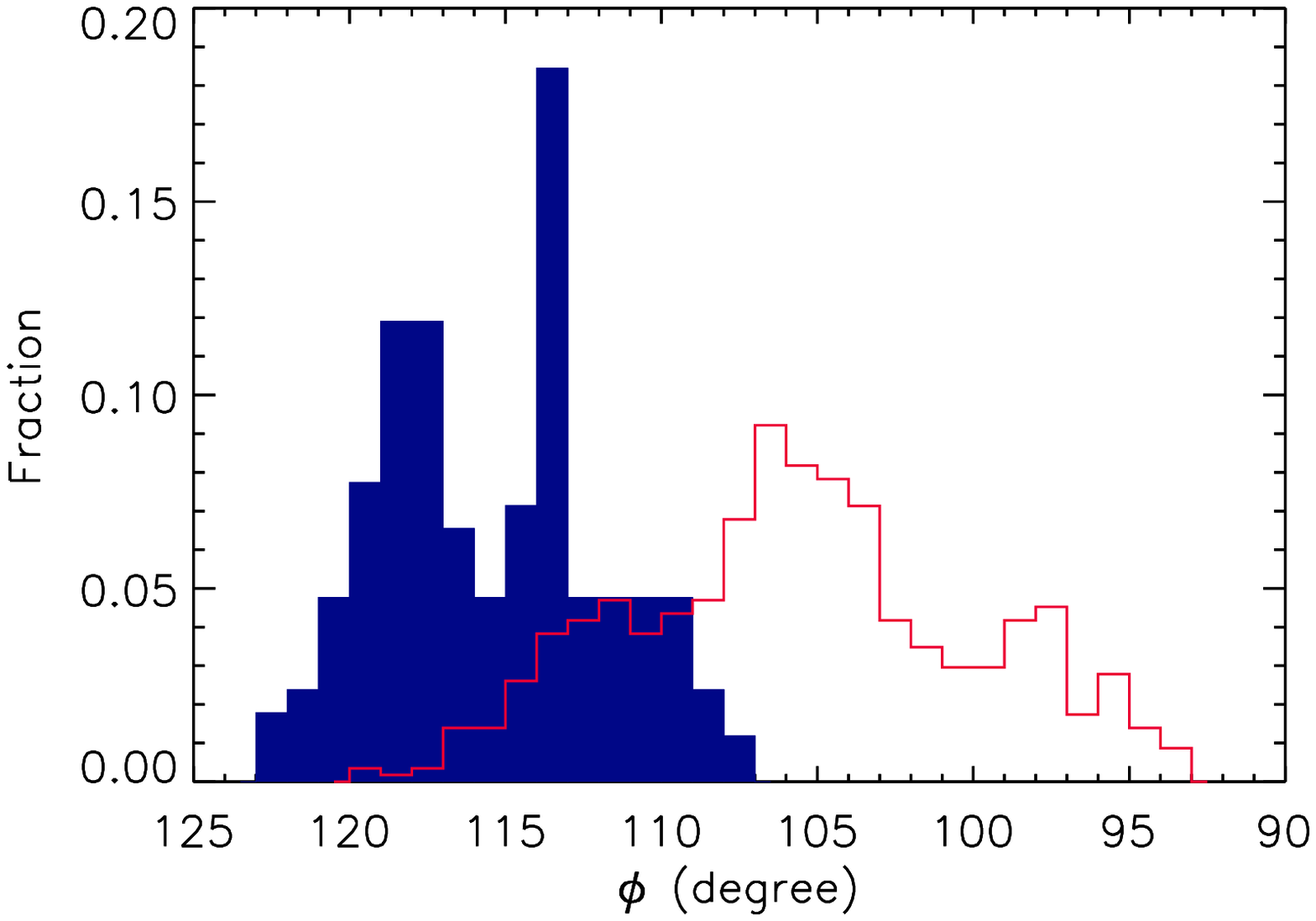}%
\includegraphics[angle=0,scale=0.37,bb=30 0 460 360]{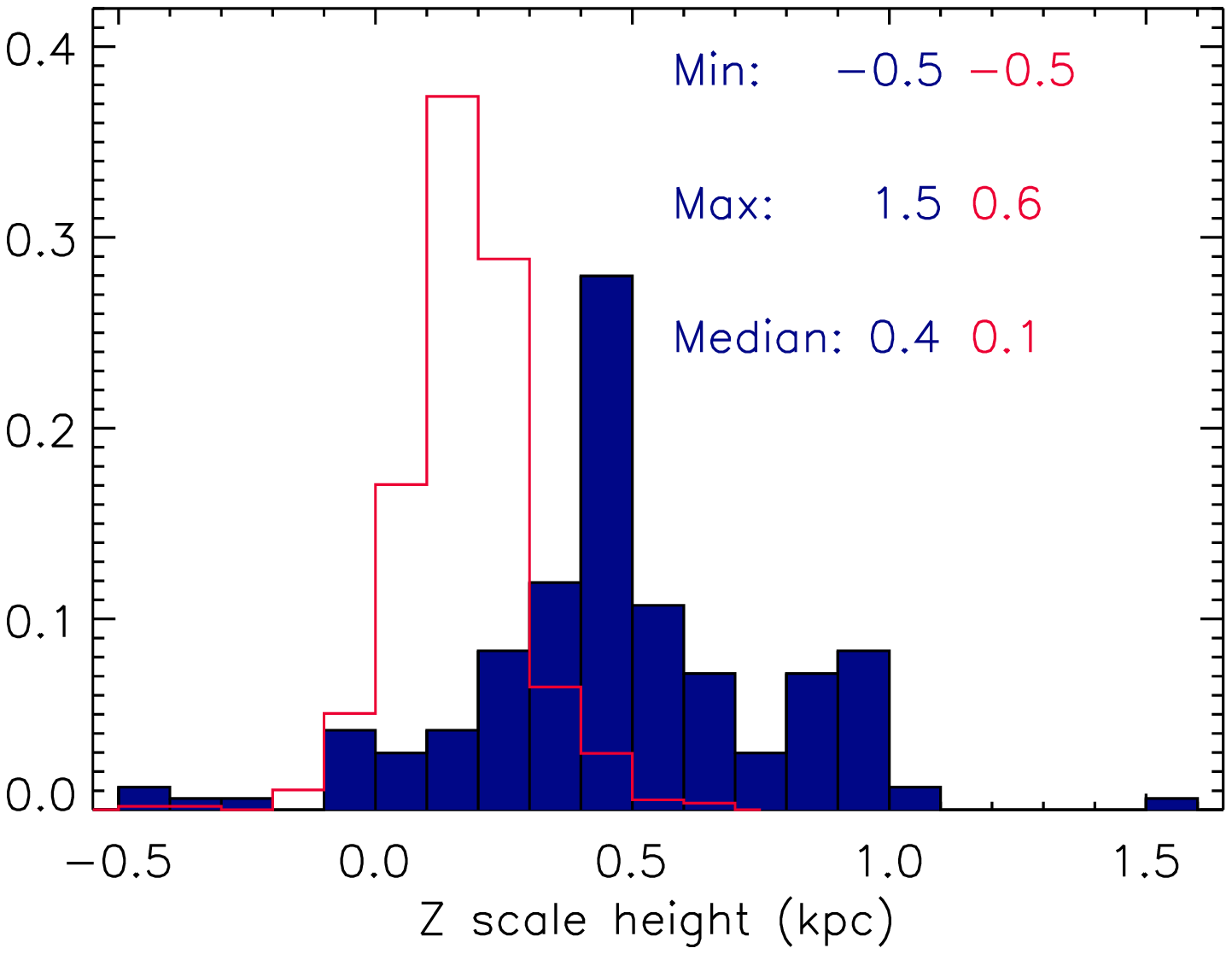}%
\includegraphics[angle=0,scale=0.37,bb=30 0 460 360]{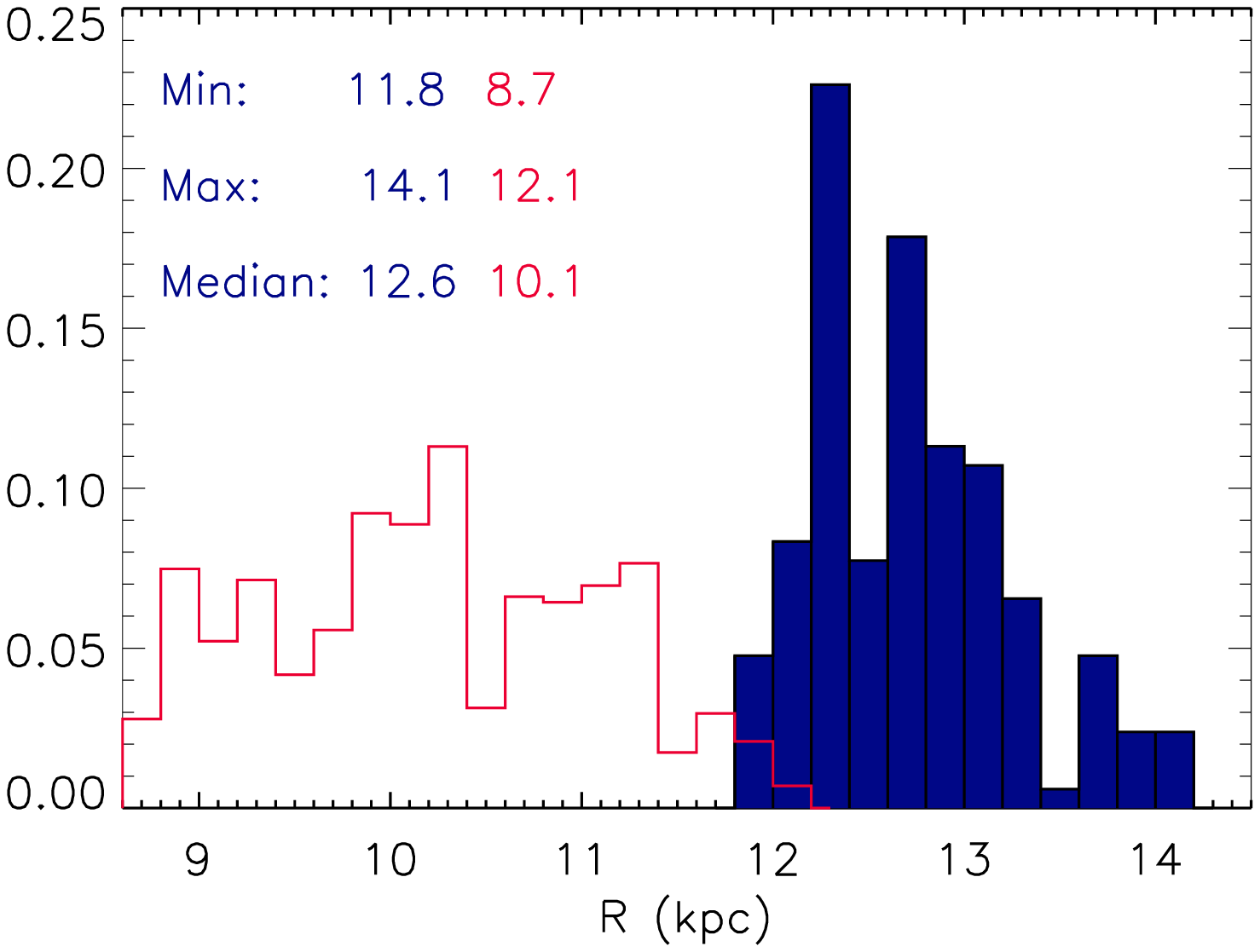}
\caption{Distributions of the MCs in both helioctocentric coordinates~($l$, $b$, $V_{\rm LSR}$),  
and Galactocentri cylindrical coordinates~($R$, $\phi$, $Z$), located in EOG region~(blue) and Outer arm~(red). 
In the middle-upper panel, the dashed blue and red lines indicate the best fit of the distributions by a 
Gaussian function for the two groups.\label{fig:dis}}
\end{figure}

Figure~\ref{fig:dis} displays the normalized number distributions of EOG clouds~(blue histogram) 
in the heliocentric coordinates, $l$, $b$, and $V_{LSR}$, as well as in the Galactocentric cylindrical 
coordinates $R$, $\phi$, and $Z$. As mentioned above, the Reid model was chosen to convert 
$(l$, $b$, $V_{LSR})$ to $(R$, $\phi$, $Z)$. The distributions of MCs in the Outer arm~(red histogram) 
are also overlaid for comparison. It appears that the spatial distributions of the EOG clouds are 
not uniform. Several distribution peaks of the MCs are clearly discerned at $l~=~$37.5$^{\circ}$, 
41$^{\circ}$, and 45$^{\circ}$, respectively. Interestingly, the distributions of the EOG clouds
in the Outer Scu-Cen arm are anti-correlated with the MCs in the Outer arm. 
We find that the EOG clouds are distributed within a broader and higher range
of Galactic latitude than the MCs in the Outer arm, and can be well fitted 
by a Gaussian function with a full width at half maximum (FWHM) of 1.45$^{\circ}$ 
and with a peak of $b~=~$1.27$^{\circ}$. At a median heliocentric distance of 17.8 kpc, 
this yields a FWHM of 450~pc, and a scale height of 400~pc above the b = 0$^{\circ}$ plane. 
These values are much higher than those of the Outer arm, which have a FWHM of 
0.7$^{\circ}$ and a peak of 0.42$^{\circ}$. The dashed blue and red lines show 
the best Gaussian fitting results for the EOG clouds and MCs within the Outer arm, respectively.

The EOG clouds are mainly distributed within a narrow range of LSR velocities, with a peak 
value of $\sim$-58~\kms. In contrast, the $V_{\rm LSR}$ distributions for the MCs within the Outer arm
exhibit a wide range of values. Since the Milky Way rotation is well approximated by an 
axisymmetric rotation curve, the LSR velocity directly relates to a Galactocentric
radius. As one would expect, the Galactocentric radial distributions of these clouds show 
similar distributions to the $V_{\rm LSR}$. 
Because most of the MCs are thought to well trace the logarithmic-spiral 
arm features~\citep[$\ln (R/R_{\mathrm{ref}})=-(\phi-\phi_{\mathrm{ref}})\tan(\psi)$; ]
[]{2003A&A...397..133R, 2009A&A...499..473H,2014A&A...569A.125H}, it is reasonable to expect that MCs within a given arm 
segment with a narrow distribution of $\phi$ should relate to a narrow distribution of
Galactocentric radii. This is the case for the EOG clouds, however, not the case for the MCs 
in the Outer arm~(see the lower-right panel of Fig.~\ref{fig:dis}). The broad distribution 
of the Galactocentric radii of the Outer arm can be understood in term of the probably 
presence of a substructure of the Outer arm near $l\sim44^{\circ}$ and $V_{\rm LSR}\sim$-20~\kms, 
as suggested by~\citet{2016ApJ...828...59S}.
  
\begin{figure}
\includegraphics[angle=0,scale=0.8,bb=0 10 566 141]{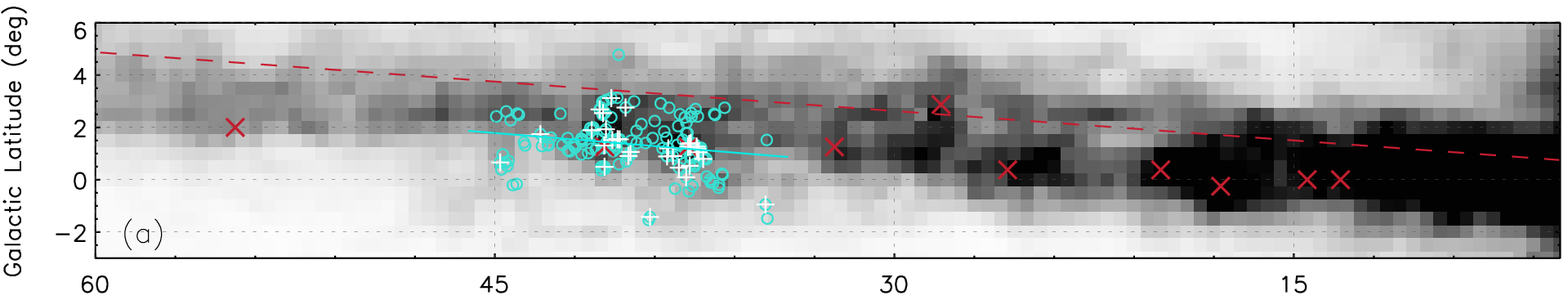}\\
\includegraphics[angle=0,scale=0.8,bb=10 0 566 210]{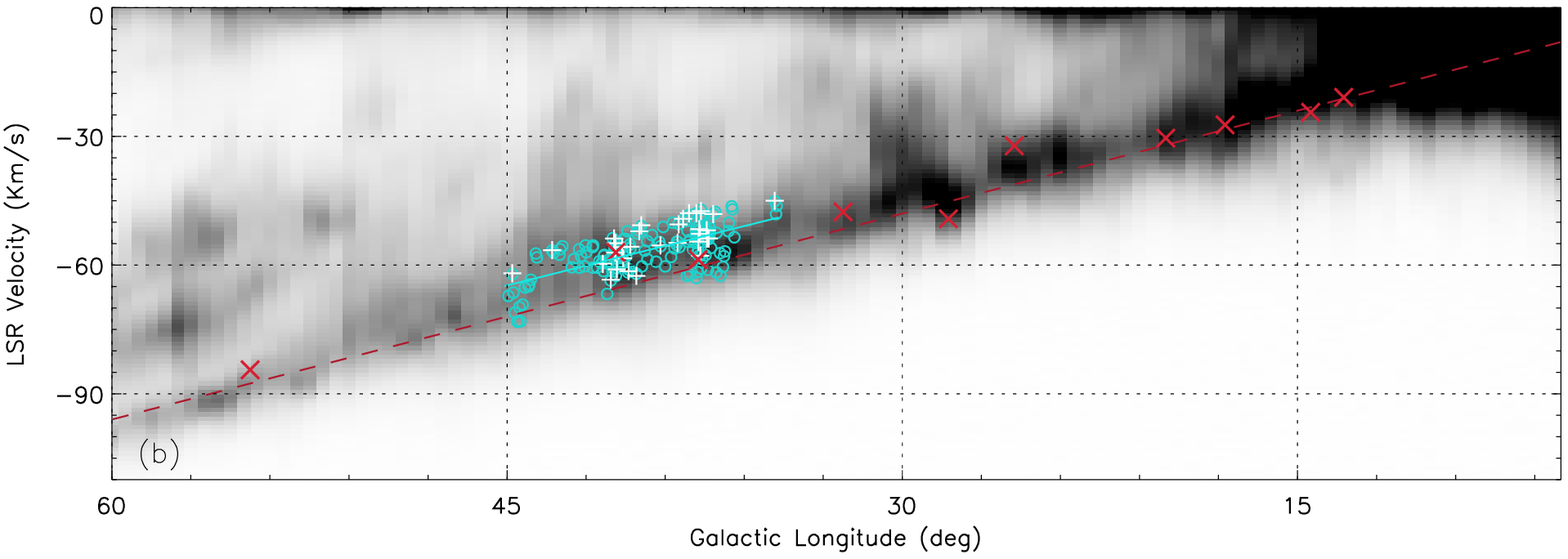}
\caption{(a) Velocity-integrated intensity map of HI emission from the Outer Scu-Cen arm, obtained by
integrating the LAB HI survey over a window $\pm$13.2~\kms~ wide that follows the arm in velocity.
(b) Longitude-velocity diagram of HI emission, obtained by integrating the 
LAB survey over a window that follows the Outer Scu-Cen arm in latitude. 
In both (a) and (b), cyan circles
and white pluses mark the MCs identified by MWISP survey in \co~ and \xco, respectively.
The red dashed and cyan solid lines indicate the Outer Scu-Cen arm model based on LAB HI survey results 
\citep{2011ApJ...734L..24D} and our new CO survey results. \label{fig:hi_co}}
\end{figure}

The distributions of all available EOG clouds are compared with the Outer Scu-Cen arm model~(red-dashed lines)
proposed by \citet{2011ApJ...734L..24D} in Figure~\ref{fig:hi_co}.
The grayscale maps of Fig.~\ref{fig:hi_co}a and Fig.~\ref{fig:hi_co}b  display the 
velocity-integrated intensity map and longitude-velocity diagram of 21 cm 
emissions from the Outer Scu-Cen arm. Similar to \citet{2011ApJ...734L..24D},
these diagrams are obtained by integrating the LAB 21 cm survey~\citep{2005A&A...440..775K} over a 
window that follows the arm in velocity and latitude.
The EOG clouds are indicated by cyan circles, and those with \xco~ detections are indicated by white pluses.

We attempted to fit the distributions of these EOG clouds by a linear function.
Their distributions can be described as $V_{\rm LSR} = -1.534~l + 4.307$, and 
$b = 0.0831^{\circ}~l - 1.9571^{\circ}$ in the $l-b-v$ space (the cyan-solid lines 
in Figure~\ref{fig:hi_co}). We find that the fitting results of atomic hydrogen~(the red dashed lines) and 
molecular gas show slightly different slope and intercept. When considering the 
larger velocity dispersion of atomic hydrogen and much higher spatial resolution of our 
observation, it appears that these EOG clouds and HI emissions trace the same coherent,  
large-scale structure, i.e., the Outer Scu-Cen arm. Broadly, our data confirm the existence of 
the Outer Scu-Cen arm proposed by \citet{2011ApJ...734L..24D}.
The total mass of the Outer Scu-Cen arm in the segment is about 2.7$\times$10$^5$ $M_{\sun}$,
which is about one order of magnitude lower than that of the Outer arm~\citep{2016ApJ...828...59S}.
It is important to note that the mass is a lower limit since the adopted CO-to-$H_{\rm 2}$ $X$ factor 
2.0$\times$10$^{\rm 20}$ ${\rm cm}^{\rm -2}{\rm (K\,km\,s^{-1})^{-1}}$ is measured in the 
solar neighborhood. However, as mentioned above, the $X$ factor might be higher 
in the EOG region than the nominal value. In addition, the beam dilution effects can not 
be neglected in the EOG region, which will lower the signal-to-noise ratio and hence 
also reduce the derived masses. 

\subsection{The Spiral Arm  and Warp of the Galactic Plane Traced by CO Emission} 

The 3D Galactic structure traced by CO emission of the Outer Scu-Cen 
arm is investigated in this section. The Galactic plane traced by the CO emission of 
Outer Scu-Cen arm is obviously offset from the $b = 0^{\circ}$ plane, and the amplitude 
evidently increases with the increasing of the Galactic longitude. It has long been recognized that both 
the effects of warping and tilted plane~(because of the Sun's offset from 
the physical mid-plane) will contribute to the offset between the 
physical midplane and the $b = 0^{\circ}$ plane~\citep[e.g.,][]{1960MNRAS.121..123B,1992ARA&A..30...51B}.  

In the outer galactic disk, warp is a common feature in spiral galaxies, with amplitude 
increasing with galactic radius~\citep[e.g.,][]{1998A&A...337....9R,2001A&A...373..402S,2002A&A...382..513R,2009ARA&A..47...27K,2014A&A...569A.125H}.
Early HI emission surveys have first revealed that the disk of our Galaxy is warped beyond the 
Solar circle~\citep[e.g.,][]{1957AJ.....62...90B,1957AJ.....62...93K,1988gera.book..295B}.
This finding was confirmed by a series of studies in the Galactic disks traced by gaseous component
~\citep[e.g.,][]{1958MNRAS.118..379O,1982ApJ...263..116H,1990A&A...230...21W,
2003A&A...397..133R,2005PASJ...57..917N,2015ApJ...798L..27S}, 
stellar component~\citep[e.g.,][]{2006MNRAS.368L..77M,2006A&A...451..515M,2009A&A...495..819R}, 
as well as dust component~\citep[e.g.,][]{2001ApJ...556..181D,2003A&A...409..205D,2006A&A...453..635M}.
However, our knowledge of the warp of the outer Galaxy to date is 
largely dependent on HI emission surveys.  
Several theories to cause and maintain warps have been proposed in the literature, 
including gravitational tidal interactions between galaxies, intergalactic magnetic 
fields, disk-halo interaction, and accretion of intergalactic medium~\citep[see the 
review by][]{1992ARA&A..30...51B}.   

Whereas, warp is not important in the inner disk of the Milky Way. The observed 
variation of Galactic latitude with repect to Galactic longitude can be understood in 
terms of the Sun's vertical displacement from the physical mid-plane~\citep[e.g.,][]
{1995MNRAS.273..206H,2016ApJ...828...59S}.
Various observations have revealed that the Sun is about 10-30~pc above the physical 
mid-plane~\citep[e.g.,][]{2001ApJ...553..184C,2006JRASC.100..146R,2009MNRAS.398..263M,
2014ApJ...797...53G,2016ApJ...828...59S,2016arXiv161008125K}.  

Here, we examine the effects of warping and tilted plane (because of the Sun's offset from 
the physical mid-plane). Four models were tested to compare with our observation, including the 
gaseous warp model based on the Galactic HI survey \citep{2006ApJ...643..881L},
the stellar model constrained by 2MASS star counts~\citep{2009A&A...495..819R}, 
the dust warp model obtained by dust extinction proposed by~\citet{2006A&A...453..635M},
as well as the tilted plane model obtained in the inner Galactic plane~\citep{1995MNRAS.273..206H}.

A Fourier mode frequency $m = 1$ was used to characterize 
the warp in the Galactic HI warp model, 
$W(\phi) = W_{\rm 0} + W_{\rm 1}~sin~(\phi-\phi_{\rm 1}) + W_{\rm 2}~sin~(2\phi-\phi_{\rm 2})$~\citep[see Eq. (12) in][]{2006ApJ...643..881L}.  
Each of the three amplitudes $W_{i}$ and two phases $\phi_{i}$ is a function of 
Galactocentric radius.
Following stellar model proposed by \citet{2009A&A...495..819R}, the height of the 
warp $z_{\rm warp}$ is calculated using, 
$z_{\rm warp} = \gamma_{\rm warp}\times(R - R_{\rm warp}\times sin~(\phi-\phi_{\rm warp}))$,
where $R_{\rm warp}$ = 8.4 kpc is taken from \citet{2001ASPC..232..229D} indicated as a best 
value for the starting Galactocentri radius of the warp, and $\gamma_{\rm warp}$ = 0.09
is a value of the warp slope~(the ratio between the displacement of the 
mid plane and the Galactocentric radius). 
Similar to the stellar warp model, the magnitude of the warp in the Galactic dust
 model can be written as,
$z_{\rm warp} = \gamma_{\rm warp}\times(R - R_{\rm warp}\times cos~(\phi - \phi_{\rm warp}))$,
where $\gamma_{\rm warp} = 0.14$, $\phi_{\rm warp}$ = 89$^{\circ}$ and 
$\gamma_{\rm warp} = 7.8$ kpc~\citep[also see Eq. (10) in][]{2006A&A...453..635M}.  
The Galactic latitude of the disk predicted by the tilted plane model can be described as, 
$b = -0.4^{\circ} sin~(l + 165^{\circ})$~\citep[see Eq. (5) in][]{1995MNRAS.273..206H}.

\begin{figure}
\includegraphics[angle=-90,scale=0.55]{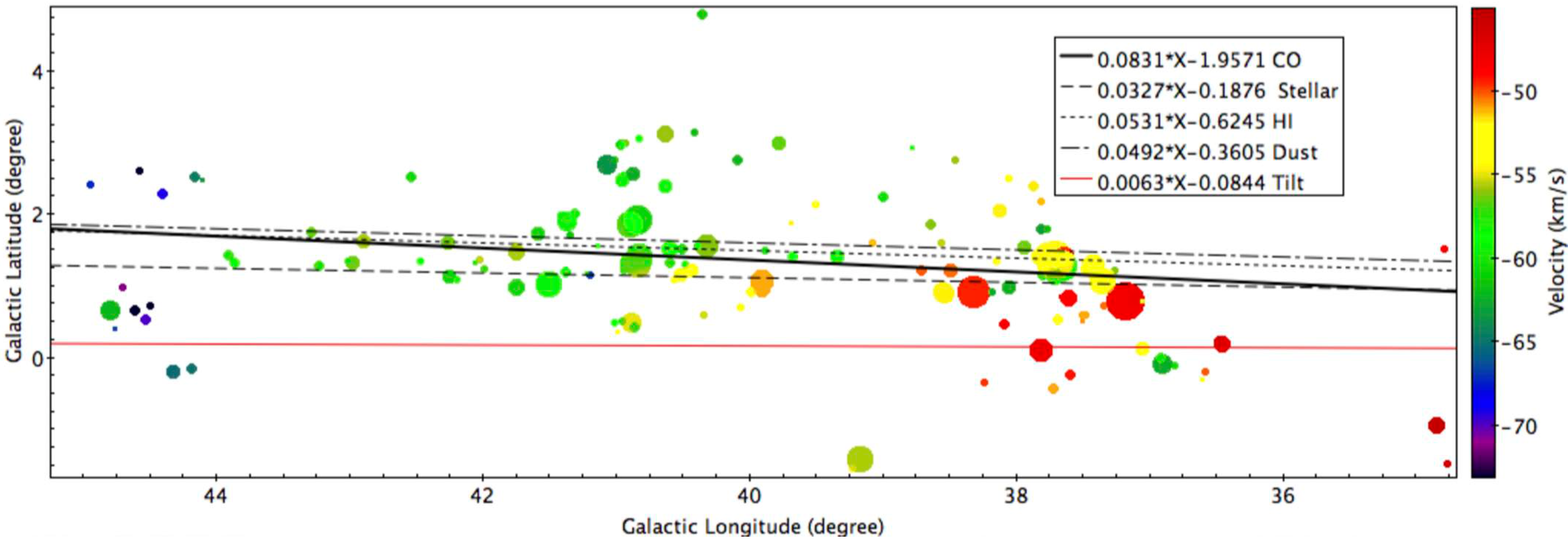}
\caption{Distributions of the 168 EOG clouds. The circle sizes are weighted with the scale of M$^{0.5}$. 
The solid line is the least-squares fit result of our CO observation. The predictions by tilted plane model 
and other warp models using different tracers are also overlaid on for comparision. Refer to Section 3.5 for details.
\label{fig:warp1}}
\end{figure}

The comparison results are given in the $l-b$ space in Figure~\ref{fig:warp1}.
The black-solid line, short-dashed line, long-dashed line, dash-dotted line, 
and red-solid line indicate our CO observation result, HI \citet{2006ApJ...643..881L} $m =$ 1 model,
stellar warp model \citep{2009A&A...495..819R}, dust warp model \citep{2006A&A...453..635M},
and tilted plane model \citep{1995MNRAS.273..206H}, respectively. 
It appears that in the region with $R_{\rm GC}\sim$12-14 kpc, the contribution of 
tilted plane is insignificant, while warp plays a dominant role. We also find that 
the dust and atomic gas HI models show very similar amplitude of warp, and are well 
consistent with our molecular gas observation results. The amplitude of the stellar
warp model is slightly smaller than our observed displacement across the present
Galactic longitude interval, as well as the other two warp models. 

\begin{figure}
\includegraphics[angle=0,scale=0.5]{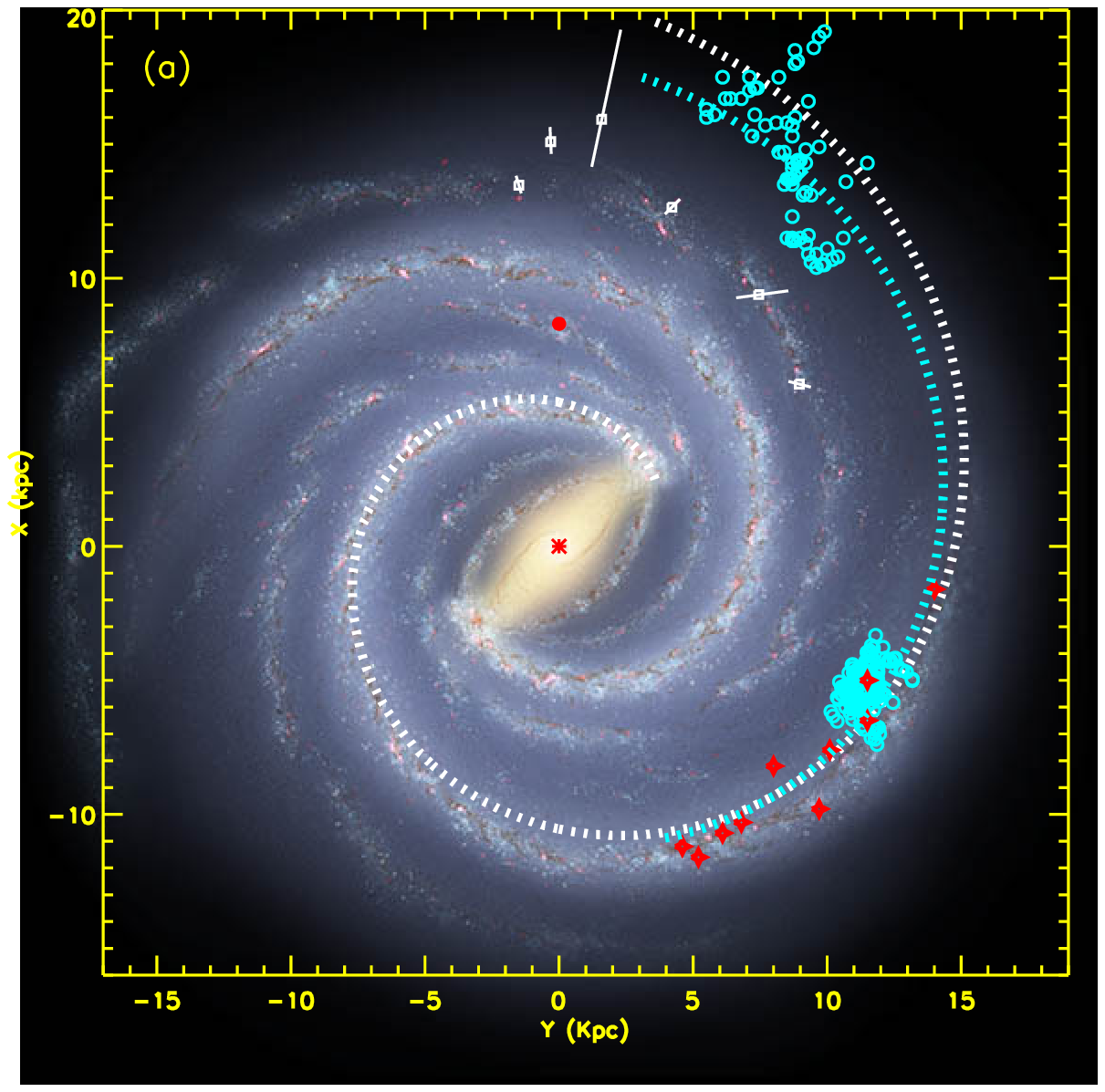}%
\includegraphics[angle=0,scale=0.5]{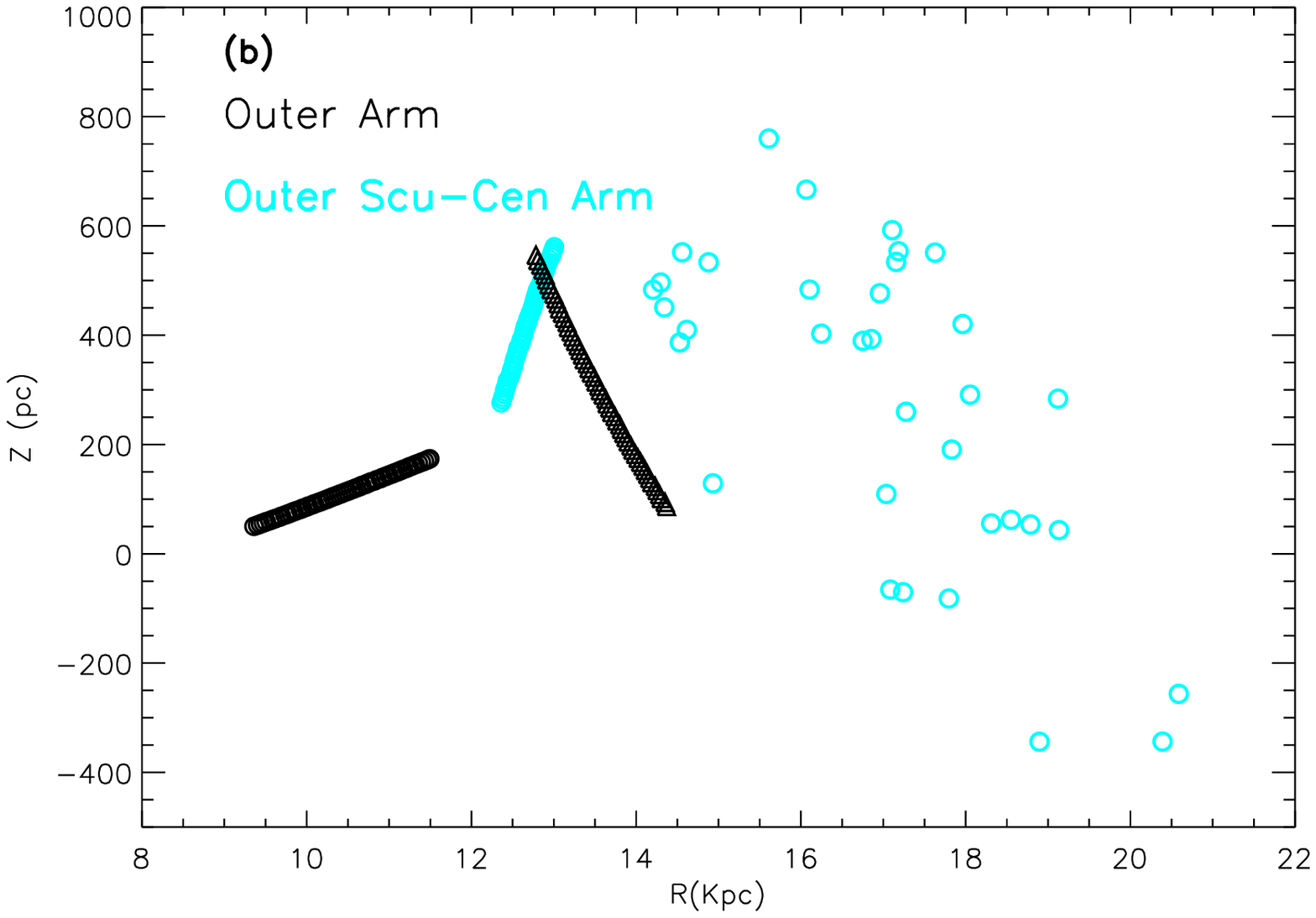}
\caption{(a) Face-on view of the Milky Way~(R. Hurt: NASA/JPL-Caltech/SSC) superposed on
all available EOG clouds. Also refer to Sec. 3.5 for details. (b) $Z$ scale height as a 
function of Galactocentric radii. The black and cyan circle-dashed lines indicate the warp fits of MCs 
between $34.75^{\circ}\le l \le 45.25^{\circ}$ in the Outer arm from \citet{2016ApJ...828...59S} and 
in the Outer Scu-Cen arm from this study, respectively. The  
black triangle-dashed line indicates the warp fit of data in the Outer arm between $100^{\circ}\le l \le150^{\circ}$ 
from \citet{2016ApJS..224....7D}. 
The cyan circles indicate data points in the Outer Scu-Cen arm between $100^{\circ} \le l \le 150^{\circ}$ from \citet{2015ApJ...798L..27S}. \label{fig:warp2}}
\end{figure}

The large-scale structure traced by CO emission is also explored in the 
Galactocentric coordinate in Figure~\ref{fig:warp2}.
The often used face-on view of the Milky Way~(R. Hurt: NASA/JPL-Caltech/SSC) superposed on 
all available EOG clouds are displayed in Figure~\ref{fig:warp2}a, 
including clouds~(filled star) identified by \citet{2011ApJ...734L..24D}, clouds~(circle) in 
the 2nd and 1st quadrants identified by \citet{2015ApJ...798L..27S}, and this study.
The white dashed line is a log spiral with a mean pitch angle of 12$^{\circ}$ that 
was fit to the Scu$-$Cen arm in the inner Galaxy \citep{2008AJ....135.1301V,2014AJ....148....5V}. And the 
cyan dashed line traces the fitting result of EOG clouds in \citet{2015ApJ...798L..27S} with pitch
angle of 9.3$^{\circ}$. Generally, our data are well in agreement with the far extension 
of the Scu$-$Cen arm in the outer Galaxy, as discussed previously. 
Figure~\ref{fig:warp2}b shows the vertical structure (in the $R_{\rm GC} - Z$ space) of the 
Outer Scu-Cen arm traced by CO emission.
We also examine the results for the Outer Scu-Cen arm in the 2nd quadrant~\citep{2015ApJ...798L..27S}, 
the Outer arm in both 1st and 2nd quadrants~\citep{2016ApJS..224....7D,2016ApJ...828...59S}.
Note that all structures are derived from the $l-b-v$ relations by applying the Reid model,
with one exception, the Outer Scu-Cen arm in the 2nd quadrant, which is indicated by data 
points directly from observations. It is obvious that the shapes of the two arms gernerally follow 
a $Sin$ function, which are in agreement with the warp models from the literature. In addition, the $Z$ scale 
heights of both the Outer Scu-Cen arm~(cyan) and the Outer arm~(black) are increased with the increasing of 
the Galactocentric radii in the same direction~(with similar value of $\phi$). 

\section{SUMMARY}
We have presented the results of an unbiased \co/\xco/\xxco~($J = 1-0$) survey in Galactic longitude range
of 34.75$^{\circ}$ to 45.25$^{\circ}$, Galactic latitude range of
-5.25$^{\circ}$ to 5.25$^{\circ}$, and velocity in the interval of
-1.6~$l~\pm$~13.2~\kms~ that might lie beyond the Outer arm.
The main results are summarized as follows.

(1) A total of 174 molecular clouds (MCs) were identified, of which 168 MCs 
most probably lie in the Extreme Outer Galacy~(EOG) region. A total of 35 cores within the MCs 
show \xco~ emission, of which 31 cores most probably lie in the EOG region.  
However, none of them show significant \xxco~ emission under current detection limit.

(2) Physical properties of the EOG clouds were examined in detail.
The revised scaling relations of Larson were tested in the EOG clouds, and compared across a
 wide range of Galactic environments. Similar to MCs in the outer Galaxy, the velocity 
dispersions of EOG clouds are also correlated with the cloud sizes, however, are well 
displaced below the scaling relationship defined by MCs in the inner Galaxy.  

(3) We compared the distributions of the EOG clouds with the MCs lie in the 
Outer arm that published in our previous paper, as well as the HI emissions from the 
Outer Scu-Cen arm defined by \citet{2011ApJ...734L..24D}.
The distributions of the EOG clouds can be described as linear features 
in the $l-b-v$ space ($V_{\rm LSR} = -1.534~l + 4.307$,
and $b = 0.0831^{\circ}~l - 1.9571^{\circ}$), which are distinctly  
different from those of the MCs in the Outer arm, while roughly follow the 
$l-b$ and $l-v$ relations of the Outer Scu-Cen arm defined by \citet{2011ApJ...734L..24D}.
All these may provide a robust evidence for the existence of the Outer Scu-Cen arm. 
The lower limit of the total molecular mass of this arm segment is about 2.7$\times$10$^5$ $M_{\sun}$, 
assuming a $X$ conversion factor of
2.0$\times$10$^{\rm 20}$ ${\rm cm}^{\rm -2}{\rm (K\,km\,s^{-1})^{-1}}$.   

(4) In the EOG region, the warp traced by CO emission is very obvious. The mean thickness of
gaseous disk is about 450~pc, and the scale height is about 400~pc above the $b = 0^{\circ}$ plane.
The comparison between our observation and warp models reveals that different components of our Galaxy
(stellar, dust, molecular gas, atomic gas) show similar amplitude of warp, in particular, the 
gaseous and dust components.

\acknowledgments
We are grateful to all of the staff members of the Qinghai Radio
Observing Station at Delingha for their support of the observations.
We would like to thank the anonymous referee for going through the paper carefully 
and appreciate a lot for many practical comments that improved this paper.
The TOPCAT tool \citep{2005ASPC..347...29T} was used while 
preparing the paper.
Research for this project is supported by the National Natural Science Foundation of 
China (grant nos. 11233007, 11133008, 11473069, and 11303097), and the Key Laboratory for Radio 
Astronomy, CAS. The work is a part of the Multi-Line Galactic Plane Survey in CO and its
Isotopic Transitions, also called the Milky Way Imaging Scroll
Painting, which is supported by the Strategic Priority Research Program,
the Emergence of Cosmological Structures of the Chinese Academy of 
Sciences~(grant No. XDB09000000).

\bibliographystyle{aasjournal}
\bibliography{references}

\begin{thebibliography}{}
\expandafter\ifx\csname natexlab\endcsname\relax\def\natexlab#1{#1}\fi
\providecommand{\url}[1]{\href{#1}{#1}}

\bibitem[{{Abdo} {et~al.}(2010){Abdo}, {Ackermann}, {Ajello}, {Baldini},
  {Ballet}, {Barbiellini}, {Bastieri}, {Baughman}, {Bechtol}, {Bellazzini},
  {Berenji}, {Bloom}, {Bonamente}, {Borgland}, {Bregeon}, {Brez}, {Brigida},
  {Bruel}, {Burnett}, {Buson}, {Caliandro}, {Cameron}, {Caraveo}, {Casandjian},
  {Cecchi}, {{\c C}elik}, {Chekhtman}, {Cheung}, {Chiang}, {Ciprini}, {Claus},
  {Cohen-Tanugi}, {Cominsky}, {Conrad}, {Dermer}, {de Palma}, {Digel}, {Silva},
  {Drell}, {Dubois}, {Dumora}, {Farnier}, {Favuzzi}, {Fegan}, {Focke},
  {Fortin}, {Frailis}, {Fukazawa}, {Funk}, {Fusco}, {Gargano}, {Gehrels},
  {Germani}, {Giavitto}, {Giebels}, {Giglietto}, {Giordano}, {Glanzman},
  {Godfrey}, {Grenier}, {Grondin}, {Grove}, {Guillemot}, {Guiriec}, {Harding},
  {Hayashida}, {Horan}, {Hughes}, {Jackson}, {J{\'o}hannesson}, {Johnson},
  {Johnson}, {Kamae}, {Katagiri}, {Kataoka}, {Kawai}, {Kerr}, {Kn{\"o}dlseder},
  {Kuss}, {Lande}, {Latronico}, {Lemoine-Goumard}, {Longo}, {Loparco}, {Lott},
  {Lovellette}, {Lubrano}, {Makeev}, {Mazziotta}, {McEnery}, {Meurer},
  {Michelson}, {Mitthumsiri}, {Mizuno}, {Monte}, {Monzani}, {Morselli},
  {Moskalenko}, {Murgia}, {Nolan}, {Norris}, {Nuss}, {Ohsugi}, {Okumura},
  {Omodei}, {Orlando}, {Ormes}, {Paneque}, {Pelassa}, {Pepe}, {Pesce-Rollins},
  {Piron}, {Porter}, {Rain{\`o}}, {Rando}, {Razzano}, {Reimer}, {Reimer},
  {Reposeur}, {Rodriguez}, {Ryde}, {Sadrozinski}, {Sanchez}, {Sander}, {Saz
  Parkinson}, {Sgr{\`o}}, {Siskind}, {Smith}, {Spandre}, {Spinelli}, {Starck},
  {Strickman}, {Strong}, {Suson}, {Takahashi}, {Tanaka}, {Thayer}, {Thayer},
  {Thompson}, {Tibaldo}, {Torres}, {Tosti}, {Tramacere}, {Uchiyama}, {Usher},
  {Vasileiou}, {Vilchez}, {Vitale}, {Waite}, {Wang}, {Winer}, {Wood}, {Ylinen},
  {Ziegler}, \& {Fermi/LAT Collaboration}}]{2010ApJ...710..133A}
{Abdo}, A.~A., {Ackermann}, M., {Ajello}, M., {et~al.} 2010, \apj, 710, 133

\bibitem[{{Ballesteros-Paredes} {et~al.}(2011){Ballesteros-Paredes},
  {Hartmann}, {V{\'a}zquez-Semadeni}, {Heitsch}, \&
  {Zamora-Avil{\'e}s}}]{2011MNRAS.411...65B}
{Ballesteros-Paredes}, J., {Hartmann}, L.~W., {V{\'a}zquez-Semadeni}, E.,
  {Heitsch}, F., \& {Zamora-Avil{\'e}s}, M.~A. 2011, \mnras, 411, 65

\bibitem[{{Bertoldi} \& {McKee}(1992)}]{1992ApJ...395..140B}
{Bertoldi}, F., \& {McKee}, C.~F. 1992, \apj, 395, 140

\bibitem[{{Binney}(1992)}]{1992ARA&A..30...51B}
{Binney}, J. 1992, \araa, 30, 51

\bibitem[{{Blaauw} {et~al.}(1960){Blaauw}, {Gum}, {Pawsey}, \&
  {Westerhout}}]{1960MNRAS.121..123B}
{Blaauw}, A., {Gum}, C.~S., {Pawsey}, J.~L., \& {Westerhout}, G. 1960, \mnras,
  121, 123

\bibitem[{{Bolatto} {et~al.}(2008){Bolatto}, {Leroy}, {Rosolowsky}, {Walter},
  \& {Blitz}}]{2008ApJ...686..948B}
{Bolatto}, A.~D., {Leroy}, A.~K., {Rosolowsky}, E., {Walter}, F., \& {Blitz},
  L. 2008, \apj, 686, 948

\bibitem[{{Bolatto} {et~al.}(2013){Bolatto}, {Wolfire}, \&
  {Leroy}}]{2013ARAA..51..207B}
{Bolatto}, A.~D., {Wolfire}, M., \& {Leroy}, A.~K. 2013, \araa, 51, 207

\bibitem[{{Brand} \& {Blitz}(1993)}]{1993A&A...275...67B}
{Brand}, J., \& {Blitz}, L. 1993, \aap, 275, 67

\bibitem[{{Brand} \& {Wouterloot}(1995)}]{1995A&A...303..851B}
{Brand}, J., \& {Wouterloot}, J.~G.~A. 1995, \aap, 303, 851

\bibitem[{{Brunt} {et~al.}(2003){Brunt}, {Kerton}, \&
  {Pomerleau}}]{2003ApJS..144...47B}
{Brunt}, C.~M., {Kerton}, C.~R., \& {Pomerleau}, C. 2003, \apjs, 144, 47

\bibitem[{{Burke}(1957)}]{1957AJ.....62...90B}
{Burke}, B.~F. 1957, \aj, 62, 90

\bibitem[{{Burton}(1988)}]{1988gera.book..295B}
{Burton}, W.~B. 1988, {The structure of our Galaxy derived from observations of
  neutral hydrogen}, ed. K.~I. {Kellermann} \& G.~L. {Verschuur}, 295--358

\bibitem[{{Chen} {et~al.}(2001){Chen}, {Stoughton}, {Smith}, {Uomoto}, {Pier},
  {Yanny}, {Ivezi{\'c}}, {York}, {Anderson}, {Annis}, {Brinkmann}, {Csabai},
  {Fukugita}, {Hindsley}, {Lupton}, {Munn}, \& {SDSS
  Collaboration}}]{2001ApJ...553..184C}
{Chen}, B., {Stoughton}, C., {Smith}, J.~A., {et~al.} 2001, \apj, 553, 184

\bibitem[{{Clark} {et~al.}(2008){Clark}, {Bonnell}, \&
  {Klessen}}]{2008MNRAS.386....3C}
{Clark}, P.~C., {Bonnell}, I.~A., \& {Klessen}, R.~S. 2008, \mnras, 386, 3

\bibitem[{{Clark} {et~al.}(2005){Clark}, {Bonnell}, {Zinnecker}, \&
  {Bate}}]{2005MNRAS.359..809C}
{Clark}, P.~C., {Bonnell}, I.~A., {Zinnecker}, H., \& {Bate}, M.~R. 2005,
  \mnras, 359, 809

\bibitem[{{Clemens} {et~al.}(1988){Clemens}, {Sanders}, \&
  {Scoville}}]{1988ApJ...327..139C}
{Clemens}, D.~P., {Sanders}, D.~B., \& {Scoville}, N.~Z. 1988, \apj, 327, 139

\bibitem[{{Clemens} {et~al.}(1986){Clemens}, {Sanders}, {Scoville}, \&
  {Solomon}}]{1986ApJS...60..297C}
{Clemens}, D.~P., {Sanders}, D.~B., {Scoville}, N.~Z., \& {Solomon}, P.~M.
  1986, \apjs, 60, 297

\bibitem[{{Dame} {et~al.}(1986){Dame}, {Elmegreen}, {Cohen}, \&
  {Thaddeus}}]{1986ApJ...305..892D}
{Dame}, T.~M., {Elmegreen}, B.~G., {Cohen}, R.~S., \& {Thaddeus}, P. 1986,
  \apj, 305, 892

\bibitem[{{Dame} {et~al.}(2001){Dame}, {Hartmann}, \&
  {Thaddeus}}]{2001ApJ...547..792D}
{Dame}, T.~M., {Hartmann}, D., \& {Thaddeus}, P. 2001, \apj, 547, 792

\bibitem[{{Dame} \& {Thaddeus}(1985)}]{1985ApJ...297..751D}
{Dame}, T.~M., \& {Thaddeus}, P. 1985, \apj, 297, 751

\bibitem[{{Dame} \& {Thaddeus}(2011)}]{2011ApJ...734L..24D}
---. 2011, \apjl, 734, L24

\bibitem[{{Derriere} \& {Robin}(2001)}]{2001ASPC..232..229D}
{Derriere}, S., \& {Robin}, A.~C. 2001, in Astronomical Society of the Pacific
  Conference Series, Vol. 232, The New Era of Wide Field Astronomy, ed.
  R.~{Clowes}, A.~{Adamson}, \& G.~{Bromage}, 229

\bibitem[{{Digel} {et~al.}(1994){Digel}, {de Geus}, \&
  {Thaddeus}}]{1994ApJ...422...92D}
{Digel}, S., {de Geus}, E., \& {Thaddeus}, P. 1994, \apj, 422, 92

\bibitem[{{Dobbs} {et~al.}(2011){Dobbs}, {Burkert}, \&
  {Pringle}}]{2011MNRAS.413.2935D}
{Dobbs}, C.~L., {Burkert}, A., \& {Pringle}, J.~E. 2011, \mnras, 413, 2935

\bibitem[{{Drimmel} {et~al.}(2003){Drimmel}, {Cabrera-Lavers}, \&
  {L{\'o}pez-Corredoira}}]{2003A&A...409..205D}
{Drimmel}, R., {Cabrera-Lavers}, A., \& {L{\'o}pez-Corredoira}, M. 2003, \aap,
  409, 205

\bibitem[{{Drimmel} \& {Spergel}(2001)}]{2001ApJ...556..181D}
{Drimmel}, R., \& {Spergel}, D.~N. 2001, \apj, 556, 181

\bibitem[{{Du} {et~al.}(2016){Du}, {Xu}, {Yang}, {Sun}, {Li}, {Zhang}, \&
  {Zhou}}]{2016ApJS..224....7D}
{Du}, X., {Xu}, Y., {Yang}, J., {et~al.} 2016, \apjs, 224, 7

\bibitem[{{Falgarone} {et~al.}(1992){Falgarone}, {Puget}, \&
  {Perault}}]{1992A&A...257..715F}
{Falgarone}, E., {Puget}, J.-L., \& {Perault}, M. 1992, \aap, 257, 715

\bibitem[{{Goodman} {et~al.}(2014){Goodman}, {Alves}, {Beaumont}, {Benjamin},
  {Borkin}, {Burkert}, {Dame}, {Jackson}, {Kauffmann}, {Robitaille}, \&
  {Smith}}]{2014ApJ...797...53G}
{Goodman}, A.~A., {Alves}, J., {Beaumont}, C.~N., {et~al.} 2014, \apj, 797, 53

\bibitem[{{Hammersley} {et~al.}(1995){Hammersley}, {Garzon}, {Mahoney}, \&
  {Calbet}}]{1995MNRAS.273..206H}
{Hammersley}, P.~L., {Garzon}, F., {Mahoney}, T., \& {Calbet}, X. 1995, \mnras,
  273, 206

\bibitem[{{Henderson} {et~al.}(1982){Henderson}, {Jackson}, \&
  {Kerr}}]{1982ApJ...263..116H}
{Henderson}, A.~P., {Jackson}, P.~D., \& {Kerr}, F.~J. 1982, \apj, 263, 116

\bibitem[{{Heyer} \& {Dame}(2015)}]{2015ARA&A..53..583H}
{Heyer}, M., \& {Dame}, T.~M. 2015, \araa, 53, 583

\bibitem[{{Heyer} {et~al.}(2009){Heyer}, {Krawczyk}, {Duval}, \&
  {Jackson}}]{2009ApJ...699.1092H}
{Heyer}, M., {Krawczyk}, C., {Duval}, J., \& {Jackson}, J.~M. 2009, \apj, 699,
  1092

\bibitem[{{Heyer} {et~al.}(1998){Heyer}, {Brunt}, {Snell}, {Howe}, {Schloerb},
  \& {Carpenter}}]{1998ApJS..115..241H}
{Heyer}, M.~H., {Brunt}, C., {Snell}, R.~L., {et~al.} 1998, \apjs, 115, 241

\bibitem[{{Heyer} {et~al.}(2001){Heyer}, {Carpenter}, \&
  {Snell}}]{2001ApJ...551..852H}
{Heyer}, M.~H., {Carpenter}, J.~M., \& {Snell}, R.~L. 2001, \apj, 551, 852

\bibitem[{{Hou} \& {Han}(2014)}]{2014A&A...569A.125H}
{Hou}, L.~G., \& {Han}, J.~L. 2014, \aap, 569, A125

\bibitem[{{Hou} {et~al.}(2009){Hou}, {Han}, \& {Shi}}]{2009A&A...499..473H}
{Hou}, L.~G., {Han}, J.~L., \& {Shi}, W.~B. 2009, \aap, 499, 473

\bibitem[{{Izumi} {et~al.}(2014){Izumi}, {Kobayashi}, {Yasui}, {Tokunaga},
  {Saito}, \& {Hamano}}]{2014ApJ...795...66I}
{Izumi}, N., {Kobayashi}, N., {Yasui}, C., {et~al.} 2014, \apj, 795, 66

\bibitem[{{Jackson} {et~al.}(2006){Jackson}, {Rathborne}, {Shah}, {Simon},
  {Bania}, {Clemens}, {Chambers}, {Johnson}, {Dormody}, {Lavoie}, \&
  {Heyer}}]{2006ApJS..163..145J}
{Jackson}, J.~M., {Rathborne}, J.~M., {Shah}, R.~Y., {et~al.} 2006, \apjs, 163,
  145

\bibitem[{{Jacq} {et~al.}(1988){Jacq}, {Despois}, \&
  {Baudry}}]{1988A&A...195...93J}
{Jacq}, T., {Despois}, D., \& {Baudry}, A. 1988, \aap, 195, 93

\bibitem[{{Kalberla} {et~al.}(2005){Kalberla}, {Burton}, {Hartmann}, {Arnal},
  {Bajaja}, {Morras}, \& {P{\"o}ppel}}]{2005A&A...440..775K}
{Kalberla}, P.~M.~W., {Burton}, W.~B., {Hartmann}, D., {et~al.} 2005, \aap,
  440, 775

\bibitem[{{Kalberla} \& {Kerp}(2009)}]{2009ARA&A..47...27K}
{Kalberla}, P.~M.~W., \& {Kerp}, J. 2009, \araa, 47, 27

\bibitem[{{Karim} \& {Mamajek}(2016)}]{2016arXiv161008125K}
{Karim}, M.~T., \& {Mamajek}, E. 2016, ArXiv e-prints, arXiv:1610.08125

\bibitem[{{Kerr}(1957)}]{1957AJ.....62...93K}
{Kerr}, F.~J. 1957, \aj, 62, 93

\bibitem[{{Kobayashi} \& {Tokunaga}(2000)}]{2000ApJ...532..423K}
{Kobayashi}, N., \& {Tokunaga}, A.~T. 2000, \apj, 532, 423

\bibitem[{{Kobayashi} {et~al.}(2008){Kobayashi}, {Yasui}, {Tokunaga}, \&
  {Saito}}]{2008ApJ...683..178K}
{Kobayashi}, N., {Yasui}, C., {Tokunaga}, A.~T., \& {Saito}, M. 2008, \apj,
  683, 178

\bibitem[{{Langer} {et~al.}(2014){Langer}, {Velusamy}, {Pineda}, {Willacy}, \&
  {Goldsmith}}]{2014A&A...561A.122L}
{Langer}, W.~D., {Velusamy}, T., {Pineda}, J.~L., {Willacy}, K., \&
  {Goldsmith}, P.~F. 2014, \aap, 561, A122

\bibitem[{{Larson}(1981)}]{1981MNRAS.194..809L}
{Larson}, R.~B. 1981, \mnras, 194, 809

\bibitem[{{Levine} {et~al.}(2006){Levine}, {Blitz}, \&
  {Heiles}}]{2006ApJ...643..881L}
{Levine}, E.~S., {Blitz}, L., \& {Heiles}, C. 2006, \apj, 643, 881

\bibitem[{{Majaess} {et~al.}(2009){Majaess}, {Turner}, \&
  {Lane}}]{2009MNRAS.398..263M}
{Majaess}, D.~J., {Turner}, D.~G., \& {Lane}, D.~J. 2009, \mnras, 398, 263

\bibitem[{{Marshall} {et~al.}(2006){Marshall}, {Robin}, {Reyl{\'e}},
  {Schultheis}, \& {Picaud}}]{2006A&A...453..635M}
{Marshall}, D.~J., {Robin}, A.~C., {Reyl{\'e}}, C., {Schultheis}, M., \&
  {Picaud}, S. 2006, \aap, 453, 635

\bibitem[{{Miville-Desch{\^e}nes} {et~al.}(2017){Miville-Desch{\^e}nes},
  {Murray}, \& {Lee}}]{2017ApJ...834...57M}
{Miville-Desch{\^e}nes}, M.-A., {Murray}, N., \& {Lee}, E.~J. 2017, \apj, 834,
  57

\bibitem[{{Moitinho} {et~al.}(2006){Moitinho}, {V{\'a}zquez}, {Carraro},
  {Baume}, {Giorgi}, \& {Lyra}}]{2006MNRAS.368L..77M}
{Moitinho}, A., {V{\'a}zquez}, R.~A., {Carraro}, G., {et~al.} 2006, \mnras,
  368, L77

\bibitem[{{Momany} {et~al.}(2006){Momany}, {Zaggia}, {Gilmore}, {Piotto},
  {Carraro}, {Bedin}, \& {de Angeli}}]{2006A&A...451..515M}
{Momany}, Y., {Zaggia}, S., {Gilmore}, G., {et~al.} 2006, \aap, 451, 515

\bibitem[{{Nakagawa} {et~al.}(2005){Nakagawa}, {Onishi}, {Mizuno}, \&
  {Fukui}}]{2005PASJ...57..917N}
{Nakagawa}, M., {Onishi}, T., {Mizuno}, A., \& {Fukui}, Y. 2005, \pasj, 57, 917

\bibitem[{{Oka} {et~al.}(2001){Oka}, {Hasegawa}, {Sato}, {Tsuboi}, {Miyazaki},
  \& {Sugimoto}}]{2001ApJ...562..348O}
{Oka}, T., {Hasegawa}, T., {Sato}, F., {et~al.} 2001, \apj, 562, 348

\bibitem[{{Oort} {et~al.}(1958){Oort}, {Kerr}, \&
  {Westerhout}}]{1958MNRAS.118..379O}
{Oort}, J.~H., {Kerr}, F.~J., \& {Westerhout}, G. 1958, \mnras, 118, 379

\bibitem[{{Pineda} {et~al.}(2013){Pineda}, {Langer}, {Velusamy}, \&
  {Goldsmith}}]{2013A&A...554A.103P}
{Pineda}, J.~L., {Langer}, W.~D., {Velusamy}, T., \& {Goldsmith}, P.~F. 2013,
  \aap, 554, A103

\bibitem[{{Reed}(2006)}]{2006JRASC.100..146R}
{Reed}, B.~C. 2006, \jrasc, 100, 146

\bibitem[{{Reid} {et~al.}(2014){Reid}, {Menten}, {Brunthaler}, {Zheng}, {Dame},
  {Xu}, {Wu}, {Zhang}, {Sanna}, {Sato}, {Hachisuka}, {Choi}, {Immer},
  {Moscadelli}, {Rygl}, \& {Bartkiewicz}}]{2014ApJ...783..130R}
{Reid}, M.~J., {Menten}, K.~M., {Brunthaler}, A., {et~al.} 2014, \apj, 783, 130

\bibitem[{{Reshetnikov} {et~al.}(2002){Reshetnikov}, {Battaner}, {Combes}, \&
  {Jim{\'e}nez-Vicente}}]{2002A&A...382..513R}
{Reshetnikov}, V., {Battaner}, E., {Combes}, F., \& {Jim{\'e}nez-Vicente}, J.
  2002, \aap, 382, 513

\bibitem[{{Reshetnikov} \& {Combes}(1998)}]{1998A&A...337....9R}
{Reshetnikov}, V., \& {Combes}, F. 1998, \aap, 337, 9

\bibitem[{{Reyl{\'e}} {et~al.}(2009){Reyl{\'e}}, {Marshall}, {Robin}, \&
  {Schultheis}}]{2009A&A...495..819R}
{Reyl{\'e}}, C., {Marshall}, D.~J., {Robin}, A.~C., \& {Schultheis}, M. 2009,
  \aap, 495, 819

\bibitem[{{Rice} {et~al.}(2016){Rice}, {Goodman}, {Bergin}, {Beaumont}, \&
  {Dame}}]{2016ApJ...822...52R}
{Rice}, T.~S., {Goodman}, A.~A., {Bergin}, E.~A., {Beaumont}, C., \& {Dame},
  T.~M. 2016, \apj, 822, 52

\bibitem[{{Roman-Duval} {et~al.}(2010){Roman-Duval}, {Jackson}, {Heyer},
  {Rathborne}, \& {Simon}}]{2010ApJ...723..492R}
{Roman-Duval}, J., {Jackson}, J.~M., {Heyer}, M., {Rathborne}, J., \& {Simon},
  R. 2010, \apj, 723, 492

\bibitem[{{Rudolph} {et~al.}(1996){Rudolph}, {Brand}, {de Geus}, \&
  {Wouterloot}}]{1996ApJ...458..653R}
{Rudolph}, A.~L., {Brand}, J., {de Geus}, E.~J., \& {Wouterloot}, J.~G.~A.
  1996, \apj, 458, 653

\bibitem[{{Russeil}(2003)}]{2003A&A...397..133R}
{Russeil}, D. 2003, \aap, 397, 133

\bibitem[{{Schwarzkopf} \& {Dettmar}(2001)}]{2001A&A...373..402S}
{Schwarzkopf}, U., \& {Dettmar}, R.-J. 2001, \aap, 373, 402

\bibitem[{{Scoville} {et~al.}(1987){Scoville}, {Yun}, {Sanders}, {Clemens}, \&
  {Waller}}]{1987ApJS...63..821S}
{Scoville}, N.~Z., {Yun}, M.~S., {Sanders}, D.~B., {Clemens}, D.~P., \&
  {Waller}, W.~H. 1987, \apjs, 63, 821

\bibitem[{{Shan} {et~al.}(2012){Shan}, {Yang}, {Shi}, {Yao}, {Zuo}, H., {Chen},
  {Zhang}, {Duan}, {Cao}, {Li}, {Li}, {Liu}, \& {Zhong}}]{Shan}
{Shan}, W.~L., {Yang}, J., {Shi}, S.~C., {et~al.} 2012, IEEE Transactions on
  Terahertz Science and Technology, 2, 593

\bibitem[{{Shetty} {et~al.}(2011){Shetty}, {Glover}, {Dullemond}, \&
  {Klessen}}]{2011MNRAS.412.1686S}
{Shetty}, R., {Glover}, S.~C., {Dullemond}, C.~P., \& {Klessen}, R.~S. 2011,
  \mnras, 412, 1686

\bibitem[{{Sodroski}(1991)}]{1991ApJ...366...95S}
{Sodroski}, T.~J. 1991, \apj, 366, 95

\bibitem[{{Solomon} \& {Rivolo}(1989)}]{1989ApJ...339..919S}
{Solomon}, P.~M., \& {Rivolo}, A.~R. 1989, \apj, 339, 919

\bibitem[{{Solomon} {et~al.}(1987){Solomon}, {Rivolo}, {Barrett}, \&
  {Yahil}}]{1987ApJ...319..730S}
{Solomon}, P.~M., {Rivolo}, A.~R., {Barrett}, J., \& {Yahil}, A. 1987, \apj,
  319, 730

\bibitem[{{Su} {et~al.}(2016){Su}, {Sun}, {Li}, {Zhang}, {Zhou}, {Fang},
  {Yang}, \& {Chen}}]{2016ApJ...828...59S}
{Su}, Y., {Sun}, Y., {Li}, C., {et~al.} 2016, \apj, 828, 59

\bibitem[{{Sun} {et~al.}(2015){Sun}, {Xu}, {Yang}, {Li}, {Du}, {Zhang}, \&
  {Zhou}}]{2015ApJ...798L..27S}
{Sun}, Y., {Xu}, Y., {Yang}, J., {et~al.} 2015, \apjl, 798, L27

\bibitem[{{Taylor}(2005)}]{2005ASPC..347...29T}
{Taylor}, M.~B. 2005, in Astronomical Society of the Pacific Conference Series,
  Vol. 347, Astronomical Data Analysis Software and Systems XIV, ed.
  P.~{Shopbell}, M.~{Britton}, \& R.~{Ebert}, 29

\bibitem[{{Vall{\'e}e}(2008)}]{2008AJ....135.1301V}
{Vall{\'e}e}, J.~P. 2008, \aj, 135, 1301

\bibitem[{{Vall{\'e}e}(2014)}]{2014AJ....148....5V}
---. 2014, \aj, 148, 5

\bibitem[{{Wolfire} {et~al.}(2010){Wolfire}, {Hollenbach}, \&
  {McKee}}]{2010ApJ...716.1191W}
{Wolfire}, M.~G., {Hollenbach}, D., \& {McKee}, C.~F. 2010, \apj, 716, 1191

\bibitem[{{Wouterloot} {et~al.}(1990){Wouterloot}, {Brand}, {Burton}, \&
  {Kwee}}]{1990A&A...230...21W}
{Wouterloot}, J.~G.~A., {Brand}, J., {Burton}, W.~B., \& {Kwee}, K.~K. 1990,
  \aap, 230, 21

\bibitem[{{Yasui} {et~al.}(2006){Yasui}, {Kobayashi}, {Tokunaga}, {Terada}, \&
  {Saito}}]{2006ApJ...649..753Y}
{Yasui}, C., {Kobayashi}, N., {Tokunaga}, A.~T., {Terada}, H., \& {Saito}, M.
  2006, \apj, 649, 753

\bibitem[{{Yasui} {et~al.}(2008){Yasui}, {Kobayashi}, {Tokunaga}, {Terada}, \&
  {Saito}}]{2008ApJ...675..443Y}
---. 2008, \apj, 675, 443

\end{thebibliography}

\clearpage
\clearpage
\clearpage
\clearpage
\clearpage
\clearpage
\startlongtable
\begin{deluxetable}{lcccccccccccc}
\tabletypesize{\tiny}
 \setlength{\tabcolsep}{0.03in}
\tablewidth{0pt} \tablecaption{Parameters of molecular clouds derived by $^{12}$CO(1-0).\label{tab:12co}}
\tablehead{ Name & $V_{\rm LSR}$&$\Delta$$V$ & $I_{\rm ^{12}CO}$ & $T_{\rm peak}$ & Area& $d$ & $R$ & $Z$ scale & Radius &  Mass& $M_{\rm vir}$&$\Sigma_{\rm gas}$\\
    & (\kms) &(\kms) & (K.km$\,$s$^{-1}$) & (K)&(arcmin$^2$)&(kpc)&(kpc)&(kpc)&(pc)&(10$^2$$M_{\odot}$)&(10$^2$$M_{\odot}$)&($M_{\odot}$ pc$^{-2}$)  \\
(1)&(2)&(3)&(4)&(5)&(6)&(7)&(8)&(9)&(10)&(11)&(12)&(13)}
\startdata
MWISP G34.767$-$1.483 & -46.2 &   1.0 &   2.0 &   2.0 &    1.2 &   17.9 &   12.1 &   -0.5 &    2.4 &    2.3 &   4.6 &  13.0 \\
MWISP G34.792$+$1.517 & -48.1 &   1.1 &   1.9 &   1.7 &    3.4 &   18.2 &   12.3 &    0.5 &    5.0 &    5.0 &  12.7 &   6.4 \\
MWISP G34.842$-$0.950 & -45.1 &   2.2 &   5.9 &   2.5 &    9.7 &   17.8 &   11.9 &   -0.3 &    8.8 &   26.5 &  92.6 &  10.9 \\
MWISP G36.375$+$2.750 & -53.5 &   1.9 &   2.9 &   1.5 &    2.2 &   18.5 &   12.8 &    0.9 &    3.9 &    5.0 &  27.9 &  10.7 \\
MWISP G36.450$+$0.183 & -47.0 &   3.1 &   5.9 &   1.8 &    7.4 &   17.6 &   11.9 &    0.1 &    7.5 &   21.8 & 155.9 &  12.2 \\
MWISP G36.475$+$0.225 & -46.3 &   2.3 &   3.6 &   1.5 &    3.0 &   17.5 &   11.8 &    0.1 &    4.4 &    7.6 &  49.8 &  12.5 \\
MWISP G36.583$-$0.192 & -50.1 &   1.8 &   2.7 &   1.4 &    2.1 &   17.9 &   12.3 &   -0.1 &    3.7 &    4.8 &  24.5 &  11.2 \\
MWISP G36.600$-$0.308 & -52.1 &   0.5 &   1.4 &   2.6 &    1.1 &   18.2 &   12.5 &   -0.1 &    2.1 &    1.9 &   1.2 &  13.0 \\
MWISP G36.733$+$2.475 & -56.8 &   1.5 &   3.3 &   2.1 &    3.4 &   18.9 &   13.2 &    0.8 &    5.2 &    8.2 &  23.9 &   9.6 \\
MWISP G36.767$+$2.517 & -57.9 &   1.1 &   1.5 &   1.4 &    0.9 &   19.0 &   13.3 &    0.8 &    1.6 &    1.3 &   3.8 &  16.0 \\
MWISP G36.808$-$0.117 & -60.4 &   1.5 &   3.1 &   1.9 &    2.6 &   19.4 &   13.7 &   -0.0 &    4.5 &    5.7 &  22.5 &   9.1 \\
MWISP G36.900$-$0.100 & -62.5 &   3.6 &   6.8 &   1.8 &    8.8 &   19.8 &   14.0 &   -0.0 &    9.3 &   37.4 & 252.2 &  13.7 \\
MWISP G36.917$+$0.000 & -57.2 &   1.5 &   4.1 &   2.7 &    3.8 &   18.9 &   13.2 &    0.0 &    5.6 &   11.0 &  25.2 &  11.2 \\
MWISP G37.050$+$0.117 & -52.5 &   2.0 &   4.9 &   2.3 &    5.0 &   18.1 &   12.5 &    0.0 &    6.3 &   15.1 &  52.8 &  12.3 \\
MWISP G37.058$+$0.783 & -61.7 &   1.1 &   1.6 &   1.4 &    1.1 &   19.5 &   13.8 &    0.3 &    2.2 &    2.0 &   6.0 &  12.9 \\
MWISP G37.100$+$0.833 & -47.6 &   2.7 &   5.1 &   1.8 &    4.5 &   17.5 &   11.9 &    0.3 &    5.7 &   13.5 &  84.6 &  13.4 \\
MWISP G37.175$+$0.792 & -48.1 &   4.9 &  13.9 &   2.6 &   23.1 &   17.5 &   12.0 &    0.2 &   13.6 &  132.6 & 690.3 &  22.7 \\
MWISP G37.258$+$1.217 & -56.0 &   1.2 &   2.5 &   1.9 &    1.1 &   18.6 &   13.0 &    0.4 &    2.2 &    2.7 &   7.2 &  17.7 \\
MWISP G37.292$+$1.158 & -51.4 &   1.3 &   1.8 &   1.3 &    1.4 &   17.9 &   12.4 &    0.4 &    2.7 &    2.7 &  10.0 &  11.7 \\
MWISP G37.308$+$1.125 & -52.7 &   1.2 &   1.6 &   1.2 &    1.9 &   18.1 &   12.5 &    0.4 &    3.4 &    2.9 &  11.1 &   8.1 \\
MWISP G37.342$+$0.708 & -50.4 &   1.2 &   2.2 &   1.8 &    1.9 &   17.8 &   12.2 &    0.2 &    3.3 &    3.7 &  10.0 &  10.8 \\
MWISP G37.350$+$1.050 & -53.9 &   2.9 &  20.4 &   6.6 &    9.7 &   18.2 &   12.7 &    0.3 &    9.0 &   70.6 & 158.9 &  27.6 \\
MWISP G37.383$+$2.417 & -61.1 &   1.5 &   2.6 &   1.6 &    1.8 &   19.3 &   13.7 &    0.8 &    3.5 &    4.8 &  17.6 &  12.2 \\
MWISP G37.408$+$1.183 & -53.8 &   1.5 &   3.5 &   2.2 &    3.6 &   18.2 &   12.6 &    0.4 &    5.1 &    8.2 &  24.3 &   9.9 \\
MWISP G37.425$+$1.267 & -52.2 &   2.1 &   8.0 &   3.6 &   12.4 &   18.0 &   12.4 &    0.4 &   10.2 &   43.8 &  92.8 &  13.5 \\
MWISP G37.433$+$1.200 & -53.6 &   0.9 &   1.9 &   2.0 &    5.0 &   18.2 &   12.6 &    0.4 &    6.3 &    8.1 &  10.8 &   6.6 \\
MWISP G37.450$+$1.300 & -51.6 &   1.8 &   2.7 &   1.4 &    1.8 &   17.9 &   12.4 &    0.4 &    3.2 &    3.1 &  20.9 &   9.8 \\
MWISP G37.467$+$1.325 & -54.0 &   1.7 &   4.4 &   2.5 &    1.8 &   18.2 &   12.7 &    0.4 &    3.3 &    5.1 &  19.4 &  14.4 \\
MWISP G37.467$+$2.700 & -57.7 &   1.9 &   4.2 &   2.0 &    3.5 &   18.8 &   13.2 &    0.9 &    5.2 &   10.0 &  41.5 &  11.6 \\
MWISP G37.475$+$0.600 & -51.0 &   1.5 &   4.2 &   2.6 &    2.3 &   17.8 &   12.3 &    0.2 &    3.8 &    4.9 &  18.7 &  10.7 \\
MWISP G37.475$+$1.242 & -53.8 &   1.6 &   2.5 &   1.4 &    4.3 &   18.2 &   12.6 &    0.4 &    5.8 &    9.0 &  31.5 &   8.6 \\
MWISP G37.492$+$1.333 & -51.4 &   0.8 &   1.5 &   1.8 &    1.3 &   17.9 &   12.3 &    0.4 &    2.4 &    2.0 &   2.9 &  10.7 \\
MWISP G37.500$+$0.508 & -50.8 &   2.0 &   2.5 &   1.2 &    1.0 &   17.8 &   12.3 &    0.2 &    1.9 &    1.4 &  16.9 &  11.9 \\
MWISP G37.500$+$0.592 & -51.0 &   1.8 &   2.3 &   1.2 &    1.7 &   17.8 &   12.3 &    0.2 &    3.1 &    3.3 &  21.4 &  10.9 \\
MWISP G37.592$-$0.242 & -49.1 &   2.1 &   4.6 &   2.1 &    3.2 &   17.5 &   12.0 &   -0.1 &    4.7 &   10.1 &  43.1 &  14.7 \\
MWISP G37.600$+$1.408 & -58.5 &   1.2 &   1.8 &   1.4 &    2.1 &   18.9 &   13.3 &    0.5 &    3.9 &    3.5 &  10.8 &   7.4 \\
MWISP G37.608$+$0.833 & -48.5 &   2.8 &   4.4 &   1.5 &    9.7 &   17.4 &   12.0 &    0.3 &    8.6 &   27.1 & 143.3 &  11.5 \\
MWISP G37.617$+$2.408 & -54.7 &   1.6 &   2.3 &   1.4 &    2.4 &   18.3 &   12.7 &    0.8 &    4.0 &    4.5 &  20.2 &   8.9 \\
MWISP G37.633$+$2.550 & -60.8 &   1.2 &   2.1 &   1.6 &    1.6 &   19.2 &   13.6 &    0.9 &    3.1 &    3.0 &   9.3 &   9.6 \\
MWISP G37.642$+$1.400$^*$ & -47.1 &   2.1 &   5.8 &   2.6 &    9.3 &   17.2 &   11.8 &    0.4 &    8.4 &   31.8 &  79.8 &  14.4 \\
MWISP G37.667$+$0.850 & -48.4 &   1.2 &   1.7 &   1.4 &    1.2 &   17.4 &   11.9 &    0.3 &    2.2 &    1.9 &   6.6 &  12.5 \\
MWISP G37.667$+$1.267 & -57.8 &   3.2 &  14.7 &   4.4 &   19.6 &   18.7 &   13.1 &    0.4 &   13.4 &  106.7 & 282.5 &  19.0 \\
MWISP G37.683$+$0.533 & -52.7 &   2.3 &   4.2 &   1.7 &    2.2 &   18.0 &   12.5 &    0.2 &    3.7 &    6.4 &  42.6 &  14.7 \\
MWISP G37.708$+$1.367 & -51.8 &   2.4 &  15.8 &   6.3 &   26.0 &   17.8 &   12.3 &    0.4 &   14.8 &  138.7 & 172.5 &  20.3 \\
MWISP G37.708$+$1.125 & -57.8 &   2.6 &   3.9 &   1.4 &    7.5 &   18.7 &   13.1 &    0.4 &    8.1 &   22.5 & 113.4 &  11.1 \\
MWISP G37.717$-$0.442 & -51.0 &   1.2 &   2.5 &   2.0 &    4.9 &   17.7 &   12.2 &   -0.1 &    6.0 &   10.2 &  17.5 &   8.9 \\
MWISP G37.742$+$1.258 & -54.8 &   1.7 &   3.5 &   2.0 &    2.9 &   18.2 &   12.7 &    0.4 &    4.5 &    7.3 &  26.3 &  11.4 \\
MWISP G37.742$+$1.183 & -56.8 &   1.1 &   1.5 &   1.3 &    4.8 &   18.5 &   13.0 &    0.4 &    6.2 &    8.6 &  15.3 &   7.1 \\
MWISP G37.758$+$1.392 & -51.8 &   2.6 &  15.3 &   5.5 &   13.9 &   17.8 &   12.3 &    0.4 &   10.7 &   80.2 & 154.4 &  22.4 \\
MWISP G37.767$+$1.783 & -61.6 &   1.4 &   2.0 &   1.4 &    1.3 &   19.3 &   13.7 &    0.6 &    2.6 &    2.3 &  10.5 &  11.0 \\
MWISP G37.792$+$1.275 & -52.9 &   1.2 &   3.3 &   2.6 &    2.5 &   18.0 &   12.5 &    0.4 &    4.1 &    6.4 &  12.5 &  12.3 \\
MWISP G37.800$+$1.142 & -57.2 &   1.0 &   2.5 &   2.4 &    5.4 &   18.6 &   13.0 &    0.4 &    6.7 &   10.7 &  13.6 &   7.7 \\
MWISP G37.808$+$2.175 & -51.1 &   1.2 &   2.0 &   1.6 &    2.4 &   17.7 &   12.3 &    0.7 &    3.9 &    3.8 &  11.3 &   8.1 \\
MWISP G37.817$+$1.792 & -63.0 &   2.3 &   3.9 &   1.6 &    2.8 &   19.5 &   13.9 &    0.6 &    4.8 &    7.9 &  54.1 &  11.1 \\
MWISP G37.817$+$0.092$^*$ & -47.9 &   1.9 &   7.4 &   3.7 &   13.8 &   17.3 &   11.9 &    0.0 &   10.3 &   54.1 &  74.8 &  16.2 \\
MWISP G37.833$+$1.392 & -52.1 &   1.7 &   3.5 &   1.9 &    5.7 &   17.8 &   12.4 &    0.4 &    6.6 &   15.3 &  38.7 &  11.2 \\
MWISP G37.842$+$0.150$^{\dag}$ & -45.9 &   3.2 &   7.3 &   2.2 &    8.7 &   17.0 &   11.6 &    0.0 &    8.0 &   35.6 & 168.6 &  17.9 \\
MWISP G37.867$+$2.383 & -54.8 &   0.9 &   1.8 &   1.8 &    4.4 &   18.2 &   12.7 &    0.8 &    5.8 &    7.8 &  10.9 &   7.3 \\
MWISP G37.942$+$1.525 & -56.3 &   2.4 &   3.5 &   1.4 &    5.9 &   18.4 &   12.9 &    0.5 &    7.0 &   14.1 &  81.4 &   9.3 \\
MWISP G38.050$+$2.500 & -54.1 &   1.3 &   2.0 &   1.4 &    2.3 &   18.0 &   12.6 &    0.8 &    3.9 &    3.6 &  14.8 &   7.4 \\
MWISP G38.058$+$0.983 & -62.2 &   2.6 &   6.0 &   2.1 &    3.9 &   19.3 &   13.7 &    0.3 &    5.7 &   16.7 &  83.8 &  16.2 \\
MWISP G38.092$+$0.458 & -48.3 &   1.9 &   3.3 &   1.6 &    3.1 &   17.3 &   11.9 &    0.1 &    4.5 &    7.3 &  34.6 &  11.4 \\
MWISP G38.117$+$2.050 & -54.6 &   2.0 &   3.9 &   1.8 &    5.5 &   18.1 &   12.6 &    0.6 &    6.6 &   14.9 &  53.8 &  11.0 \\
MWISP G38.150$+$1.333 & -52.0 &   0.9 &   1.8 &   2.0 &    4.1 &   17.7 &   12.3 &    0.4 &    5.5 &    6.0 &   8.6 &   6.4 \\
MWISP G38.175$+$0.917 & -62.5 &   2.4 &   3.2 &   1.3 &    2.2 &   19.3 &   13.7 &    0.3 &    4.0 &    5.1 &  46.3 &  10.2 \\
MWISP G38.233$-$0.342 & -49.7 &   0.8 &   2.2 &   2.5 &    1.7 &   17.4 &   12.0 &   -0.1 &    2.9 &    2.3 &   4.1 &   8.6 \\
MWISP G38.317$+$0.908 & -49.4 &   2.7 &  12.4 &   4.3 &   15.3 &   17.4 &   12.0 &    0.3 &   10.9 &  109.5 & 166.3 &  29.2 \\
MWISP G38.458$+$2.750 & -55.8 &   1.5 &   4.0 &   2.5 &    2.8 &   18.2 &   12.7 &    0.9 &    4.4 &    6.1 &  21.5 &   9.9 \\
MWISP G38.492$+$1.200 & -50.4 &   1.1 &   4.9 &   4.2 &    5.0 &   17.4 &   12.1 &    0.4 &    6.0 &   13.2 &  14.9 &  11.7 \\
MWISP G38.533$+$0.892 & -54.6 &   2.3 &   4.7 &   2.0 &   11.5 &   18.0 &   12.6 &    0.3 &    9.8 &   36.6 & 105.9 &  12.2 \\
MWISP G38.550$+$0.908 & -54.5 &   2.6 &   4.9 &   1.8 &   11.3 &   18.0 &   12.6 &    0.3 &    9.6 &   37.1 & 137.9 &  12.8 \\
MWISP G38.558$+$1.600 & -55.3 &   0.9 &   1.7 &   1.9 &    4.2 &   18.1 &   12.7 &    0.5 &    5.7 &    6.1 &   8.9 &   6.1 \\
MWISP G38.642$+$1.842 & -56.1 &   3.0 &   4.6 &   1.5 &    2.3 &   18.2 &   12.8 &    0.6 &    3.8 &    6.4 &  70.9 &  13.7 \\
MWISP G38.708$+$1.200 & -50.1 &   1.7 &   2.6 &   1.4 &    3.3 &   17.3 &   12.0 &    0.4 &    4.7 &    6.7 &  28.1 &   9.5 \\
MWISP G38.775$+$2.917 & -58.1 &   1.6 &   2.4 &   1.4 &    1.1 &   18.4 &   13.0 &    0.9 &    2.1 &    1.7 &  11.3 &  12.2 \\
MWISP G39.000$+$2.242 & -60.3 &   3.8 &   5.4 &   1.3 &    4.5 &   18.6 &   13.3 &    0.7 &    6.1 &   11.3 & 181.1 &   9.8 \\
MWISP G39.075$+$1.592 & -51.1 &   1.0 &   2.0 &   1.9 &    2.2 &   17.4 &   12.1 &    0.5 &    3.6 &    3.8 &   7.3 &   9.2 \\
MWISP G39.175$-$1.425 & -55.6 &   3.3 &  11.0 &   3.1 &   19.2 &   17.9 &   12.6 &   -0.4 &   12.7 &   69.9 & 290.8 &  13.8 \\
MWISP G39.225$-$1.542 & -54.6 &   1.6 &   2.1 &   1.2 &    2.8 &   17.8 &   12.5 &   -0.5 &    4.3 &    4.8 &  23.8 &   8.3 \\
MWISP G39.342$+$1.400 & -59.2 &   2.0 &   4.1 &   1.9 &    6.4 &   18.4 &   13.1 &    0.4 &    7.3 &   15.1 &  61.7 &   9.1 \\
MWISP G39.492$+$0.667$^{*\dag}$ & -49.1 &   2.3 &  11.6 &   4.8 &   13.7 &   17.0 &   11.8 &    0.2 &   10.1 &   63.3 & 110.2 &  19.7 \\
MWISP G39.500$+$2.125 & -53.0 &   1.7 &   2.5 &   1.4 &    1.8 &   17.5 &   12.3 &    0.6 &    3.2 &    3.4 &  18.5 &  10.7 \\
MWISP G39.675$+$1.400 & -58.6 &   1.0 &   2.2 &   2.1 &    4.2 &   18.2 &   12.9 &    0.4 &    5.7 &    7.5 &  11.3 &   7.3 \\
MWISP G39.683$+$1.867 & -54.4 &   1.2 &   2.1 &   1.7 &    0.9 &   17.6 &   12.4 &    0.6 &    1.6 &    1.4 &   4.7 &  18.6 \\
MWISP G39.775$+$2.992 & -56.6 &   1.4 &   3.1 &   2.1 &    6.9 &   17.9 &   12.7 &    0.9 &    7.4 &   14.9 &  29.4 &   8.7 \\
MWISP G39.883$+$1.492 & -60.3 &   1.3 &   2.3 &   1.6 &    2.2 &   18.4 &   13.1 &    0.5 &    3.9 &    4.3 &  14.2 &   9.1 \\
MWISP G39.908$+$0.958 & -50.9 &   2.5 &   5.9 &   2.2 &    8.8 &   17.1 &   12.0 &    0.3 &    8.1 &   24.3 & 103.5 &  11.9 \\
MWISP G39.908$+$1.058$^*$ & -50.9 &   3.2 &   9.9 &   2.9 &   11.0 &   17.1 &   12.0 &    0.3 &    9.1 &   46.0 & 192.8 &  17.8 \\
MWISP G39.983$+$0.908 & -52.0 &   1.9 &   4.5 &   2.3 &    2.2 &   17.2 &   12.1 &    0.3 &    3.6 &    6.3 &  26.9 &  15.0 \\
MWISP G40.067$+$0.700 & -52.6 &   0.8 &   1.9 &   2.1 &    1.8 &   17.3 &   12.1 &    0.2 &    3.1 &    2.5 &   4.8 &   8.1 \\
MWISP G40.092$+$2.758 & -62.0 &   2.1 &   4.4 &   1.9 &    3.2 &   18.5 &   13.3 &    0.9 &    4.9 &   10.2 &  46.6 &  13.7 \\
MWISP G40.292$+$1.150$^{*\dag}$ & -50.6 &   3.1 &  12.5 &   3.8 &    4.8 &   17.0 &   11.9 &    0.3 &    5.7 &   20.1 & 111.7 &  19.8 \\
MWISP G40.317$+$1.542 & -56.1 &   2.7 &  10.1 &   3.5 &   13.3 &   17.7 &   12.5 &    0.5 &   10.3 &   49.2 & 160.0 &  14.7 \\
MWISP G40.342$+$0.592 & -55.2 &   1.7 &   4.5 &   2.5 &    1.5 &   17.5 &   12.4 &    0.2 &    2.7 &    4.9 &  16.6 &  21.7 \\
MWISP G40.350$+$4.775 & -61.1 &   1.9 &   3.4 &   1.7 &    4.5 &   18.3 &   13.1 &    1.5 &    5.9 &    7.7 &  42.7 &   7.0 \\
MWISP G40.392$+$1.550 & -61.0 &   1.7 &   3.6 &   2.0 &    2.0 &   18.3 &   13.1 &    0.5 &    3.5 &    4.5 &  21.7 &  11.5 \\
MWISP G40.408$+$3.133 & -61.7 &   1.7 &   2.7 &   1.5 &    2.8 &   18.4 &   13.2 &    1.0 &    4.5 &    5.3 &  27.8 &   8.4 \\
MWISP G40.433$+$1.208 & -52.1 &   3.3 &   5.1 &   1.4 &    7.3 &   17.1 &   12.0 &    0.4 &    7.3 &   17.5 & 171.8 &  10.4 \\
MWISP G40.442$+$1.150$^{*\dag}$ & -50.0 &   3.1 &   6.2 &   1.9 &    4.0 &   16.9 &   11.8 &    0.3 &    5.1 &   12.4 & 104.8 &  15.0 \\
MWISP G40.483$+$1.292 & -58.0 &   2.0 &   2.9 &   1.4 &    2.8 &   17.9 &   12.7 &    0.4 &    4.4 &    4.1 &  36.2 &   6.9 \\
MWISP G40.500$+$1.517 & -60.4 &   2.1 &   3.7 &   1.7 &    2.4 &   18.2 &   13.0 &    0.5 &    4.1 &    7.3 &  37.2 &  14.1 \\
MWISP G40.517$+$1.150 & -53.9 &   4.3 &   5.9 &   1.3 &    5.1 &   17.3 &   12.2 &    0.3 &    6.0 &   12.9 & 239.5 &  11.3 \\
MWISP G40.533$+$1.267$^{\dag}$& -50.2 &   5.3 &  14.3 &   2.6 &   13.8 &   16.9 &   11.8 &    0.4 &   10.0 &   57.7 & 585.3 &  18.2 \\
MWISP G40.558$+$1.092 & -53.9 &   2.8 &   4.2 &   1.4 &    2.3 &   17.3 &   12.2 &    0.3 &    3.7 &    5.8 &  63.1 &  13.5 \\
MWISP G40.592$+$1.517 & -58.7 &   2.7 &   6.7 &   2.4 &    6.3 &   17.9 &   12.8 &    0.5 &    7.0 &   25.0 & 105.4 &  16.2 \\
MWISP G40.600$+$1.325 & -57.1 &   1.4 &   3.3 &   2.3 &    3.8 &   17.7 &   12.6 &    0.4 &    5.2 &    8.0 &  20.7 &   9.3 \\
MWISP G40.633$+$3.117 & -55.9 &   2.4 &   4.3 &   1.7 &    9.5 &   17.5 &   12.5 &    1.0 &    8.6 &   25.0 & 103.4 &  10.8 \\
MWISP G40.633$+$2.383 & -59.2 &   2.6 &   5.0 &   1.8 &    4.8 &   18.0 &   12.9 &    0.7 &    6.1 &   12.3 &  89.1 &  10.7 \\
MWISP G40.767$+$1.308 & -57.8 &   1.5 &   2.2 &   1.3 &    7.8 &   17.8 &   12.7 &    0.4 &    7.8 &   10.6 &  38.3 &   5.5 \\
MWISP G40.825$+$1.400 & -57.1 &   3.6 &   7.9 &   2.1 &   15.4 &   17.6 &   12.6 &    0.4 &   11.1 &   58.3 & 298.2 &  14.9 \\
MWISP G40.825$+$3.042 & -58.6 &   1.9 &   3.2 &   1.6 &    1.8 &   17.8 &   12.8 &    0.9 &    3.3 &    3.4 &  24.3 &  10.2 \\
MWISP G40.833$+$1.300 & -55.7 &   2.6 &   8.8 &   3.2 &   20.5 &   17.5 &   12.4 &    0.4 &   12.8 &   58.6 & 184.2 &  11.4 \\
MWISP G40.833$+$1.925 & -60.7 &   2.3 &   6.0 &   2.4 &   25.0 &   18.1 &   13.0 &    0.6 &   14.7 &   80.6 & 168.2 &  11.8 \\
MWISP G40.858$+$0.425 & -58.5 &   3.7 &   5.8 &   1.5 &    5.1 &   17.8 &   12.7 &    0.1 &    6.2 &    9.6 & 178.4 &   7.9 \\
MWISP G40.867$+$2.558 & -62.5 &   2.4 &   4.6 &   1.8 &    3.7 &   18.4 &   13.3 &    0.8 &    5.4 &   12.7 &  64.7 &  14.1 \\
MWISP G40.883$+$0.492 & -54.9 &   3.8 &   9.0 &   2.2 &    9.4 &   17.4 &   12.3 &    0.1 &    8.5 &   30.7 & 250.6 &  13.7 \\
MWISP G40.892$+$1.308 & -57.1 &   2.2 &   8.6 &   3.6 &    8.9 &   17.6 &   12.6 &    0.4 &    8.3 &   34.5 &  86.5 &  15.9 \\
MWISP G40.900$+$1.842 & -56.8 &   1.8 &   4.7 &   2.4 &   27.0 &   17.6 &   12.5 &    0.6 &   14.8 &   67.9 & 102.9 &   9.8 \\
MWISP G40.925$+$1.267 & -56.6 &   0.7 &   2.1 &   2.8 &    3.9 &   17.5 &   12.5 &    0.4 &    5.2 &    6.9 &   5.7 &   7.9 \\
MWISP G40.925$+$2.983 & -56.1 &   1.5 &   2.3 &   1.4 &    1.7 &   17.5 &   12.4 &    0.9 &    3.0 &    2.7 &  14.7 &   9.3 \\
MWISP G40.942$+$1.208 & -56.9 &   1.6 &   2.7 &   1.6 &    2.2 &   17.6 &   12.5 &    0.4 &    3.7 &    3.4 &  19.9 &   8.2 \\
MWISP G40.942$+$2.542 & -59.4 &   1.1 &   1.6 &   1.4 &    2.0 &   17.9 &   12.8 &    0.8 &    3.4 &    3.0 &   8.8 &   8.0 \\
MWISP G40.958$+$2.483 & -59.3 &   1.9 &   7.1 &   3.5 &    4.2 &   17.9 &   12.8 &    0.8 &    5.6 &   17.8 &  42.6 &  18.0 \\
MWISP G40.958$+$0.500 & -57.5 &   1.5 &   1.9 &   1.2 &    2.3 &   17.7 &   12.6 &    0.2 &    3.8 &    4.5 &  18.3 &  10.1 \\
MWISP G40.958$+$2.975 & -58.4 &   1.7 &   2.5 &   1.3 &    2.9 &   17.8 &   12.7 &    0.9 &    4.4 &    4.5 &  28.1 &   7.4 \\
MWISP G40.967$+$2.958 & -60.8 &   1.6 &   3.4 &   2.0 &    3.5 &   18.1 &   13.0 &    0.9 &    5.1 &    7.7 &  26.2 &   9.6 \\
MWISP G40.992$+$0.358 & -53.7 &   1.0 &   1.7 &   1.5 &    1.0 &   17.2 &   12.2 &    0.1 &    1.9 &    1.4 &   4.1 &  12.0 \\
MWISP G41.008$+$0.492 & -58.6 &   2.4 &   4.0 &   1.6 &    2.5 &   17.8 &   12.7 &    0.2 &    4.0 &    4.8 &  48.6 &   9.4 \\
MWISP G41.008$+$2.758 & -61.5 &   1.5 &   2.3 &   1.4 &    1.6 &   18.2 &   13.1 &    0.9 &    3.0 &    2.8 &  15.0 &   9.7 \\
MWISP G41.067$+$2.683 & -63.6 &   2.1 &   5.9 &   2.6 &    9.7 &   18.5 &   13.4 &    0.9 &    9.1 &   36.5 &  83.7 &  14.0 \\
MWISP G41.133$+$1.550 & -58.7 &   1.1 &   2.3 &   1.9 &    1.2 &   17.8 &   12.7 &    0.5 &    2.3 &    1.8 &   6.0 &  10.7 \\
MWISP G41.200$+$1.142 & -66.7 &   1.5 &   3.0 &   1.9 &    1.6 &   18.9 &   13.8 &    0.4 &    3.1 &    3.1 &  14.0 &  10.6 \\
MWISP G41.217$+$1.167 & -59.6 &   0.9 &   1.4 &   1.5 &    1.4 &   17.9 &   12.8 &    0.4 &    2.7 &    2.0 &   4.2 &   9.2 \\
MWISP G41.308$+$2.000 & -57.6 &   1.6 &   6.1 &   3.5 &    2.9 &   17.6 &   12.6 &    0.6 &    4.3 &    8.6 &  24.7 &  14.4 \\
MWISP G41.325$+$1.925 & -57.8 &   1.2 &   2.0 &   1.6 &    2.1 &   17.6 &   12.6 &    0.6 &    3.5 &    3.1 &  10.3 &   7.8 \\
MWISP G41.342$+$1.708 & -61.4 &   2.3 &   3.8 &   1.5 &    2.6 &   18.1 &   13.0 &    0.5 &    4.2 &    5.9 &  47.3 &  10.7 \\
MWISP G41.367$+$1.892 & -59.4 &   1.7 &   4.3 &   2.4 &   10.8 &   17.8 &   12.8 &    0.6 &    9.3 &   31.9 &  54.3 &  11.7 \\
MWISP G41.375$+$1.183 & -59.2 &   1.5 &   4.3 &   2.8 &    3.1 &   17.8 &   12.8 &    0.4 &    4.6 &    8.7 &  21.3 &  13.0 \\
MWISP G41.417$+$1.975 & -59.6 &   1.8 &   3.0 &   1.6 &    2.7 &   17.8 &   12.8 &    0.6 &    4.3 &    5.2 &  28.0 &   9.0 \\
MWISP G41.500$+$1.017 & -59.1 &   2.8 &   6.2 &   2.1 &   21.1 &   17.7 &   12.7 &    0.3 &   13.2 &   60.0 & 221.1 &  11.0 \\
MWISP G41.583$+$1.725 & -60.6 &   1.6 &   4.7 &   2.8 &    7.8 &   17.9 &   12.9 &    0.5 &    7.9 &   20.1 &  40.6 &  10.3 \\
MWISP G41.733$+$1.517 & -55.4 &   1.5 &   2.5 &   1.5 &    1.5 &   17.2 &   12.3 &    0.5 &    2.7 &    3.3 &  12.9 &  15.0 \\
MWISP G41.742$+$1.458 & -55.6 &   2.6 &   7.7 &   2.8 &    6.7 &   17.2 &   12.3 &    0.4 &    7.0 &   25.3 &  96.4 &  16.5 \\
MWISP G41.750$+$0.967 & -60.6 &   2.2 &   6.5 &   2.7 &    7.5 &   17.8 &   12.9 &    0.3 &    7.7 &   25.3 &  80.4 &  13.6 \\
MWISP G41.758$+$1.567 & -55.8 &   0.8 &   1.4 &   1.8 &    1.6 &   17.2 &   12.3 &    0.5 &    2.9 &    1.9 &   3.5 &   7.2 \\
MWISP G41.992$+$1.233 & -57.0 &   1.8 &   2.5 &   1.3 &    2.9 &   17.3 &   12.4 &    0.4 &    4.3 &    4.3 &  30.2 &   7.3 \\
MWISP G42.025$+$1.358 & -55.4 &   3.3 &   3.8 &   1.1 &    2.7 &   17.1 &   12.2 &    0.4 &    4.0 &    6.2 &  92.4 &  12.1 \\
MWISP G42.058$+$1.325 & -60.3 &   1.5 &   2.2 &   1.4 &    1.2 &   17.7 &   12.8 &    0.4 &    2.3 &    1.9 &  10.3 &  11.5 \\
MWISP G42.192$+$1.083 & -57.9 &   2.1 &   2.7 &   1.2 &    3.1 &   17.4 &   12.5 &    0.3 &    4.5 &    5.8 &  41.8 &   9.1 \\
MWISP G42.258$+$1.125 & -60.6 &   2.7 &   5.5 &   1.9 &    5.7 &   17.7 &   12.8 &    0.3 &    6.6 &   17.2 &  98.1 &  12.6 \\
MWISP G42.267$+$1.592 & -56.3 &   2.1 &   3.7 &   1.6 &    6.8 &   17.1 &   12.3 &    0.5 &    7.0 &   15.1 &  64.4 &   9.8 \\
MWISP G42.467$+$1.333 & -58.6 &   1.2 &   2.1 &   1.6 &    1.9 &   17.4 &   12.5 &    0.4 &    3.2 &    2.7 &  10.3 &   8.3 \\
MWISP G42.542$+$2.525 & -60.5 &   1.4 &   2.6 &   1.7 &    4.1 &   17.6 &   12.8 &    0.8 &    5.4 &    6.8 &  23.4 &   7.5 \\
MWISP G42.900$+$1.608$^*$ & -55.6 &   1.6 &   4.2 &   2.5 &    8.2 &   16.9 &   12.2 &    0.5 &    7.6 &   20.2 &  39.2 &  11.0 \\
MWISP G42.983$+$1.317 & -56.3 &   2.3 &   4.9 &   2.0 &    7.2 &   16.9 &   12.2 &    0.4 &    7.1 &   15.8 &  76.6 &   9.9 \\
MWISP G43.008$+$1.342 & -57.6 &   1.8 &   3.2 &   1.6 &    4.1 &   17.1 &   12.4 &    0.4 &    5.3 &    6.0 &  37.2 &   6.9 \\
MWISP G43.233$+$1.283 & -57.1 &   1.6 &   3.4 &   2.0 &    3.5 &   17.0 &   12.3 &    0.4 &    4.8 &    7.7 &  25.6 &  10.8 \\
MWISP G43.292$+$1.742 & -56.6 &   1.3 &   4.0 &   2.8 &    3.9 &   16.9 &   12.2 &    0.5 &    5.0 &    9.5 &  18.8 &  12.0 \\
MWISP G43.867$+$1.308 & -58.2 &   2.0 &   5.9 &   2.8 &    2.7 &   16.9 &   12.3 &    0.4 &    4.1 &    7.4 &  33.6 &  14.1 \\
MWISP G43.908$+$1.425 & -57.3 &   2.5 &   6.4 &   2.4 &    2.8 &   16.8 &   12.2 &    0.4 &    4.1 &    8.2 &  52.1 &  15.6 \\
MWISP G44.100$+$2.483 & -63.4 &   1.1 &   2.3 &   2.0 &    1.2 &   17.5 &   12.9 &    0.8 &    2.3 &    2.0 &   5.5 &  11.8 \\
MWISP G44.158$+$2.525 & -64.5 &   2.4 &   4.7 &   1.9 &    2.3 &   17.6 &   13.0 &    0.8 &    3.8 &    6.3 &  45.7 &  13.8 \\
MWISP G44.183$-$0.167 & -65.0 &   0.8 &   1.9 &   2.1 &    3.7 &   17.7 &   13.1 &   -0.1 &    5.1 &    7.5 &   7.2 &   9.3 \\
MWISP G44.325$-$0.208 & -65.2 &   2.0 &   4.2 &   2.0 &    7.0 &   17.7 &   13.1 &   -0.1 &    7.3 &   16.3 &  59.1 &   9.6 \\
MWISP G44.408$+$2.275 & -69.2 &   2.2 &   4.4 &   1.9 &    2.8 &   18.2 &   13.6 &    0.7 &    4.5 &    8.5 &  43.5 &  13.5 \\
MWISP G44.492$+$0.717 & -73.1 &   1.8 &   3.2 &   1.7 &    1.4 &   18.8 &   14.1 &    0.2 &    2.8 &    3.4 &  19.2 &  14.2 \\
MWISP G44.533$+$0.525 & -69.8 &   1.4 &   3.0 &   2.1 &    2.8 &   18.3 &   13.7 &    0.2 &    4.4 &    6.5 &  17.3 &  10.6 \\
MWISP G44.575$+$2.608 & -73.2 &   1.0 &   1.5 &   1.4 &    1.7 &   18.8 &   14.1 &    0.9 &    3.2 &    2.6 &   6.7 &   8.1 \\
MWISP G44.608$+$0.650 & -73.2 &   3.1 &   6.3 &   1.9 &    2.4 &   18.8 &   14.1 &    0.2 &    4.2 &    9.6 &  85.0 &  17.5 \\
MWISP G44.708$+$0.967 & -71.1 &   1.0 &   2.5 &   2.3 &    2.0 &   18.4 &   13.8 &    0.3 &    3.5 &    3.2 &   7.4 &   8.2 \\
MWISP G44.758$+$0.392 & -66.4 &   1.3 &   2.4 &   1.7 &    1.3 &   17.7 &   13.2 &    0.1 &    2.5 &    1.9 &   9.2 &   9.8 \\
MWISP G44.800$+$0.658 & -62.1 &   2.6 &  14.4 &   5.2 &    6.0 &   17.1 &   12.7 &    0.2 &    6.5 &   31.1 &  91.8 &  23.2 \\
MWISP G44.950$+$2.417 & -67.2 &   1.4 &   2.6 &   1.8 &    2.6 &   17.8 &   13.3 &    0.7 &    4.1 &    4.2 &  15.9 &   8.0 \\
MWISP G45.208$+$0.958$^{\dag}$ & -58.9 &   1.9 &   4.4 &   2.2 &    5.5 &   16.6 &   12.3 &    0.3 &    6.1 &    9.2 &  46.7 &   8.0 \\
\enddata
\tablecomments{
Column (1): source named by the MWISP project and the peak position of 
$I_{\rm ^{12}CO}$ in Galactic Coordinates.
Sources marked with * are also listed in \citet{2016ApJ...828...59S}. Sources marked with $\dag$ 
with $V_{\rm LSR}>$-1.6 $l$ + 13.2~\kms probably lie in the Outer arm. 
Columns (2)-(5): the single Gaussian fitting results of the brightest CO spectrum of the MCs.
Column (6): the solid angle defined by the 3$\sigma$ limits. Columns (7)-(8): heliocentric distance and Galactocentric radius
of the cloud, assuming the rotation curve and Galactic parameters obtained by \citet{2014ApJ...783..130R}.
Column (9): scale height, $Z$, $Z~=~D~sin(b)$. Column (10): equivalent radii of the molecular clouds corrected by the beam size of the telescope. 
Column (11): cloud mass calculated using $X=2.0\times10^{20}$cm$^{\rm -2}$(K km s$^{\rm -1}$)$^{\rm -1}$ \citep{2013ARAA..51..207B}.
Column (12): Viral mass. Column (13): Mass surface density.}
\end{deluxetable}
\begin{deluxetable}{rrr}
\tabletypesize{\tiny}
 \setlength{\tabcolsep}{0.03in}
 \tablewidth{0pt} \tablecaption{Molecular clouds marginally detected by this study.\label{tab:margin}}
 \tablehead{Name ($V_{\rm lsr}$)&Name ($V_{\rm lsr}$)&Name ($V_{\rm lsr}$)\\
            \kms & \kms & \kms }
\startdata
G35.400$+$0.225(-56.0)& G36.667$-$0.133(-59.5)& G37.600$+$1.450(-52.2)\\
G37.933$+$1.308(-48.1)& G37.942$+$1.308(-49.4)& G38.042$+$1.325(-48.9)\\
G38.050$+$1.275(-48.9)& G39.133$+$3.442(-57.6)& G39.725$+$1.317(-55.6)\\
G39.900$+$2.683(-60.0)& G40.558$+$1.342(-55.4)& G40.875$+$1.233(-57.0)\\
G40.917$+$1.150(-57.9)& G40.917$+$0.492(-56.2)& G41.158$+$0.233(-57.5)\\
G41.433$+$2.017(-59.1)& G41.525$+$1.133(-57.8)& G41.533$+$1.058(-57.6)\\
G41.717$+$0.925(-58.6)& G42.892$+$2.242(-67.8)& G43.008$+$2.017(-66.4)\\
G44.308$-$0.825(-68.9)& G44.725$+$1.983(-68.6)&\\
\enddata
\end{deluxetable}
\begin{deluxetable}{lcccccc}
\tabletypesize{\tiny}
 \setlength{\tabcolsep}{0.03in}
\tablewidth{0pt} \tablecaption{Parameters of molecular clouds derived by $^{13}$CO(1-0).\label{tab:13co}}
\tablehead{ Name & $V_{\rm LSR}$  &$\Delta$$V$ & $I_{\rm ^{13}CO}$   & $T_{\rm peak}$  &$I_{\rm ^{12}CO}$/$I_{\rm ^{13}CO}$ & $\sigma_{rms}$ \\
    & (\kms) &(\kms) & (K.km$\,$s$^{-1}$) & (K)& &(K)  \\
(1)&(2)&(3)&(4)&(5)&(6)&(7)}
\startdata
MWISP G34.842$-$0.950 & -45.0(0.09)&   2.3(0.44)&   1.3(0.25)&   0.6 &   4.4 &   0.29\\
MWISP G37.175$+$0.792$\lhd$ & -48.1(0.15)&   3.6(0.35)&   1.1(0.11)&   0.3 &   ... &   0.10\\
MWISP G37.225$+$0.767$\lhd$ & -50.0(0.10)&   2.1(0.29)&   1.5(0.20)&   0.7 &   9.2 &   0.24\\
MWISP G37.350$+$1.050 & -53.8(0.08)&   1.9(0.23)&   2.7(0.25)&   1.3 &   7.7 &   0.27\\
MWISP G37.408$+$1.183 & -53.8(0.10)&   0.9(0.29)&   0.6(0.14)&   0.6 &   5.8 &   0.22\\
MWISP G37.425$+$1.267 & -51.9(0.13)&   1.4(0.31)&   1.1(0.21)&   0.7 &   7.4 &   0.28\\
MWISP G37.642$+$1.400 & -47.4(0.06)&   0.9(0.21)&   0.3(0.07)&   0.3 &   ... &   0.12\\
MWISP G37.667$+$1.267 & -57.8(0.08)&   2.3(0.22)&   2.8(0.25)&   1.1 &   5.3 &   0.26\\
MWISP G37.683$+$0.533 & -52.7(0.10)&   1.8(0.38)&   1.2(0.22)&   0.6 &   3.4 &   0.27\\
MWISP G37.708$+$1.367 & -51.6(0.04)&   1.7(0.17)&   2.2(0.20)&   1.3 &   7.1 &   0.26\\
MWISP G37.742$+$1.258 & -54.5(0.08)&   1.3(0.40)&   0.9(0.20)&   0.6 &   4.0 &   0.25\\
MWISP G37.817$+$0.092 & -47.9(0.05)&   1.6(0.29)&   1.1(0.18)&   0.7 &   6.7 &   0.23\\
MWISP G38.092$+$0.458 & -47.9(0.08)&   2.5(0.48)&   0.7(0.13)&   0.3 &   ... &   0.14\\
MWISP G38.317$+$0.908 & -49.2(0.06)&   1.7(0.40)&   1.2(0.21)&   0.6 &  10.7 &   0.25\\
MWISP G38.492$+$1.200 & -50.5(0.10)&   0.5(0.13)&   0.7(0.13)&   1.2 &   7.0 &   0.27\\
MWISP G38.533$+$0.892 & -54.8(0.14)&   0.9(0.21)&   0.6(0.13)&   0.6 &   8.4 &   0.24\\
MWISP G39.175$-$1.425 & -55.4(0.09)&   3.7(0.43)&   2.6(0.25)&   0.7 &   4.3 &   0.21\\
MWISP G39.492$+$0.667 & -49.0(0.07)&   0.9(0.28)&   0.8(0.19)&   0.9 &  14.3 &   0.28\\
MWISP G39.908$+$1.058 & -50.7(0.09)&   2.0(0.41)&   0.6(0.10)&   0.3 &   ... &   0.12\\
MWISP G39.983$+$0.908 & -52.1(0.05)&   0.9(0.16)&   0.8(0.14)&   0.8 &   5.6 &   0.25\\
MWISP G40.092$+$2.758 & -62.5(0.16)&   1.1(0.18)&   0.9(0.15)&   0.8 &   4.7 &   0.26\\
MWISP G40.292$+$1.150 & -50.5(0.10)&   1.8(0.27)&   1.5(0.22)&   0.8 &   8.2 &   0.27\\
MWISP G40.317$+$1.542 & -55.7(0.12)&   3.5(0.59)&   2.4(0.33)&   0.7 &   4.2 &   0.28\\
MWISP G40.392$+$1.550 & -61.4(0.04)&   2.3(0.52)&   1.3(0.25)&   0.5 &   2.8 &   0.26\\
MWISP G40.533$+$1.267 & -50.0(0.07)&   4.5(0.68)&   1.4(0.18)&   0.3 &   ... &   0.13\\
MWISP G40.592$+$1.517 & -58.4(0.12)&   1.6(0.32)&   1.2(0.21)&   0.7 &   5.4 &   0.27\\
MWISP G40.633$+$3.117 & -55.8(0.04)&   0.9(0.18)&   0.4(0.07)&   0.4 &   ... &   0.13\\
MWISP G40.833$+$1.925 & -61.0(0.09)&   1.0(0.22)&   1.0(0.19)&   0.9 &   6.2 &   0.29\\
MWISP G40.883$+$0.492 & -54.9(0.04)&   2.7(0.45)&   1.7(0.27)&   0.6 &   5.2 &   0.28\\
MWISP G40.892$+$1.308 & -57.1(0.18)&   1.5(0.34)&   1.1(0.23)&   0.7 &   7.6 &   0.30\\
MWISP G40.942$+$2.542 & -53.8(0.12)&   0.9(0.22)&   0.3(0.06)&   0.3 &   ... &   0.11\\
MWISP G41.067$+$2.683 & -63.4(0.13)&   1.5(0.32)&   0.6(0.10)&   0.4 &   ... &   0.13\\
MWISP G41.367$+$1.892 & -59.7(0.09)&   1.0(0.14)&   0.4(0.06)&   0.3 &   ... &   0.11\\
MWISP G43.292$+$1.742 & -56.5(0.15)&   0.8(0.16)&   0.4(0.07)&   0.5 &   ... &   0.12\\
MWISP G44.800$+$0.658 & -61.9(0.05)&   2.5(0.35)&   2.4(0.29)&   0.9 &   6.1 &   0.29\\
\enddata
\tablecomments{Column (1): source named by the MWISP project and the peak position of 
$I_{\rm ^{13}CO}$ in Galactic Coordinates.
Sources marked with $\lhd$ are belong to the same cloud MWISP G37.175$+$0.792. 
Columns (2)-(5): the single Gaussian fitting results of the brightest \xco~ spectrum 
of the MCs. Column (6): integrated intensity ratio of \co~ and \xco.
Column (7): the rms level of each detection with a velocity resolution of 0.166~\kms~
(those with values larger than 0.2~K are the rms value of a single pixel,  otherwise
the rms was averaged over nine pixels).}
\end{deluxetable}
\end{document}